\documentclass[11pt,a4paper]{article}
\usepackage[T1]{fontenc}

\usepackage{amsmath,amssymb,amsthm,bm}
\usepackage{microtype}
\usepackage[margin=25mm]{geometry}
\usepackage{booktabs,tabularx,enumitem,multirow,longtable}
\usepackage{multicol,needspace}
\usepackage[round,authoryear]{natbib}
\usepackage{xcolor}
\usepackage{algorithm,algpseudocode}
\usepackage[colorlinks=true,linkcolor=blue!45!black,citecolor=blue!45!black,urlcolor=blue!45!black]{hyperref}
\usepackage{graphicx}
\usepackage[section]{placeins}

\newcommand{\ii}{\mathrm{i}}
\newcommand{\E}{\mathbb E}

\newcommand{\Qbar}{\bar Q}

\newcommand{\WSname}{ANS}
\newcommand{\WSsuffix}{+ANS}
\newcommand{\vEtaNullSLong}{0.050}
\newcommand{\vEtaNullSShort}{0.100}
\newcommand{\vEtaNullTSX}{0.066}

\newcommand{\vPsCmpFaMax}{0.60}
\newcommand{\vPsCmpFaMin}{0.40}
\newcommand{\vRtChainSaveMax}{32}
\newcommand{\vRtChainSaveMin}{14}
\newcommand{\vRtCpu}{Intel Xeon E5-2697 v3}
\newcommand{\vRtCpuB}{Intel Core i7-9700}
\newcommand{\vRtCpuC}{AMD Ryzen 7 4800H}
\newcommand{\vRtMaskVispCentres}{75}
\newcommand{\vRtMaskVispSeconds}{0.35}

\newcommand{\vSceneCohHigh}{0.95}
\newcommand{\vSceneCohLow}{0.50}
\newcommand{\vSceneHeightPhase}{0.37}
\newcommand{\vSceneHeightPhaseRatio}{3.10}
\newcommand{\vaAbsExcessMax}{29}
\newcommand{\vaAbsExcessMin}{13}
\newcommand{\vaAbsGain}{11}
\newcommand{\vaEtaGsqDev}{0.06}

\newcommand{\vaccCmpVisp}{19}
\newcommand{\vaccFaVisp}{42}

\newcommand{\vcCoreSOne}{55}
\newcommand{\vcCoreTsx}{31}

\newcommand{\vcEvalCentresMax}{135,460}
\newcommand{\vcEvalCentresMin}{134,480}
\newcommand{\vcEvalScenes}{2000}
\newcommand{\vcFigCmpFaMaxDuo}{8.1}
\newcommand{\vcFigFullFaMaxDuo}{15}

\newcommand{\vcGuardSOne}{277}
\newcommand{\vcGuardTsx}{105}

\newcommand{\vcKappa}{0.5}
\newcommand{\vcLambda}{3}

\newcommand{\vcNqHeldBetweenDuoMax}{56}
\newcommand{\vcNqHeldBetweenDuoMin}{42}

\newcommand{\vcNqRealN}{16}

\newcommand{\vcRejFaPubHighMax}{27}
\newcommand{\vcRejFaPubHighMin}{10}
\newcommand{\vcRejFaPubHighSLong}{27}
\newcommand{\vcRejFaPubHighSShort}{10}

\newcommand{\vcRejHomFaMax}{1.8}
\newcommand{\vcRejHomFaMin}{0.6}
\newcommand{\vcRejHomFaPubMax}{27}
\newcommand{\vcRejHomFaPubMin}{1.9}
\newcommand{\vcSettingsMaxDuo}{6.7}

\newcommand{\vcTuneCentresMax}{40,520}
\newcommand{\vcTuneCentresMin}{39,950}

\newcommand{\vcTuneScenes}{600}
\newcommand{\vcUnitScenes}{55}

\newcommand{\vdCompSceneSurfNei}{19}

\newcommand{\vdEstSoneAzMax}{0.61}
\newcommand{\vdEstSoneAzMin}{0.50}
\newcommand{\vdEstVispAz}{0.68}
\newcommand{\vdEstVispRg}{0.62}

\newcommand{\vdNatFloorMed}{0.32}

\newcommand{\vdPredSoneAzOne}{0.66}

\newcommand{\vdPredSoneRgOne}{0.82}

\newcommand{\vdSimFloor}{0.43}
\newcommand{\vdSimFloorIndep}{0.05}
\newcommand{\vdSoneLayMaskMax}{63}
\newcommand{\vdSoneLayMaskMin}{31}

\newcommand{\vdSoneNeiMaskMax}{13.8}
\newcommand{\vdSoneNeiMaskMin}{6.3}
\newcommand{\vdSoneShMaskMax}{61}
\newcommand{\vdSoneShMaskMin}{33}

\newcommand{\vdTownMaskedInSurfSh}{63}

\newcommand{\vdTownSurfShArea}{31}

\newcommand{\vdVispLayMaskTwentyFive}{56}

\newcommand{\vdVispMasked}{3.7}
\newcommand{\vdVispMode}{0.43}

\newcommand{\vdVispShMaskTwentyFive}{3.9}

\newcommand{\vdVispZero}{0.9}
\newcommand{\vfDispSLong}{27.7}
\newcommand{\vfDispSShort}{27.7}
\newcommand{\vfDispTSX}{15.5}
\newcommand{\vfHalfBoxAz}{3}
\newcommand{\vfHalfBoxRg}{27}
\newcommand{\vfLandLongDays}{288}
\newcommand{\vfLandShortDays}{12}
\newcommand{\vfRateSLong}{9.4}
\newcommand{\vfRateSShort}{35.2}
\newcommand{\vfRateTSX}{3.0}
\newcommand{\vfSOneMetAz}{98}
\newcommand{\vfSOneMetRg}{59}
\newcommand{\vfSOnePixAz}{7}
\newcommand{\vfSOnePixRg}{29}
\newcommand{\vfSOneRingMax}{12}
\newcommand{\vfSOneRingMin}{4}
\newcommand{\vfSlowBandMed}{1.6}

\newcommand{\vfSpanSLong}{2.96}
\newcommand{\vfSpanSShort}{0.79}
\newcommand{\vfSpanTSX}{5.24}
\newcommand{\vfTsxMetAz}{26}
\newcommand{\vfTsxMetRg}{25}
\newcommand{\vfTsxPixAz}{13}
\newcommand{\vfTsxPixRg}{13}
\newcommand{\vfTsxRingMax}{4.1}
\newcommand{\vfTsxRingMin}{3}

\newcommand{\vgEtaMidTSX}{0.22}
\newcommand{\vgEtaSqrtNMax}{0.45}
\newcommand{\vgEtaSqrtNMin}{0.44}

\newcommand{\vgRmsGMid}{0.28}

\newcommand{\vhRealBig}{19}

\newcommand{\vhTruncFrDuoMax}{8.46}
\newcommand{\vhTruncFrDuoMin}{2.57}

\newcommand{\vkHRatioMax}{0.58}
\newcommand{\vkHRatioMin}{0.56}

\newcommand{\vkNullSLongMed}{0.54}
\newcommand{\vkNullSLongPixels}{230,010}
\newcommand{\vkNullTSXBelowSci}{4\times10^{-4}}
\newcommand{\vkNullTSXMed}{0.43}
\newcommand{\vkNullTSXQmil}{0.26}
\newcommand{\vkNullTSXQone}{0.31}

\newcommand{\vmSelAdvWSDuoMax}{3.9}
\newcommand{\vmSelAdvWSDuoMin}{0.7}

\newcommand{\vnMin}{100}

\newcommand{\vpPredTsxAz}{0.695}
\newcommand{\vpPredTsxRg}{0.606}
\newcommand{\vpUsedSOneAz}{0.664}
\newcommand{\vpUsedSOneRg}{0.817}
\newcommand{\vpUsedTsxAz}{0.695}
\newcommand{\vpUsedTsxRg}{0.606}

\newcommand{\vpaPerBin}{120}

\newcommand{\vpaScenes}{30}

\newcommand{\vrDecMedWindow}{3835}
\newcommand{\vrDecMedianEta}{0.13}

\newcommand{\vrSlowMedFa}{3258}
\newcommand{\vrSlowMedFaWS}{1218}

\newcommand{\vrVispBelow}{61}

\newcommand{\vsAdjRiseMax}{0.061}
\newcommand{\vsAdjZeroMin}{0.25}
\newcommand{\vsAdjZeroSLong}{0.51}
\newcommand{\vsPsCentresMax}{13}
\newcommand{\vsPsCentresMin}{12}
\newcommand{\vsPsWeight}{0.4}

\newcommand{\vtAbsFrFullDuoLow}{42}
\newcommand{\vtAbsFrSDuoUp}{10}

\newcommand{\vtChainFrGainDuoMax}{4.8}
\newcommand{\vtChainFrGainDuoMin}{2.2}
\newcommand{\vtChainGainAllDuoMax}{31}
\newcommand{\vtChainGainAllDuoMin}{21}

\newcommand{\vtGatedZeroDuoMax}{90}
\newcommand{\vtGatedZeroDuoMin}{74}

\newcommand{\vtPerHigherDuoMax}{17}
\newcommand{\vtPerHigherDuoMin}{13}

\newcommand{\vtRhoMin}{0.025}

\newcommand{\vtSGainAllDuoMax}{30}
\newcommand{\vtSGainAllDuoMin}{22}
\newcommand{\vtScaledGainAllDuoMax}{20}
\newcommand{\vtScaledGainAllDuoMin}{14}

\newcommand{\vtTailWSSLong}{9.7}
\newcommand{\vtTailWSTSX}{13}
\newcommand{\vtTie}{0.1}

\newcommand{\vtWinExcessPcNSLongMin}{13}
\newcommand{\vtWinExcessPcNTSXMax}{31}

\newcommand{\vuCostSUp}{4}

\newcommand{\vwMainWSBestNDuoMax}{11}
\newcommand{\vwMainWSBestNDuoMin}{9.6}

\newcommand{\vwSelSpreadHigh}{4.6}
\newcommand{\vwSelSpreadPsNDuoLow}{129}

\newcommand{\vwWinSpreadLow}{13}
\newcommand{\vwWinSpreadPsNDuoUp}{27}

\newcommand{\vxDistCalPcMax}{3}
\newcommand{\vxDistIndDuoMax}{1.94}
\newcommand{\vxDistIndDuoMin}{1.20}

\newcommand{\vxNminChangeDuoMax}{0.4}

\newcommand{\vxScoreScenes}{100}

\let\oldappendix\appendix
\renewcommand{\appendix}{\oldappendix
  \counterwithin{equation}{section}\counterwithin{table}{section}%
  \counterwithin{figure}{section}\counterwithin{algorithm}{section}%
  \renewcommand{\theHtable}{app.\thetable}\renewcommand{\theHfigure}{app.\thefigure}%
  \renewcommand{\theHequation}{app.\theequation}\renewcommand{\theHalgorithm}{app.\thealgorithm}}

\title{Spatial phase coherence and its applications to masking and adaptive neighborhood sizing in multitemporal InSAR}
\author{Magnus Heimpel\textsuperscript{a,*}, Irena Hajnsek\textsuperscript{a,b},
and Othmar Frey\textsuperscript{a,c}\\[0.6em]
\small\textsuperscript{a}Institute of Environmental Engineering, ETH Zurich,\\
\small Laura-Hezner-Weg 7, 8093 Zurich, Switzerland\\
\small\textsuperscript{b}Microwaves and Radar Institute, German Aerospace Center (DLR),\\
\small M\"unchener Stra\ss e 20, 82234 We\ss ling, Germany\\
\small\textsuperscript{c}Gamma Remote Sensing AG, Worbstrasse 225, 3073 G\"umligen, Switzerland\\[0.3em]
\small\textsuperscript{*}Corresponding author. E-mail: \href{mailto:mheimpel@ethz.ch}{mheimpel@ethz.ch} (M.~Heimpel),\\
\small \href{mailto:hajnsek@ifu.baug.ethz.ch}{hajnsek@ifu.baug.ethz.ch} (I.~Hajnsek), \href{mailto:frey@ifu.baug.ethz.ch}{frey@ifu.baug.ethz.ch} (O.~Frey)}
\date{November 15, 2026}

\begin{document}
\maketitle

\begin{abstract}
We examine spatial phase coherence (SPC), a measure of the correlation between the wrapped phase time series of two
pixels in a stack of single-look synthetic aperture radar acquisitions. A pixel and its adjacent neighbors share
speckle through the point spread function, so that decorrelation alone does not lower their SPC below a floor. We find
that masking pixels whose mean SPC to their eight neighbors is well below that floor, as in parts of layover and shadow
areas, is an inexpensive complement to geometry-derived layover and shadow masks. For distributed scatterer pixels,
averaging SPC over a fixed set of neighbors that share no speckle yields a proxy for the local coherence, the phase
coherence indicator (PCI), that is available before the selection of statistically homogeneous pixels (SHPs). In phase
estimation, averaging more pixels reduces noise but may also increase the bias in heterogeneous areas. To relax this
trade-off, we propose to enhance SHP selection algorithms by scaling the search window size according to the PCI and
truncating the set of selected pixels where a smaller nested set agrees clearly better with the phases at the center.
In simulations of a TerraSAR-X and a Sentinel-1 stack with 44 and 79 acquisitions, we apply this adaptive neighborhood
sizing (ANS) to established amplitude- and covariance-based SHP selection algorithms with thresholds that adapt to the
PCI. Without ANS, the largest tested search window size has \vaAbsExcessMin{} to \vaAbsExcessMax\,\% more phase
linking error than the best of the tested sizes. With ANS, the largest window has the lowest error of the tested sizes,
\vaAbsGain\,\% lower than that of the same selection at its best tested size without ANS. On a real TerraSAR-X stack,
the mask based on SPC flags pixels in building shadows that a terrain model misses. On real Sentinel-1 stacks, ANS
preserves dense fringes on a slowly deforming slope and in part on a fast landslide.
\end{abstract}

\section{Introduction}
\label{sec:introduction}

In multitemporal interferometric synthetic aperture radar (InSAR), phase estimation and phase linking over distributed scatterers rely on averaging observations from selected
neighboring pixels \citep{minh2023interferometric}. This raises three questions for each pixel to be estimated, namely whether an average over its
neighbors can represent it at all, which neighbors are statistically homogeneous with it, and how large the set of
averaged pixels should be. We call the pixel to be estimated the center pixel, or center, and the set of averaged
pixels the ensemble.

The first question arises in particular for pixels in layover and shadow areas, which are typically masked. In
layover, scatterers at several heights are projected onto the same pixel in the synthetic aperture radar (SAR)
image, so that subtracting a topographic phase may introduce
systematic phase inconsistencies. In shadow, the phase is often dominated by noise that is less correlated between
neighboring pixels than speckle. Geometric layover and shadow maps derived from the acquisition geometry and a
digital elevation model (DEM) \citep{kropatsch1990generation} can identify these areas, but they depend on the
accuracy of the DEM. Layover and shadow caused by objects that are not included in the model, such as buildings and
trees, are another limitation. Layover has also been detected from the data itself.
\citet{tupin1996improving} use the amplitude, \citet{zhu2017highrise} the local fringe frequency of an interferogram
over high-rise buildings, and \citet{rossi2015highresolution} the number of times each SAR pixel occurs in the
geocoding of an elevation model generated from the interferometric data, without external data.
\citet{tupin1996improving} also detect areas without topographic phase from the filtered coherence and the confidence
of the local fringe frequency estimates of an interferogram. These areas include shadow, decorrelated areas and
surfaces without backscatter.

Established processing frameworks such as SqueeSAR \citep{ferretti2011new} identify statistically
homogeneous pixels (SHPs) by tests on the temporal amplitude distributions
\citep{even2018insar,parizzi2011adaptive}. Later tests also use the interferometric phases
\citep{dong2018unified,song2017statistically,zhao2022statistically}. Other approaches collect pixels of similar
intensity by region growing \citep{vasile2006intensity} or weight pixels by the similarity of their surrounding
patches \citep{deledalle2011nlinsar,deledalle2015nl}. SHP tests that compare the temporal mean
amplitudes or intensities
of two pixels lose their significance level when the speckle is temporally correlated. On a coherent target the
speckle changes little between acquisitions, so that the means of two pixels of the same surface keep the difference
of their speckle. A threshold set for independent samples or for one assumed coherence therefore holds only at that
coherence.

Averaging reduces noise, but its extent determines how much spatial
detail is retained. Strongly decorrelated targets need larger ensembles
to reach a given precision \citep[cf.][Eqs.~(5.22) to (5.25)]{rocca2007modeling}.
Where the phase varies rapidly in space, as in areas with strong spatial
gradients of line-of-sight deformation, a smaller ensemble may suffice if the coherence is high enough,
whereas a large one blurs fringes and lowers the estimated coherence
\citep{zebker2005accurate}. The averaging neighborhood has been adapted to this in shape and size. For single
interferograms, directional windows follow the local fringe orientation \citep{lee1998new}, and the intersection of
confidence intervals (ICI) rule chooses, for each pixel, the largest
window whose local polynomial estimate stays consistent with those of
smaller windows \citep{katkovnik2002adaptive,katkovnik2008phase}. For stacks,
\citet{pepe2021adaptive} adapt a fixed multilook window to the phase by removing the
pixels whose phases deviate from the multilooked phases of that
window and comparing the deviations of the remaining pixels with those of the pixel that deviates least. In other SHP-based processing, however, the
coherence matrix is usually estimated from all pixels that the SHP selection accepts \citep[e.g.,][]{jiang2015fast,zhao2022statistically}, also where the phase varies within the search window.
\citet[Section~3.2 and Eq.~(13)]{chi2022distributed} instead shrink the search window of an SHP selection at each pixel until a coefficient of variation of the phase of the
time-averaged differential interferogram falls below a threshold, which requires unwrapped interferograms.

In this paper, we examine a simple quantity, the squared magnitude of the mean phasor of the relative phase of two pixels across acquisitions, which we refer to as spatial phase coherence (SPC), and its use for all
three questions. The quantity itself is known under
several names (Section~\ref{sec:related}). It is used in directional statistics
\citep{mardia2000directional} and in InSAR for persistent scatterers \citep{ferretti2000nonlinear}, and in related forms for phase stability
\citep{hooper2004new,hooper2007persistent}, point coherence estimation \citep{vecchioli2023novel} and SHP selection
\citep{narayan2018novel,pepe2021adaptive,shi2020potential}.
Less attention has so far been paid to the behavior of the statistic on
single-look data with spatially correlated speckle, and to the uses of its
properties in the processing chain, from masking to the size of the ensemble that an SHP selection returns. Our contributions are as follows.
\begin{enumerate}[leftmargin=*,itemsep=1pt]
\item A statistical model of SPC on single-look data, with (a) its
expectation under the circular complex Gaussian (CCG) model, (b) its floor
under speckle that adjacent pixels share through the system
point spread function, predicted from the processing parameters,
and (c) an identity that links pairwise SPC to the mean squared phase-only
coherence of an ensemble.
\item Two maps, the neighborhood SPC (NSPC) and the phase coherence indicator (PCI), computed from fixed sets of
neighbors at the cost of only one inner product over the
acquisitions per neighbor, and available directly from the single-look complex (SLC) stack. A threshold on NSPC, the average SPC of a pixel with its eight direct neighbors, flags
pixels where this average lies below the floor that shared speckle sets, and
complements geometric layover and shadow maps. The PCI, an average of SPC over
neighbors that share no speckle with the pixel, rises with its coherence and steers the SHP selection and the sizing of the ensemble, as described below.
\item Thresholds for SHP tests that compare temporal mean amplitudes or intensities. They depend on the PCI and
keep the false rejection rate of the order of the significance level at every coherence, also where the
speckle of single-look data is temporally correlated.
\item Adaptive neighborhood sizing (\WSname) in two parts: a search window scaled with the
PCI (S), and a truncation step (T) that reduces a selected ensemble to a smaller nested
one only where that one agrees clearly better with the phases of the center and of the pixels that share its
speckle. The agreement is measured with a generalization of SPC, and \WSname{} can be combined with any SHP selection.
\end{enumerate}
The two maps, the thresholds and \WSname{} are evaluated in simulations with known phase and demonstrated on
TerraSAR-X data over Visp,
Switzerland, and Sentinel-1 data over Gilgit-Baltistan, Pakistan. \WSname{} is combined with three kinds of SHP selection: (a) the
amplitude-based fast SHP selection (FaSHPS) of \citet{jiang2015fast}, which tests the temporal amplitudes of two pixels, (b) the intensity- and phase-based covariance matrix patch (CMP) selection of
\citet{zhao2022statistically} and (c), as their reference, all pixels of the search window (the full window). In Section~\ref{sec:theory}, we develop
the theory, and in Section~\ref{sec:selection}, we define the mask, the PCI, the SHP selections, the two parts of
\WSname{} and the estimator of the coherence matrix. In Sections~\ref{sec:simulation} and~\ref{sec:realdata}, we
present the simulations and the real data, including the runtime. In Section~\ref{sec:discussion}, we discuss the
applicability of the two maps and of \WSname, the variants not adopted, and limitations.

\section{Spatial phase coherence}
\label{sec:theory}

\subsection{Definition}
\label{sec:definition}
\begin{figure}[tbp]
\centering
\includegraphics[width=\linewidth]{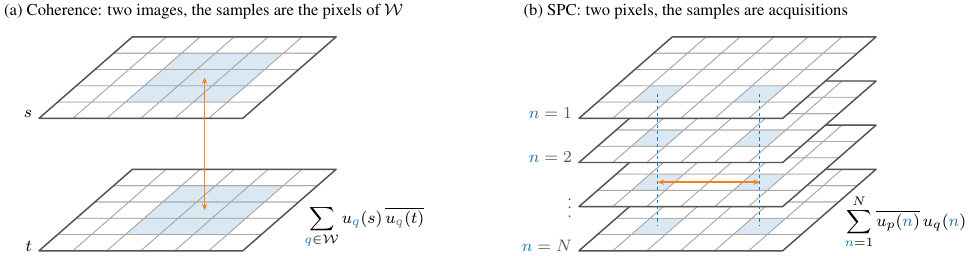}
\caption{Illustration of phase-only coherence and SPC on a stack of images.
(a) The phase-only coherence (Eq.~\eqref{eq:phaseonlycoh}) of two acquisitions
$s,t$ is built from the sum of the products $u_q(s)\overline{u_q(t)}$ over
the pixels $q$ of a neighborhood $\mathcal W$, the same in both images.
(b) SPC (Eq.~\eqref{eq:spc}) of two pixels $p,q$ is built from the sum of the
products $\overline{u_p(n)}u_q(n)$ over the $N$ acquisitions of the stack.}
\label{fig:roles}
\end{figure}

Consider a stack of $N>1$ coregistered SLC images in radar coordinates, and let $x_p(n)$ denote the complex value of pixel $p$ in
acquisition $n$. A pixel is identified with its integer image
coordinates (azimuth, range), so that $q-p$ denotes the spatial lag
from pixel $p$ to pixel $q$. A model phase, such as a simulated topographic
phase, may have been subtracted beforehand
(Section~\ref{sec:model}). For nonzero observations, define
\begin{equation}
 u_p(n)=\frac{x_p(n)}{|x_p(n)|}=e^{\ii\phi_p(n)},\qquad
 \bm u_p=(u_p(1),\ldots,u_p(N))^{\mathsf T}.
\end{equation}
The coherence of two acquisitions $s,t$ is the normalized covariance
of their complex values \citep[Eq.~(2)]{even2018insar}. It is estimated
by the sample coherence over a neighborhood
$\mathcal W$ \citep[Eq.~(8)]{even2018insar}, whose phase-only
counterpart is
\begin{equation}\label{eq:phaseonlycoh}
 \widehat\gamma^{\circ}_{st}(\mathcal W)
 =
 \left|\frac1{|\mathcal W|}
 \sum_{q\in\mathcal W}u_q(s)\overline{u_q(t)}\right|.
\end{equation}
Exchanging the roles of pixels and acquisitions in
Eq.~\eqref{eq:phaseonlycoh}, i.e., fixing two pixels $p,q$ and averaging
over the acquisitions (Fig.~\ref{fig:roles}), gives the SPC of the two pixels, which we define as the square of the
resulting magnitude,
\begin{equation}\label{eq:spc}
 S_{pq}=\left|\frac1N\sum_{n=1}^N \overline{u_p(n)}u_q(n)\right|^2
 =\frac{|\bm u_p^H\bm u_q|^2}{N^2}.
\end{equation}
It satisfies $0\le S_{pq}\le1$, with $S_{pq}=1$ precisely when the two
phase vectors differ by a constant offset. SPC is invariant to
constant phase rotations of either pixel and to any
acquisition-dependent rotation common to both. Referencing all
phases to one acquisition is a rotation of the first kind and
leaves SPC unchanged, provided that the reference acquisition,
whose phase is then zero at every pixel, is kept as one of the $N$
terms of the sum. Like Eq.~\eqref{eq:phaseonlycoh}, SPC uses the phases only, as do the estimates of the coherence
matrix in this paper (Section~\ref{sec:estimate}). Exchanging
the roles in the sample coherence instead would give the correlation coefficient of the two complex time series.
Bright samples and acquisitions would dominate it, and its expectation, the floor of Section~\ref{sec:null} and the
thresholds derived from it would depend on amplitude changes over the acquisitions, e.g., from soil moisture or
vegetation. SPC does not change under any positive scaling of the samples, and the amplitudes enter the processing
only through the SHP tests (Sections~\ref{sec:amptests} and~\ref{sec:pairwise}).

If the relative phases $\phi_q(n)-\phi_p(n)$ are independent across the acquisitions and uniformly distributed,
$\E[S_{pq}]=1/N$. Removing this bias and rescaling gives the
corrected SPC and its clipped version
\begin{equation}\label{eq:q}
 Q_{pq}=\frac{NS_{pq}-1}{N-1},\qquad Q_{pq}^+=\max(Q_{pq},0),
\end{equation}
respectively. Where the distinction does not matter, we also call $Q_{pq}$, $Q_{pq}^+$ and their square roots SPC.

\subsection{Relation to existing measures}
\label{sec:related}
The quantities in Eqs.~\eqref{eq:spc} and \eqref{eq:q} are known in InSAR
and in other scientific fields under different names. Section~\ref{app:related-measures} of the supplement gives the
details of the relations below. In directional statistics, testing $S_{pq}$ against a
threshold is the Rayleigh test of uniformity, applied here to the
relative phases \citep[Section~6.3.1, Eqs.~(6.3.3) and (6.3.4)]{mardia2000directional}. In InSAR, with the phase model of the multi-image coherence set to zero, $\sqrt{S_{pq}}$ has the form
of the multi-image phase coherence of a pair of neighboring persistent
scatterers $p$ and $q$ \citep[Eq.~(2.7)]{ferretti2000nonlinear}, summed
over the acquisitions instead of single-reference interferograms. \citet{vecchioli2023novel} estimate the coherence
of each point of a stack from such coherences of pairs of nearby points, computed after a motion model is fitted to
each pair. They assume that the phase noise of two points is independent, which may require leaving out the nearest
pixels of oversampled images. The expected coherence of a pair is then the product of the coherences of its two
points, and the logarithms of these products form an overdetermined linear system for the coherence of every point.
The phase-stability measure of \citet[Eq.~(20)]{hooper2007persistent} applies the same construction, the magnitude of a mean phasor, to the phases of one pixel relative to the
filtered phase of the surrounding pixels, after a fitted look angle error is removed. For SHP selection, the zero-lag phase cross-correlation of \citet[Eq.~(6)]{narayan2018novel} has magnitude
$N_I\sqrt{S_{pq}}$ when $S_{pq}$ is computed over the $N_I$ single-reference interferograms that it uses,
instead of over the acquisitions. The magnitude of the centered complex correlation of \citet[Eq.~(1)]{shi2020potential} differs from
$\sqrt{S_{pq}}$ in general and coincides with it when both temporal phasor means vanish. The circular variance used for adaptive multilooking by
\citet[Eqs.~(9) and (10)]{pepe2021adaptive} is one minus the magnitude
of a mean phasor, taken over the interferograms and computed between the interferometric phases of a pixel
and the multilooked phases of the window rather than those of a
second pixel. The SPC formula is therefore not a contribution of this paper. The
contributions concern its behavior on single-look SAR data and its
use for masking, for an indicator of local coherence that steers the SHP selection and for the size
of an ensemble. Its statistics have also been examined in neuroscience: $Q_{pq}$ is the pairwise phase consistency
of \citet[Eqs.~(14) and (15)]{vinck2010pairwise}, an unbiased estimator of squared phase synchrony between neuronal signals recorded over
repeated trials, and $\sqrt{S_{pq}}$ is the phase-locking value of \citet[p.~195]{lachaux1999measuring}. Acquisitions take
the place of trials here. The unbiasedness of the pairwise phase consistency assumes independent, identically
distributed trials. Acquisitions are neither, and $\E[Q_{pq}]$ is then the mean expected cosine agreement
(Eq.~\eqref{eq:agreement}).

\subsection{Agreement across interferograms and coherence of an ensemble}
\label{sec:agreement}
Phase linking uses the coherence of an ensemble, whereas SPC compares pixels pairwise. On average the two agree:
an ensemble is coherent over the interferograms exactly when its pixels agree in phase pairwise over the acquisitions,
and the coherence of an ensemble is maximal in every interferogram exactly when the SPC of each pair of its pixels
is maximal. Write $\Delta\phi_p^{st}=\phi_p(s)-\phi_p(t)$,
$v_p^{st}=e^{\ii\Delta\phi_p^{st}}$ and $M=\binom N2$. Expanding
Eq.~\eqref{eq:spc} gives
\begin{equation}\label{eq:agreement}
 Q_{pq}=\frac1M\sum_{s<t}
 \cos\!\left(\Delta\phi_p^{st}-\Delta\phi_q^{st}\right).
\end{equation}
Corrected SPC is thus the mean cosine of the difference between the interferometric phases of the two pixels over all
$M$ interferograms (Section~\ref{app:proofs} of the supplement), computed in $O(N)$ operations without forming the
interferograms or a coherence matrix. The identity with the coherence of an ensemble holds for any set $\mathcal E$ of $L=|\mathcal E|\ge2$
pixels,
\begin{equation}\label{eq:coherence}
 \frac1M\sum_{s<t}\frac{L\,\widehat\gamma^{\circ}_{st}(\mathcal E)^2-1}{L-1}
 =\Qbar(\mathcal E)=\frac1{L(L-1)}\sum_{\substack{q,q'\in\mathcal E\\ q\ne q'}}Q_{qq'}.
\end{equation}
The left side corrects the squared coherence magnitude of each
interferogram for its bias $1/L$ at zero coherence for $L$ independent
looks \citep[cf.][Eq.~(5)]{touzi1999coherence}, as Eq.~\eqref{eq:q} corrects $S_{pq}$. The
mean corrected SPC over all pairs of an ensemble is thus the mean
corrected squared phase-only coherence of that ensemble (Section~\ref{app:proofs} of the supplement).
Unlike this mean over all pairs, which includes pairs that do not contain the center, the maps of Sections~\ref{sec:nspcmask} and~\ref{sec:evidence} are built from the pairs of each
pixel with fixed neighbors and describe that pixel and its
immediate surroundings. The two maps, NSPC and the PCI,
need eight and 24 inner products of length $N$ per pixel, against $O(LN^2)$ operations for the coherence
of an ensemble of $L$ pixels.

\subsection{Expected value under the CCG model}
\label{sec:expectation}
The expected value of SPC shows how much of a low SPC comes from decorrelation and how much from different true phase
histories. It follows from the expectation of each cosine in Eq.~\eqref{eq:agreement}, which depends on the
distribution of the single-look interferometric phase $\Delta\phi$ of an interferogram with coherence $\gamma$. Let
$g$ denote the mean cosine of its phase error $\Delta\phi-\arg\gamma$ (\citealp[cf.][Eq.~(15)]{just1994phase}; \citealp[Eq.~(21)]{lee1994intensity}), in the
closed form of \citet[Eq.~(8)]{lopezmartinez2002modeling}, with ${}_2F_1$ the Gauss hypergeometric function,
\begin{equation}\label{eq:g}
 g(|\gamma|)=\E[\cos(\Delta\phi-\arg\gamma)]
 =\frac\pi4\,|\gamma|\,{}_2F_1\!\left(\tfrac12,\tfrac12;2;|\gamma|^2\right).
\end{equation}
For example, $g(0.2)=0.16$, $g(0.5)=0.41$, $g(0.8)=0.70$.
Suppose pixels $p$ and $q$ are independent and each follows a
zero-mean CCG distribution with true coherence matrix entries
$\gamma^p_{st}$, $\gamma^q_{st}$ (not sample estimates) and true phase histories
$\theta_p,\theta_q$, with $\arg\gamma^p_{st}=\Delta\theta_p^{st}=\theta_p(s)-\theta_p(t)$. Since $p$ and $q$ are independent and the phase error of a CCG pixel is symmetric about zero
(Section~\ref{app:proofs} of the supplement), taking expectations in Eq.~\eqref{eq:agreement} gives
\begin{equation}\label{eq:EQccg}
 \E[Q_{pq}]=\frac1M\sum_{s<t} g(|\gamma^p_{st}|)\,g(|\gamma^q_{st}|)\,
 \cos\!\left(\Delta\theta_p^{st}-\Delta\theta_q^{st}\right).
\end{equation}
Decorrelation lowers $\E[Q_{pq}]$ through $g^2$, and differences
in phase history through the cosine. Even two pixels with the same true phase history and $|\gamma_{st}|=0.5$ for all acquisition pairs give
$\E[Q_{pq}]\approx0.17$, so that where the coherence is low, a single value of SPC says little about whether two
pixels share a true phase history.

When the relative phasors $z_n=\overline{u_p(n)}u_q(n)$ are
independent across acquisitions with means $\xi_n$, the
expectation (Section~\ref{app:proofs} of the supplement) is
\begin{equation}\label{eq:EQindep}
 \E[Q_{pq}]=\frac{\left|\sum_n\xi_n\right|^2-\sum_n|\xi_n|^2}{N(N-1)},
\end{equation}
which reduces to $|\xi|^2$ when all $\xi_n$ equal a common mean $\xi$
\citep[cf.][Eqs.~(13) and (19)]{vinck2010pairwise}. We use Eq.~\eqref{eq:EQindep} for the null
hypothesis below.

\subsection{SPC under correlated speckle}
\label{sec:null}
If the relative phases are independent across the acquisitions and uniformly distributed,
$\E[Q_{pq}]=0$. This does not hold for nearby pixels in SLC data. The SAR point spread function spreads each
scatterer over several samples, so that neighboring pixels share
speckle. Let $\rho_{\bm\ell}$ denote the complex correlation of
the speckle at spatial lag $\bm\ell$. Under the null hypothesis
that the pixels carry no common signal beyond this shared
speckle, and if the observations are independent across acquisitions,
the relative phasors have mean magnitude $g(|\rho_{\bm\ell}|)$ (Section~\ref{app:proofs} of the supplement), and
Eq.~\eqref{eq:EQindep} gives
\begin{equation}\label{eq:psfnull}
 \E_{\rho}[Q_{pq}]=g(|\rho_{\bm\ell}|)^2.
\end{equation}
We call the level that SPC reaches under this null hypothesis the correlated-noise floor. We refer to distributed
scattering without point scatterers as clutter, the name that persistent scatterer processing gives to the
background of a stable scatterer \citep[p.~11]{ferretti2001permanent}, and to clutter under this null
hypothesis as zero-coherence clutter. We predict the correlation from the processing parameters of the SLC
data and check the prediction with an estimate from the stack itself (Section~\ref{sec:realspeckle}). For a
spectral weighting $W(f)$ over the processed bandwidth $B$ and a
sampling rate $f_{\mathrm s}$, the speckle correlation at a lag of $\ell$
samples along one image axis is
(\citealp[Eqs.~(8) to (10), (12) and (16)]{just1994phase};
\citealp[Eq.~(5.8)]{goodman2020speckle})
\begin{equation}\label{eq:rhopred}
 \rho(\ell)=\frac{\bigl|\int_{-B/2}^{B/2}W(f)^2\,e^{\ii2\pi f\ell/f_{\mathrm s}}\,df\bigr|}
 {\int_{-B/2}^{B/2}W(f)^2\,df},
\end{equation}
and the correlation at a two-dimensional lag is the product of the
range and azimuth values. For example, the processing parameters of the TerraSAR-X stack of
Section~\ref{sec:stacks} (Table~\ref{tab:sensor} of the supplement) predict a lag-one correlation of \vpPredTsxAz{} in
azimuth and \vpPredTsxRg{} in range. We treat two pixels as sharing no
speckle, and refer to them as lying beyond the speckle correlation
length, if the predicted $|\rho_{q-p}|$ is below $\rho_{\min}=\vtRhoMin$, a bound tuned with the
truncation step (Section~\ref{sec:simdesign}).

\subsection{SPC under differential motion and subtracted model phases}
\label{sec:model}
SPC responds to signal differences as well as noise. For a
noise-free relative phase that changes by $\delta$ from one acquisition to the next,
\begin{equation}\label{eq:ramp}
 S_{pq}=\left|\frac{\sin(N\delta/2)}{N\sin(\delta/2)}\right|^2 .
\end{equation}
Low SPC therefore does not imply noise, because two coherent
pixels with different deformation can have low SPC.

A related case is important in practice. Suppose a model phase
$\psi_p(n)$, such as a simulated topographic phase, is subtracted
from every pixel before SPC is computed, and the model differs
between neighbors by $\psi_{pq}(n)=\psi_q(n)-\psi_p(n)$. Under
the null hypothesis of Section~\ref{sec:null}, the means in
Eq.~\eqref{eq:EQindep} become $\xi_n=g(|\rho|)e^{-\ii\psi_{pq}(n)}$,
and
\begin{equation}\label{eq:attenuation}
 \E[Q_{pq}]=g(|\rho|)^2\;Q\!\left[e^{\ii\psi_{pq}}\right],
\end{equation}
where $Q[e^{\ii\psi_{pq}}]$ is the corrected SPC of the
deterministic model difference alone. Where the model phase
changes strongly between neighbors, as for simulated topographic
phase in layover, the subtraction itself reduces SPC below the
correlated-noise floor. Section~\ref{sec:realmask} shows this effect on
real data.

\section{Applications to masking and adaptive neighborhood sizing}
\label{sec:selection}

\begin{figure}[tbp]
\centering
\includegraphics[width=\linewidth]{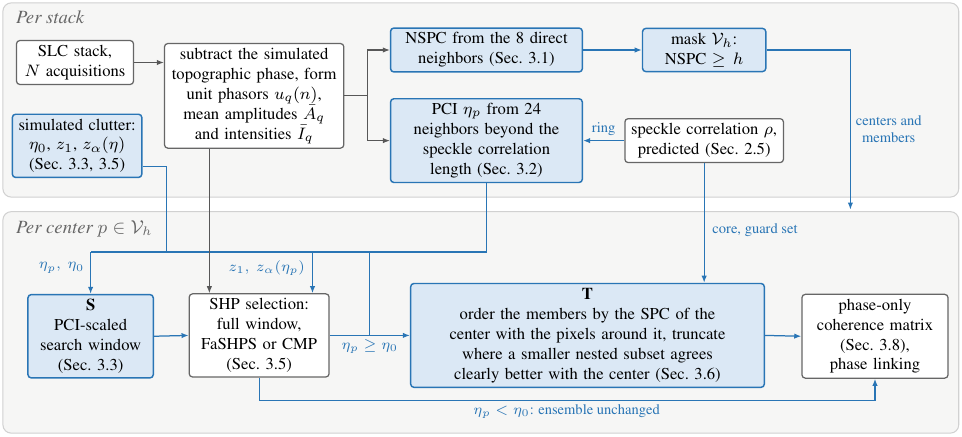}
\caption{Processing of a stack. Per stack, the mask and the PCI are computed from fixed neighbors of each pixel,
and the null quantile $\eta_0$ of the PCI and the threshold tables of FaSHPS and CMP from simulated clutter. Per
center retained by the mask, the PCI scales the search window (S), sets the second-pass threshold of FaSHPS and the thresholds of CMP from these
tables and gates the truncation step (T), which compares nested subsets of the ensemble with the phases of the core around the center. S and T together form \WSname. The shaded steps are those introduced in
this paper.
Appendix~\ref{app:algorithm} gives the processing as an algorithm.}
\label{fig:processing}
\end{figure}

This section describes the applications of SPC, and the steps they are combined with, in the order of the
processing of a stack (Fig.~\ref{fig:processing}). Two per-pixel maps are computed first, before any SHP selection, the NSPC mask (Section~\ref{sec:nspcmask}) and the PCI
(Section~\ref{sec:evidence}). For each center pixel, the search window is then scaled with the PCI
(Section~\ref{sec:scaledwindow}). Within it, the ensemble is collected by an SHP selection, either a pairwise one with thresholds that depend on the
PCI (Sections~\ref{sec:amptests} and~\ref{sec:pairwise}) or the full window, which takes all eligible pixels of the
window (Section~\ref{sec:scaledwindow}) and serves as a reference for the pairwise ones. We count the full window among
the SHP selections, although it tests no pixel. An adaptive \emph{truncation step}, which can follow any SHP selection, may reduce the ensemble to a
nested subset. In \WSname{} it acts only where a gate on the PCI passes (Section~\ref{sec:adaptive}). Together, the scaled window and the gated truncation step form the adaptive neighborhood sizing proposed here, \WSname. The truncation step reduces the ensemble
where a smaller nested subset agrees clearly better in phase with the center, and the scaled window
gives a coherent center fewer candidates, since the number of looks needed for a given precision decreases with the coherence. The coherence matrix is then estimated from the returned ensemble in the same way for every SHP selection, with
or without these steps (Section~\ref{sec:estimate}).

Figure~\ref{fig:processing} shows this processing with the three SHP selections compared here, the full
window, FaSHPS of \citet{jiang2015fast} and CMP of \citet{zhao2022statistically}, both described in
Section~\ref{sec:pairwise}. The mask step of
Fig.~\ref{fig:processing} applies to the real stacks, whereas the simulations apply no mask
(Section~\ref{sec:simdesign}). The simulations of Section~\ref{sec:simulation} mirror the real stacks in three
configurations, named by sensor and number of acquisitions, TSX-44 for the TerraSAR-X stack with 44 acquisitions and
S1-79 and S1-20 for the Sentinel-1 stack with 79 acquisitions and its substack of 20. Appendix~\ref{app:algorithm} gives the processing as an algorithm.

\subsection{Neighborhood SPC mask}
\label{sec:nspcmask}
Let $\mathcal V$ contain the pixels that have finite, nonzero
observations in all acquisitions and that have not been excluded by
any prior mask, and let $\mathcal N_p=(\mathcal W_p\cap\mathcal V)\setminus\{p\}$
be the valid neighbors of $p$ in a small neighborhood $\mathcal W_p$
centered on $p$.

We define the neighborhood SPC as the mean of
the square roots of the clipped corrected SPC,\footnote{Compared with averaging $Q_{pq}$ before taking the square root, this mean
favors broad agreement among the neighbors over strong agreement with a few of them
(Section~\ref{app:neighborhood-aggregation} of the supplement).}
\begin{equation}\label{eq:local}
 \operatorname{NSPC}(p)=\frac1{|\mathcal N_p|}\sum_{q\in\mathcal N_p}
 \sqrt{Q_{pq}^+},
\end{equation}
with $\operatorname{NSPC}(p)=0$ for $p\notin\mathcal V$ or without
valid neighbors. We use a $3\times3$ neighborhood throughout, so that
all eight neighbors, the direct neighbors, are at lag one. A larger neighborhood would add neighbors
at lags of two and more, where the SPC of zero-coherence clutter, which is correlated
only through the point spread function, is much lower. In the TerraSAR-X
stack of Section~\ref{sec:stacks}, the mean corrected SPC at lag two is a
small fraction of that at lag one. NSPC would then lie much closer to zero in
areas dominated by such clutter, which leaves little room for the values below the floor that the mask detects.

\paragraph{Value ranges}
Because the $3\times3$ neighborhood contains only lag-one neighbors,
Eq.~\eqref{eq:psfnull} sets a floor that zero-coherence clutter alone reaches. In the
TerraSAR-X stack of Section~\ref{sec:stacks}, the NSPC values fall into
three ranges. Their limits depend on the speckle
correlation and on the number of acquisitions and are specific to that
stack.
\begin{itemize}[leftmargin=*,itemsep=1pt]
\item \emph{Above about 0.6.} Agreement beyond shared speckle,
i.e., coherent signal common to the pixel and its neighbors.
\item \emph{Around \vdVispMode.} The correlated-noise floor
(Section~\ref{sec:realspeckle}).
\item \emph{Below about 0.3.} Agreement \emph{below} the floor.
The simulations use a separable model of speckle correlation and temporal coherence, in
which the covariance is the product of the temporal coherence and the speckle correlation. Under this model,
decorrelation alone does not produce agreement below the floor, since the expected corrected SPC of direct
neighbors is then at least $g(|\rho|)^2$ at any coherence (Section~\ref{app:proofs} of the supplement).
Within this model, it requires systematic differences between the
phase histories of neighbors, namely differential motion, mixing of
scatterers at different heights, or a subtracted model phase that the
data does not share. Outside the separable model, noise that is less correlated than speckle,
such as thermal noise, and the sidelobes of a bright scatterer, which reach adjacent samples with
opposite signs (Section~\ref{sec:realmask}), also produce it.
\end{itemize}

\paragraph{Threshold and mask}
For a threshold $h$, the mask retains the pixels
$\mathcal V_h=\{p\in\mathcal V:\operatorname{NSPC}(p)\ge h\}$. We set $h$ by inspection of the NSPC histogram (Section~\ref{sec:realspeckle}), below the correlated-noise
mode, at $h=0.25$ for this stack. To check this choice, we simulated clutter with zero
coherence between any two acquisitions, for
the speckle correlation and the 44 acquisitions of the TerraSAR-X stack
(TSX-44). In this clutter,
NSPC has a median of \vkNullTSXMed, its
1\,\% and 0.1\,\% quantiles are \vkNullTSXQone{} and \vkNullTSXQmil, respectively, and
it falls below 0.25 with a probability of $\vkNullTSXBelowSci$.

\paragraph{Sensor dependence}
The floor depends on the speckle correlation, and hence on the
sampling of the SLC data relative to its resolution. In simulated homogeneous clutter at zero coherence, the mean
corrected SPC of neighbors at lag one along an image axis is lowest for range neighbors on TerraSAR-X, at \vsAdjZeroMin, and highest for range
neighbors on Sentinel-1, at \vsAdjZeroSLong, since that data is oversampled by a factor of two in range
(Section~\ref{sec:stacks}). A fixed
$h$ therefore does not transfer between sensors, and we recommend
setting $h$ relative to the floor simulated for the measured or
predicted $\rho$. The thresholds set here by inspection, $h=0.25$ on TerraSAR-X and $h=0.30$ on Sentinel-1, are
\vkHRatioMin{} to \vkHRatioMax{} times the median NSPC of zero-coherence clutter simulated for the TerraSAR-X and Sentinel-1 stacks, and half
this median can serve as a rough starting point. The prediction of $\rho$ holds for SLCs on their original grid with known processing parameters, including a
band-limited oversampling at its effective sampling rate. Coregistration, however, resamples every acquisition except
the reference, which changes the correlation slightly (Section~\ref{supp:sensor} of the supplement). The floor can also vary within a scene. Thermal
noise is not shaped by the azimuth antenna pattern and is less
correlated in azimuth than speckle, so pixels dominated by thermal
noise have a lower floor than pixels dominated by zero-coherence clutter, a floor closer to the threshold. On Sentinel-1, a large part of the radar shadow falls below the threshold, with an
NSPC lower still than that simulated for thermal noise (Section~\ref{sec:realmask}).

\paragraph{Scope of the mask}
The NSPC mask flags pixels that agree with their neighbors less than zero-coherence
clutter with the speckle correlation does. The cause is either a systematic phase
difference between neighbors, such as a subtracted topographic phase
that does not match the phase of the data in layover, or noise that is
less correlated than speckle, such as thermal noise in radar shadow. On both sensors the mask covers a large
part of the geometric layover, and on Sentinel-1 also of the shadow (Section~\ref{sec:realmask}). It provides complementary information where the DEM is
incomplete, needs no data beyond the stack after topographic phase removal, and is computed with eight inner
products of length $N$ per pixel, one per neighbor. However, isolated
coherent targets and sharp deformation gradients can also have low
NSPC (Eq.~\eqref{eq:ramp}). Their retention should be checked when NSPC
is used as a hard mask.

\paragraph{Treatment of masked pixels}
Where the mask precedes the SHP selection, as in the processing of the real stacks in this paper, masked pixels are not
processed as centers, receive no estimate and enter no ensemble. Appendix~\ref{app:algorithm} gives their treatment in
the PCI and in the truncation step.

\subsection{Phase coherence indicator}
\label{sec:evidence}
The NSPC of a pixel averages SPC over its direct neighbors, which share speckle with the pixel. This gives NSPC the correlated-noise floor on which the mask relies,
but it also makes NSPC respond weakly to low and moderate
coherence. In simulated homogeneous clutter, the mean corrected SPC of adjacent pixels rises by at most \vsAdjRiseMax{} from zero coherence to a coherence of 0.5 between consecutive acquisitions, on both sensors and in
both directions. An indicator of local coherence at a
pixel therefore uses neighbors that share no speckle with it. Let
\begin{equation}\label{eq:eta}
 \tilde\eta_p=\frac1{|\mathcal R_p|}\sum_{q\in\mathcal R_p}\sqrt{Q^+_{pq}},
\end{equation}
where the ring $\mathcal R_p$ contains the 24 pixels nearest to $p$ beyond the speckle correlation length, i.e., as many pixels as a $5\times5$
window holds besides its center, though none of them shares speckle with $p$. The region beyond the correlation length
is $\mathcal F_p=\{q:\ |\rho_{q-p}|<\rho_{\min}\}$ with the predicted correlation $\rho$, a set that excludes $p$ itself. Nearness is measured by the normalized distance of
Eq.~\eqref{eq:normdist} (Section~\ref{sec:scaledwindow}), and the order it gives the offsets depends only on the
ratio of the semiaxes $b_1$ and $b_2$ of the search window. Offsets tied in this distance at the
cut are taken in point-symmetric pairs, so that the ring is symmetric about $p$. In the experiments of this paper, the ring pixels lie at \vfTsxRingMin{} to \vfTsxRingMax{}
pixels from $p$ on TerraSAR-X and, mostly along range, at \vfSOneRingMin{} to \vfSOneRingMax{}
pixels on Sentinel-1, where the data is oversampled in range and the search window is elongated in range.
The \emph{phase coherence indicator} (PCI) $\eta_p$ is the mean of $\tilde\eta_q$ over the 25 pixels $q$ of the $5\times5$ window centered on $p$. The point coherence estimation of \citet{vecchioli2023novel} also rests on pairs
of points with independent phase noise (Section~\ref{sec:related}) and solves for the coherence of every point,
whereas the PCI averages over a fixed ring and a $5\times5$ neighborhood and thus describes the coherence of $p$
together with that of its surroundings. These
surroundings span an area of $\vfTsxPixAz\times\vfTsxPixRg$ pixels
(about $\vfTsxMetAz\times\vfTsxMetRg$\,m in azimuth and ground range) around $p$ on TerraSAR-X and
$\vfSOnePixAz\times\vfSOnePixRg$ pixels (about $\vfSOneMetAz\times\vfSOneMetRg$\,m) on Sentinel-1. In the processing of this paper (Algorithm~\ref{alg:adaptive}), $\eta$
is computed once for every pixel of the image, and all later uses of the
PCI, including Table~\ref{tab:eta}, refer to this value.

To relate $\eta$ to the coherence, consider neighbors with the same coherences $|\gamma_{st}|$ and the same
true phase history, for which Eq.~\eqref{eq:EQccg}
gives $\E[Q_{pq}]$ as the mean of $g(|\gamma_{st}|)^2$ over the pairs.
Since $\eta$ averages the roots $\sqrt{Q^+_{pq}}$ instead of taking the root of the mean of the unclipped $Q_{pq}$, it differs from $\sqrt{\E[Q_{pq}]}$. In the simulations (Section~\ref{sec:simdesign}), the coherence
between consecutive acquisitions is $c$, and it decays toward $c^2$ as the
interval grows (Eq.~\eqref{eq:cohfamily}). In TSX-44 at $c=0.5$, $\sqrt{\E[Q_{pq}]}$ is
\vgRmsGMid, and the weighted median of $\eta$ is \vgEtaMidTSX.
Table~\ref{tab:eta} gives the weighted median of $\eta$ at the distributed centers of the simulations
(Section~\ref{sec:simdesign}), together with $g(c^2)$, the value of
$g$ in Eq.~\eqref{eq:g} at the long-term coherence. The median rises monotonically with the
coherence between consecutive acquisitions, its value at zero coherence decreases with the number of acquisitions,
and from $c=0.5$ on it lies within \vaEtaGsqDev{} of $g(c^2)$. The root $\sqrt{\E[Q_{pq}]}$
lies above $g(c^2)$, since the shorter intervals have higher coherences,
and the median of $\eta$ lies below the root at every $c>0$. At a given coherence, however, $\eta$ is lower in dense
fringes, since the phase difference between the pixel and its ring neighbors then changes over the stack
(Eq.~\eqref{eq:EQccg}). In this paper, $\eta$ sets the thresholds of the pairwise SHP selections
(Section~\ref{sec:pairwise}), scales the search window (Section~\ref{sec:scaledwindow}) and gates the truncation step (Section~\ref{sec:adaptive}), and
it is shown in the maps of Section~\ref{sec:realdata}.

\begin{table}[tbp]
\centering\small
\setlength{\tabcolsep}{4pt}
\begin{tabular}{lcccc}
\toprule
 & & \multicolumn{3}{c}{Weighted median of $\eta$} \\
\cmidrule(lr){3-5}
Coherence & $g(c^2)$ & TSX-44 & S1-79 & S1-20 \\
\midrule
$c=0$ & 0.00 & 0.05 & 0.04 & 0.08 \\
$c=0.2$ & 0.03 & 0.07 & 0.06 & 0.11 \\
$c=0.35$ & 0.10 & 0.13 & 0.12 & 0.17 \\
$c=0.5$ & 0.20 & 0.22 & 0.21 & 0.26 \\
$c=0.65$ & 0.34 & 0.35 & 0.33 & 0.38 \\
$c=0.8$ & 0.53 & 0.53 & 0.50 & 0.54 \\
$c=0.95$ & 0.82 & 0.81 & 0.78 & 0.80 \\
\addlinespace[2pt]
$\eta_0$ & & 0.066 & 0.050 & 0.100 \\
\bottomrule
\end{tabular}
\caption{PCI $\eta$ (Section~\ref{sec:evidence}) at the simulated distributed centers (evaluation scenes, Section~\ref{sec:simdesign}) by the coherence $c$ between consecutive acquisitions, for the configurations TSX-44, S1-79 and S1-20 (44 TerraSAR-X, 79 and 20 Sentinel-1 acquisitions, Table~\ref{tab:configs}): weighted median, $g$ of Eq.~\eqref{eq:g}, the mean cosine of the single-look phase error, at the long-term coherence $c^2$ of Eq.~\eqref{eq:cohfamily}, and the null quantile $\eta_0$, the 99th percentile of $\eta$ in simulated homogeneous clutter with zero coherence (calibration scenes).}
\label{tab:eta}
\end{table}

\subsection{PCI-scaled search window}
\label{sec:scaledwindow}
For an eligible center $p$, let $\mathcal B_p$ contain the eligible
pixels in a bounded search window around $p$, including $p$.
Eligible pixels belong to $\mathcal V_h$ if masking is applied,
and to $\mathcal V$ otherwise. The search window has the same size at every center of a stack, and $\mathcal B_p$ holds fewer pixels only where
the window extends beyond the image or contains pixels that are not eligible. The user chooses its size large enough
for the decorrelated parts of a scene, which need many looks. A coherent center needs fewer looks for
the same precision, and the distant members of an SHP selection in the unscaled search
window are those most likely to carry a different phase. The truncation step
(Section~\ref{sec:adaptive}) reduces such an ensemble only where a smaller nested subset agrees clearly better with
the center, and the scaled window complements it by limiting the number of candidates at a coherent center before the SHP selection. Since the PCI (Section~\ref{sec:evidence}) rises with the coherence of the
center (Table~\ref{tab:eta}), it is used here to shrink the search window
of each center.

The window is shrunk only above the null quantile $\eta_0$ of the PCI, the 99th percentile of $\eta$ in
simulated homogeneous clutter with zero coherence, computed for each
configuration on the calibration scenes (Section~\ref{sec:simdesign}). The null quantile is \vEtaNullTSX{} for TSX-44, and
\vEtaNullSLong{} and \vEtaNullSShort{} for S1-79 and S1-20, respectively
(Table~\ref{tab:eta}). For the point spread functions of TerraSAR-X and Sentinel-1, $\eta_0\sqrt N$ is
\vgEtaSqrtNMin{} to \vgEtaSqrtNMax{} in all three configurations, i.e., $\eta_0$ is nearly proportional to $1/\sqrt N$.

For a pixel $q$, let $\bm d_q=q-p$ be its offset from the center, with
the components $d_{q,1}$ in azimuth and $d_{q,2}$ in range, in pixels. Its
normalized distance from the center is given by
\begin{equation}\label{eq:normdist}
 \|\bm d_q\|_b^2=\Bigl(\frac{d_{q,1}}{b_1}\Bigr)^2+\Bigl(\frac{d_{q,2}}{b_2}\Bigr)^2,
\end{equation}
where $b_1$ and $b_2$ are the semiaxes of the search window in azimuth
and range, i.e., half the side lengths of the rectangle in
Table~\ref{tab:configs}, so that $\mathcal B_p$ contains the eligible pixels
with $\|\bm d_q\|_b\le1$. With the null quantile $\eta_0$ of
the PCI, the relative PCI and the scale of the window
are
\begin{equation}\label{eq:scale}
 \eta'_p=\min\Bigl(1,\max\Bigl(0,\frac{\eta_p-\eta_0}{1-\eta_0}\Bigr)\Bigr),
 \qquad \tau_p=1-\eta'_p\,(1-\tau_{\min}),
\end{equation}
respectively, where $\tau_{\min}$ is the normalized distance of the $n_{\min}$-th nearest
pixel. The \emph{PCI-scaled window}
$\mathcal B'_p$ contains the pixels of $\mathcal B_p$ with $\|\bm d_q\|_b\le \tau_p$,
and at least the $n_{\min}$ nearest ones. Its semiaxes thus decrease
linearly with the PCI, from those of $\mathcal B_p$ at
$\eta_p\le\eta_0$ to those of the $n_{\min}$ nearest pixels at $\eta_p=1$, so that the number of its pixels falls about quadratically.

The value $n_{\min}=\vnMin$ is fixed without tuning. Since no center reaches the window of $\eta_p=1$, in the
simulations or on the real stacks, $n_{\min}$ acts only through the slope of $\tau_p$ in
$\eta_p$. With $n_{\min}=50$ or 200 the error of the full window and FaSHPS with \WSname{} changes by at most
\vxNminChangeDuoMax\,\% over all centers (TSX-44 and S1-79, Section~\ref{supp:sweep} of the supplement).

An SHP selection then works within $\mathcal B'_p$ in place of
$\mathcal B_p$.

\subsection{Amplitude tests under temporal correlation}
\label{sec:amptests}
Amplitude tests for statistically homogeneous pixels treat the $N$
amplitudes of a pixel as independent samples. Distribution tests compare
the amplitudes of two pixels as samples of one distribution
\citep{ferretti2011new,parizzi2011adaptive}, and the test of
\citet[Eq.~(7)]{jiang2015fast} compares their temporal means with a confidence
interval derived for $N$ independent samples, which they scale with one coefficient of variation per scene
(Section~\ref{sec:pairwise}). For a distributed scatterer, independence holds only if the speckle decorrelates
between acquisitions. Otherwise, the amplitude of a pixel
changes little from one acquisition to the next. Two pixels of the same homogeneous surface then
have different, nearly constant amplitudes, and a test that assumes
independent samples wrongly rejects them.

Under the CCG model with coherence matrix $\bm\Gamma$ and equal
mean intensity in all acquisitions, the intensities of one pixel in
acquisitions $s$ and $t$ have correlation coefficient
$|\gamma_{st}|^2$ \citep[p.~53, Eq.~(3.97)]{goodman2020speckle}.
For the mean of correlated observations, equating its variance to
that of the mean of independent samples defines the effective
sample size \citep[pp.~800--801, Eq.~(2.1)]{thiebaux1984interpretation}.
For the temporal mean intensity it is
\begin{equation}\label{eq:neff}
 N_{\mathrm{eff}}=\frac{N^2}{\sum_{s,t}|\gamma_{st}|^2} ,
\end{equation}
where the sum includes the terms $s=t$ with $|\gamma_{ss}|=1$.
\citet[Eq.~(4.106)]{goodman2020speckle} gives the continuous counterpart of Eq.~\eqref{eq:neff} for temporally
integrated speckle. For the temporal mean amplitude, which the tests
below use, Eq.~\eqref{eq:neff} is a close approximation. We have $N_{\mathrm{eff}}=N$
for a fully decorrelated target and $N_{\mathrm{eff}}\to1$ for a
fully coherent one. A test calibrated for $N$ samples attains its
significance level only in the first case. The more coherent the target,
the more often such a test rejects homogeneous neighbors, which is
contrary to the desired behavior, because the homogeneity of a
target does not depend on its coherence.
The thresholds of FaSHPS, and those of CMP, whose statistic also contains the ratio of the temporal
mean intensities, are therefore set in temporally correlated clutter, and the second-pass threshold of FaSHPS and the
thresholds of CMP depend on the PCI of the center, which rises with its coherence (Section~\ref{sec:pairwise}).

\subsection{Pairwise SHP selection and connectivity}
\label{sec:pairwise}
Besides the full window, this paper combines \WSname{} with two pairwise SHP selections. FaSHPS compares each
candidate $q$ in the search window with $p$ on the temporal mean amplitudes, CMP on the temporal mean intensities and
the single-look phase differences, and we call these two tests the SHP tests. Phase-based pixel selection has also been formulated as tests on covariance matrices
\citep{dong2018unified,song2017statistically} and, without reference to
the center, as a clustering by the trigonometric moments of residual
phases \citep{yao2024phase}, which groups pixels by their decorrelation
rather than by the agreement of their phases with those of the center. These methods are not compared here.

\paragraph{Amplitude test}
We use the statistic of \citet{jiang2015fast} as the representative of
the amplitude tests of SqueeSAR-type frameworks \citep{ferretti2011new}. It
shares their assumption of independent samples (Section~\ref{sec:amptests}) and
requires a few operations per candidate, independent of $N$ (Section~\ref{sec:cost}). Its authors report improvements
over previous methods, among them the Kolmogorov--Smirnov SHP selection of SqueeSAR, in edge preservation, retention
of resolution, noise reduction, bias mitigation and computational efficiency (their Section~VII). Following their
Sections~III-A and IV, let $\bar A_q$ be the temporal mean amplitude of pixel $q$ after the removal of outlying
amplitudes with a revised boxplot, and
\begin{equation}\label{eq:fastat}
 T_{pq}=\frac{|\bar A_q-\mu_p|}{c_A\,\mu_p/\sqrt N},\qquad c_A=0.52,
\end{equation}
where $c_A$ is the coefficient of variation of a Rayleigh amplitude,
so that $T_{pq}\le z$ is the confidence interval of their Eq.~(7).

The test has two passes, of which the first only gives the second a reference mean from more pixels than the center
alone. In the first pass, $\mu_p=\bar A_p$, and the pixels with $T_{pq}\le z_1$ form, with the center, the
\emph{reference set} of $p$. In the second pass, $\mu_p$ is the mean of $\bar A_q$ over the reference set, every
candidate of the window is tested against it, and a candidate passes if $T_{pq}\le z_\alpha$, where $\alpha$ is the
significance level of the test. The pixels that pass form the ensemble $\mathcal O_p$ through their eight-connected
component that contains $p$, whereas the original test accepts any pixel of the window that passes.

The original thresholds $z_1$ and $z_\alpha$ are normal quantiles of a 50\,\% and a $1-\alpha$ interval,
respectively, for $N$ independent samples, with $\alpha=5\,\%$ in their test, and both are scaled with a
coefficient of variation of the temporal mean estimated once per scene (their Section~III-B). We use them only for
comparison, at the significance level of 1\,\% used in this paper (Section~\ref{app:amptests} of the
supplement).

By Section~\ref{sec:amptests}, a threshold for $N$ independent samples attains its significance level only at zero
coherence, and a threshold set at one coherence loses it at another. A single value per scene cannot hold the level
at every center of a scene of mixed coherence. The outlier removal and, on oversampled data, the shared speckle of
adjacent pixels change the distribution of $T_{pq}$ as well. We therefore set both thresholds by simulating
homogeneous clutter with the stack size, the search window and the processing parameters of the SLC data
(Section~\ref{sec:null}), once per stack and before the centers are processed. The first-pass threshold $z_1$ admits
half of the homogeneous candidates to the reference set at a coherence of 0.5, in the middle of the simulated
coherences, more at lower and fewer at higher coherence. It is the same at every center, since the significance level
and its dependence on the coherence lie in the second pass. For the second pass, the $1-\alpha$ quantile of $T_{pq}$
is tabulated against the median PCI of the clutter at eight coherences from 0 to 0.95. Each center takes the threshold
$z_\alpha(\eta_p)$ interpolated linearly at its own PCI (Section~\ref{sec:evidence}), which rises with the coherence
(Table~\ref{tab:eta}). The threshold table contains no real data, which enters only through the PCI of the center
(Section~\ref{app:amptests} of the supplement). The statistic keeps the theoretical $c_A$ instead of a value estimated
from the scene. With simulated thresholds, $c_A$ only fixes the scale of $T_{pq}$, so that the test is the original
one with other thresholds. At a significance level of $\alpha=1\,\%$, the test then rejects \vcRejHomFaMin{} to
\vcRejHomFaMax\,\% of the homogeneous candidates at every simulated coherence, including coherences between those of
the calibration. The original thresholds reject \vcRejHomFaPubMin{} to \vcRejHomFaPubMax\,\%, most at $c=0.95$ with
\vcRejFaPubHighMin{} to \vcRejFaPubHighMax\,\% (three configurations, Fig.~\ref{fig:rejection} and
Table~\ref{tab:rejection} of the supplement). We keep the name FaSHPS for this adjusted SHP selection.

\paragraph{Test of CMP}
CMP \citep{zhao2022statistically} compares two pixels by a symmetrized Kullback--Leibler divergence over the
single-reference interferograms, which combines the ratio of their temporal mean intensities with a weighted mean
cosine of their single-look phase differences, with weights from an assumed coherence of each interferogram. Of the terms in the statistics of the SHP tests, this phase term is the one most closely related to SPC, but it requires a
choice of interferograms, here the single-reference ones, and a coherence model, whereas SPC requires neither. As in FaSHPS, the test has two passes, the second against the mean
intensity and the multilooked phases of the reference set formed by the first, and the ensemble is the eight-connected
component of the center \citep[Section~2.2.3]{zhao2022statistically}. Both of its thresholds are tabulated over the
PCI from the same simulated clutter as the second-pass threshold of FaSHPS, at the same significance level, with
the decay constant of its coherence model fitted to the stack (Section~\ref{supp:cmp} of the supplement).
That section also gives the statistic and the coherence model, of which we leave out the term of
the perpendicular baseline.

\begin{figure}[tbp]
\centering
\includegraphics{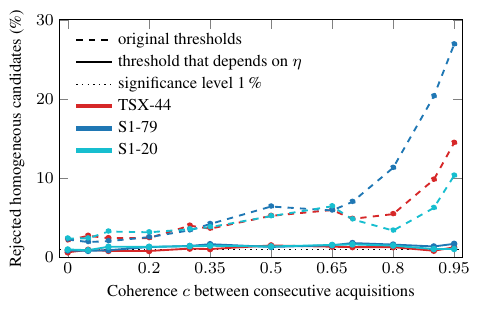}
\caption{With the threshold that depends on the PCI, FaSHPS rejects homogeneous candidates at close to the
significance level at every coherence, and with the original thresholds more often, most at the highest coherence.
Share of the homogeneous candidates outside the guard set that
FaSHPS rejects with the original thresholds (dashed) and with the threshold that depends on the PCI $\eta$ (solid), at a
significance level of 1\,\% (dotted line), in simulated homogeneous clutter by the coherence
$c$ between consecutive acquisitions (Eq.~\eqref{eq:cohfamily}). The guard set holds the pixels that share speckle
with the center or with a pixel that does (Section~\ref{sec:adaptive}). The shares are means over scenes generated
from the seeds of the evaluation scenes, for each of the configurations TSX-44, S1-79 and S1-20 (44 TerraSAR-X, 79 and 20 Sentinel-1
acquisitions, Section~\ref{sec:simdesign}). The original thresholds are those of \citet{jiang2015fast} at this significance level instead of their 5\,\%, with a
coefficient of variation estimated per scene, which in these scenes of a single coherence fits every center (Section~\ref{app:amptests} of the
supplement). Table~\ref{tab:rejection} of the supplement gives
the values, also for CMP and at coherence steps.}
\label{fig:rejection}
\end{figure}

\paragraph{Limitation at high coherence}
At high coherence, the speckle of a pixel changes little between acquisitions, and its temporal mean amplitude
scatters about the mean amplitude of its surface nearly as much as a single amplitude. A test on the mean amplitudes
must then either accept candidates over a wide range of brightness or reject homogeneous candidates more often than
its significance level, and this applies to any test that uses only the amplitudes. The phase term of CMP on its own
rejects practically no candidate (Section~\ref{supp:accept} of the supplement), so that
candidates of similar brightness but different phase are left to the truncation step.

\paragraph{Minimum size}
If $\mathcal O_p$ has fewer than $L_{\min}=25$ members, the center is
insufficient, and its processing stops. The center keeps its
single-look data. In the comparisons below, the error of such a center is computed from the
single-look matrix $\bm u_p\bm u_p^H$,
so that all errors refer to the same pixels, and the share of
insufficient centers is reported for the SHP selections with and without \WSname{}
(Section~\ref{supp:selmore} of the supplement).

\subsection{Truncation step}
\label{sec:adaptive}
An SHP selection returns every compatible pixel it finds, which is the desired behavior in a homogeneous area, where all members share the
true phase history of the center, because there the estimate improves
with the number of looks at any coherence. Where coherence is high and the phase varies
rapidly in space, however, members far from the center carry a
different true phase and bias the estimate, and where the center is
dominated by a single stable scatterer, its own phase is more accurate
than an average over hundreds of pixels with other heights and
histories (Section~\ref{sec:bright}). A single value of $Q_{pq}$ cannot distinguish these
situations from decorrelation, because a decorrelated neighbor and a
coherent neighbor on a different fringe both have low SPC
(Eqs.~\eqref{eq:EQccg} and \eqref{eq:ramp}). The spatial pattern of $Q_{pq}$
around the center can. Under decorrelation, $Q_{pq}$ is low at all
offsets beyond the speckle correlation length, and under dense fringes
it is high near the center and decays with distance, fastest along the
phase gradient.

The adaptive truncation step takes an ensemble selected beforehand and
returns either the ensemble itself or a subset of it. It orders the
members using the SPC of the center with the surrounding pixels, scores nested subsets of
increasing size up to the whole ensemble by how well they describe the
center, and then decides which size to return. In a homogeneous area, every added member reduces the noise of
the multilooked phases, so that the score rises with the size up to estimation noise, and the ensemble is truncated only where this
noise happens to favor a smaller subset.
In dense fringes, the score attains its maximum at a small size and then decreases, and the ensemble is truncated.
The three stages (ordering, score and decision) are described below.
The step can follow any SHP selection, and in
Section~\ref{sec:simulation} it is applied to the ensembles of the full window, FaSHPS and CMP.

\paragraph{Ordering of the members}
Members are ordered so that those most likely to share the true
phase of the center come first. The simplest such order, the
\emph{distance order}, sorts them by the normalized distance from the center (Eq.~\eqref{eq:normdist}). The corrected SPC $Q_{pq}$ of the center with the pixels $q$ of $\mathcal F_p$ within the search window
($\mathcal F_p$ of Section~\ref{sec:evidence}, $\mathcal B'_p$ of Section~\ref{sec:scaledwindow} with the PCI-scaled
window) suggests an alternative. It stays high along a
fringe and falls quickly across it. With the offsets $\bm d_q$ of Eq.~\eqref{eq:normdist}, the second-moment matrix
of the offsets, weighted with these values, is
\begin{equation}\label{eq:moment}
 \bm M_p=\frac{\sum_{q\in\mathcal F_p\cap\mathcal B'_p}Q_{pq}\,\bm d_q\bm d_q^{T}}
 {\sum_{q\in\mathcal F_p\cap\mathcal B'_p}Q_{pq}} .
\end{equation}
Like a covariance matrix of the offsets, $\bm M_p$ describes an
ellipse, long in the direction in which the center retains a high SPC
with distant pixels and short in the direction in which the SPC
decreases quickly. Sorting the members by $\bm d_q^{T}\bm M_p^{-1}\bm d_q$,
the squared Mahalanobis distance \citep[cf.][Eq.~(2)]{akyilmaz2017similarity}, places pixels along a fringe before pixels at the same Euclidean
distance across it. We call this the \emph{ellipse order}. Where $\bm M_p$ is not positive definite or the
SPC of the center carries too little weight, the distance order is used (Section~\ref{supp:truncdetails} of the supplement). The ellipse order
is fixed by design, and Section~\ref{app:settings} of the supplement compares it with the distance order, which has a
higher error in dense fringes. In either case, $\nu(q)$ denotes the position of $q$ in the sorted
ensemble, counted from one.

\paragraph{Score of a candidate ensemble}
The score assesses nested subsets of the ensemble $\mathcal O_p$ by
how well their multilooked phases agree with those of the core, the pixels that share their speckle with the
center. Here $\mathcal O_p$ denotes the ensemble passed to the
truncation step, i.e., the ensemble of any SHP selection, sorted as described above.

The core is
$\mathcal C_p=\{q:\ |\rho_{q-p}|\ge\rho_{\min}\}$, the pixels within the
speckle correlation length of $p$, including $p$, i.e., the complement of $\mathcal F_p$
(Section~\ref{sec:evidence}). This gives a core of \vcCoreTsx{} pixels on TerraSAR-X and of \vcCoreSOne{} pixels on Sentinel-1
(Table~\ref{tab:configs}). Taking the core instead of $p$ alone lowers the
noise of the score, on the assumption that the true phase histories of the core pixels, which are not known, are
similar to that of $p$.

Pixels whose speckle is correlated with that of a core pixel agree
with the core whatever their true phase. If they were included in
the multilook, small subsets would agree with the core because they contain the core and the pixels whose speckle is correlated with it, and the ensemble would
be truncated in zero-coherence clutter. These pixels form the guard set
$\mathcal G_p=\{q:\ |\rho_{q-p'}|\ge\rho_{\min}\ \text{for some}\ p'\in\mathcal C_p\}\supset\mathcal C_p$,
and they are excluded from the score.

Let $q_1,q_2,\dots$ be the pixels of $\mathcal O_p\setminus\mathcal
G_p$ in the sorted order. The score is evaluated
for the nested subsets
\begin{equation}\label{eq:prefix}
 \mathcal E_k=\{q_1,\dots,q_{L_k}\},\qquad k=1,\dots,K,
\end{equation}
with sizes $L_k=\lfloor L_1\,1.5^{k-1}\rfloor$ from $L_1=L_{\min}=25$ and, as the last size,
$L_K=|\mathcal O_p\setminus\mathcal G_p|$. Each size $L_k$ with $k<K$ lies at least a factor 1.5 below $L_K$
(Section~\ref{supp:truncdetails} of the supplement), so that with fewer than $\lceil1.5\,L_{\min}\rceil=38$ members
outside the guard set there is no smaller candidate and the truncation step is skipped. To form the multilooked phases of these
subsets, let $\mathcal A$ be a set of $R=8$ reference acquisitions spread evenly
over the stack, and let $\mathcal I$ be the set of unordered pairs of
distinct acquisitions of which at least one belongs to $\mathcal A$. It
has $|\mathcal I|=R(N-1)-R(R-1)/2$ elements, so that the cost of the multilooks grows with $RN$ instead of $N^2/2$.
With a single reference acquisition, the outcome of the truncation step would change with the choice of that
acquisition, and $R=8$ is a trade-off between this cost and the phase linking error (Sections~\ref{supp:runtime}
and~\ref{app:options} of the supplement). We write each pair as $(n,r)$ with $r\in\mathcal A$. For $(n,r)\in\mathcal I$, the multilooked interferometric phasor of acquisitions $n$ and $r$ over $\mathcal E_k$
and the multilooked interferogram of the core are
\begin{equation}\label{eq:multilook}
 \hat e_k(n,r)=\frac{\sum_{q\in\mathcal E_k}u_q(n)\,\overline{u_q(r)}}
 {\bigl|\sum_{q\in\mathcal E_k}u_q(n)\,\overline{u_q(r)}\bigr|},\qquad
 \bar v_{\mathcal C}(n,r)=\frac1{|\mathcal C_p|}\sum_{p'\in\mathcal C_p}u_{p'}(n)\,\overline{u_{p'}(r)},
\end{equation}
respectively. The score is the agreement of the two in the pair form of corrected SPC
(Eq.~\eqref{eq:agreement}), over the pairs of $\mathcal I$ instead of all pairs, as the mean of the terms $y_k(n,r)$,
\begin{equation}\label{eq:score}
 a_k=\frac1{|\mathcal I|}\sum_{(n,r)\in\mathcal I}y_k(n,r),\qquad
 y_k(n,r)=\operatorname{Re}\bigl[\,\overline{\bar v_{\mathcal C}(n,r)}\,\hat e_k(n,r)\bigr],
\end{equation}
i.e., the mean over the pairs of the cosine of the difference between the multilooked phases of $\mathcal E_k$ and of
the core, each multiplied by $|\bar v_{\mathcal C}(n,r)|$, the phase-only coherence of the core in that pair
(Eq.~\eqref{eq:phaseonlycoh} with $\mathcal W=\mathcal C_p$). Pairs in which the core has decorrelated thus weigh
less.

For a single core pixel $p'$, a one-pixel ensemble
$\mathcal E_k=\{q\}$ and all acquisitions in $\mathcal A$, Eq.~\eqref{eq:agreement} gives $a_k=Q_{p'q}$, the mean
cosine agreement of the interferometric phases of $p'$ and $q$. The score generalizes this to the multilooked
interferograms of $\mathcal E_k$ and of the core, which need not be consistent with a single phase history, and to
the pairs of $\mathcal I$. Like $Q_{pq}$, the score has zero mean when the phases of the core
are unrelated to those of the ensemble, so, unlike $S_{pq}$, it needs no bias correction.
Section~\ref{supp:scorebias} of the supplement gives its expectation over the speckle of the core and shows that, in the
simulations, SHP selections and orders that use the phases of the center do not bias it.

\paragraph{Decision}
The decision compares the gain of the highest score (Eq.~\eqref{eq:score}) over the score of the whole ensemble
with the noise of that gain, and the highest score with its own noise. Let $a_{\max}=\max_ka_k$, attained at
$k=k_{\max}$. The truncation step retains the whole ensemble unless a smaller subset scores higher than the whole
ensemble by more than the margin $\zeta$ and its score exceeds $\lambda$ times its noise $\sigma_{k_{\max}}$ (Eq.~\eqref{eq:snoise} below):
\begin{equation}\label{eq:stop}
 \mathcal E^\star=
 \begin{cases}
 \mathcal O_p & \text{if } a_{\max}-a_K\le \zeta \text{ or } a_{\max}\le\lambda\,\sigma_{k_{\max}},\\[2pt]
 \{q\in\mathcal O_p:\ \nu(q)\le\nu(q_{L_{k^\star}})\} & \text{otherwise,}
 \end{cases}
\end{equation}
with $k^\star=\max\{k:\ a_k\ge a_{\max}-\zeta/3\}$, the largest size whose score lies within the tolerance $\zeta/3$
of the maximum. A truncated ensemble keeps the guard pixels that come before $q_{L_{k^\star}}$ and at least $L_1$
members outside the guard set, so that it never falls below the minimum size $L_{\min}$ of
Section~\ref{sec:pairwise}.

The margin is proportional to the noise of the gain $\Delta a=a_{k_{\max}}-a_K$, the mean of the term differences
$\Delta y(n,r)=y_{k_{\max}}(n,r)-y_K(n,r)$ over $\mathcal I$. Where the core carries information about the true phase,
terms that share an acquisition are correlated, so that the standard deviation $\sigma_{\Delta a}$ of $\Delta a$ is
estimated by a jackknife over acquisitions \citep{efron1981jackknife}, given in Eq.~\eqref{eq:jackknife} of the
supplement. The margin is $\zeta=\kappa\,\sigma_{\Delta a}$, with the margin factor $\kappa=\vcKappa$. The
decision does not explicitly weigh the looks that truncation gives up. Instead, the margin lets a smaller ensemble trigger truncation only if its gain in agreement with the core exceeds $\kappa$
times the standard deviation of that gain.

The second condition of Eq.~\eqref{eq:stop} limits truncation where the core carries almost no information about
the true phase. All scores then scatter about zero, and their maximum over the $K$ sizes can exceed the score of the
whole ensemble by more than $\zeta$ by chance. In the limiting case of a core with zero coherence, every term $y_k(n,r)$ has zero expectation and
different terms are uncorrelated, so that the standard deviation of $a_k$ is estimated by
\begin{equation}\label{eq:snoise}
 \sigma_k=\frac1{|\mathcal I|}\Bigl(\sum_{(n,r)\in\mathcal I}y_k(n,r)^2\Bigr)^{1/2},
\end{equation}
and $a_{\max}$ must exceed $\lambda\,\sigma_{k_{\max}}$, with the noise multiplier $\lambda=\vcLambda$. With \WSname, the gate on the PCI (below) skips the truncation
step at the majority of such centers, so that the choice of $\lambda$ changes the error little (Table~\ref{tab:variants7} of the
supplement).

The rule (Eq.~\eqref{eq:stop}) has no stated significance level, since the
terms of the score are not independent where the core is coherent and the maximum is taken over the
sizes.

\paragraph{Gate on the PCI}
The gate is a one-sided test of $\eta_p$ against $\eta_0$ (Section~\ref{sec:scaledwindow}). In
homogeneous clutter with zero coherence it passes 1\,\% of the pixels, by
the definition of $\eta_0$. At a center below $\eta_0$ the PCI shows no coherence, and a
truncation would respond mostly to noise. In \WSname, the truncation step is therefore
applied only at centers with $\eta_p\ge\eta_0$. At the other centers the ensemble of the SHP selection in $\mathcal B_p$ is returned, since $\mathcal B'_p=\mathcal B_p$
there. In \WSname{} the SHP selection and the truncation step are otherwise unchanged.

\paragraph{Relation to existing adaptive neighborhoods}
Three properties distinguish the truncation step from the adaptive neighborhoods and the nonlocal estimators cited in
Section~\ref{sec:introduction}.
First, the reference is the center. The adaptive multilook of \citet{pepe2021adaptive}, whose agreement measure is
of the same kind as Eq.~\eqref{eq:score} (Section~\ref{sec:related}), judges every pixel against the multilook of a
fixed window, which no longer describes the center where the phase varies within the window. NL-SAR
\citep{deledalle2015nl} selects among nonlocal estimates by their equivalent number of looks, so that phase
differences enter only through the similarity of the patches. The truncation step compares the multilooks of nested
subsets with the phases of the core around the center and thus
responds directly to a bias of the multilooked phases relative to the center. Second, the
decision pools all acquisitions and requires neither a phase model nor a separate noise estimate, whereas the ICI
rule \citep{katkovnik2002adaptive} with local polynomial fits \citep{katkovnik2008phase} compares estimates of one
interferogram at a time through confidence intervals. Third, the guard set keeps the pixels whose speckle is
correlated with that of the core out of the score, so that shared speckle cannot make a small subset agree with the
core. Section~\ref{app:options} of the supplement applies the local
polynomial fit with ICI to every interferogram and compares its error with that of FaSHPS followed by the truncation
step at every center and that of FaSHPS\WSsuffix.

\subsection{Point scatterers}
\label{sec:bright}
A pixel dominated by a strong, stable scatterer requires a different
ensemble, namely the few pixels that carry the response of the same
scatterer, which have the highest SPC with the center but need not be
the nearest. The methods compared below have no such treatment. Point
scatterers are left to the processing of persistent scatterers, and
their errors are reported separately (Section~\ref{sec:simresults}).

\subsection{Estimator of the coherence matrix}
\label{sec:estimate}
The coherence matrix of an ensemble $\mathcal E$ is estimated from
amplitude-normalized samples,
\begin{equation}\label{eq:unitcov}
 \widehat{\bm\Gamma}=\frac1{|\mathcal E|}\sum_{q\in\mathcal E}\bm u_q\bm u_q^{H},
\end{equation}
whose entries are the complex phase-only coherences, i.e.,
Eq.~\eqref{eq:phaseonlycoh} without the magnitude.
Its entries are the first trigonometric moments of the single-look
interferometric phases over the ensemble, so the amplitudes do not
enter. Phase linking based on circular statistics uses the same
phase-only information \citep{yao2024phase,yao2025cltc}. The phases
of $\widehat{\bm\Gamma}$ estimate those of $\bm\Gamma$, and its magnitudes are
reduced by the known, monotone function $g$ of Eq.~\eqref{eq:g}
and can be corrected by inverting it. We use Eq.~\eqref{eq:unitcov}
for every method in this paper, because a few
bright pixels in a large ensemble dominate the ordinary sample
covariance, whereas with Eq.~\eqref{eq:unitcov} they receive the same
weight as any other pixel. In homogeneous Gaussian clutter without texture, however, the ordinary
sample covariance is the efficient estimator, and Eq.~\eqref{eq:unitcov} gives up the
amplitude information there. Over the random scenes of the simulations (Section~\ref{sec:simdesign}), which contain
texture and point scatterers, the ordinary sample covariance has a higher
error over the distributed centers for every SHP selection with the truncation step or with \WSname, most for the
full window (Section~\ref{supp:raw} of the
supplement).

\subsection{Computational cost}
\label{sec:cost}
Per candidate and pass, the amplitude test (Eq.~\eqref{eq:fastat}) requires a few operations, independent of $N$,
since the temporal mean amplitudes are computed once per pixel, and the test of CMP a sum over the $N-1$
single-reference interferograms. The second pass also needs the reference statistics, i.e., the mean amplitude over
the reference set in FaSHPS, and the mean intensity and the multilooked phases over it in CMP. \WSname{} uses the PCI $\eta$, computed
once per image pixel (Section~\ref{sec:evidence}). FaSHPS and CMP need it
for their thresholds in any case (Section~\ref{sec:pairwise}), and for the
full window it is an additional preparation. The truncation step costs considerably more than the PCI and the SHP
tests (Table~\ref{tab:steps} of the supplement). It adds the SPC of the center with the pixels of $\mathcal F_p$ within the search window for the ordering (Eq.~\eqref{eq:moment}), one complex
inner product of length $N$ each, a sort of the members, $R$ running sums of length $N$ per member, which give the multilooks of all $K$ nested
subsets in one pass over the ensemble, and the
evaluation of Eq.~\eqref{eq:score} at $K$ sizes, where $K$ grows with
the logarithm of the ensemble size. The PCI-scaled window reduces the number of candidates at coherent centers, and together with the truncated
ensembles it lowers the cost of the coherence matrix itself.
Section~\ref{sec:runtime} reports the measured times.

\section{Simulation study}
\label{sec:simulation}

\subsection{Design}
\label{sec:simdesign}
\paragraph{Questions}
The simulations ask
\begin{enumerate}[leftmargin=*,itemsep=1pt]
\item how much \WSname{} changes the phase linking error after each SHP selection, and under which local conditions,
\item what its two parts and the gate contribute,
\item how much the choice of SHP selection, and the settings of its tests, matter once \WSname{} is applied,
\item how \WSname{} and the truncation step change the dependence of this error on the size of the search window
that the user chooses,
\item how the results change with fewer acquisitions.
\end{enumerate}

\begin{figure}[tbp]
\centering
\includegraphics[width=\linewidth]{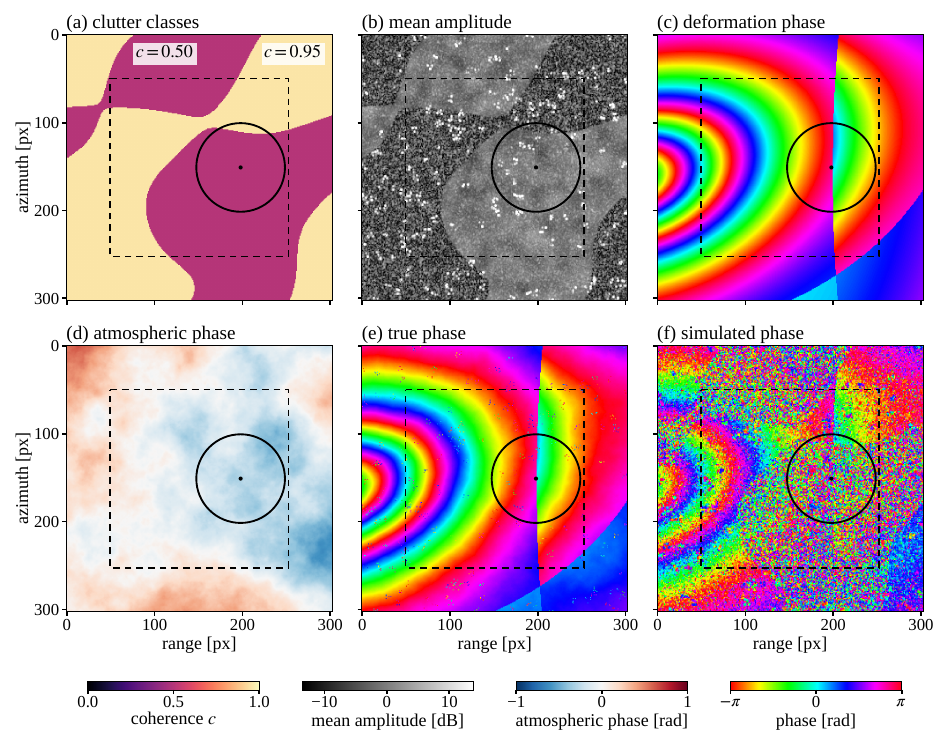}
\caption{One of the random scenes of the TSX-44 configuration and its components, for the interferogram of the last acquisition
with the first. (a) Two clutter classes with coherence $c=\vSceneCohHigh$ and $c=\vSceneCohLow$
(Eq.~\eqref{eq:cohfamily}). (b) Temporal mean amplitude of the simulated stack, with the point scatterers of both
classes. In the coherent class the speckle persists in the temporal mean. (c) Deformation phase, discontinuous along
a fault (vertical) and at the edge of a moving region (lower right). (d) Atmospheric phase, unwrapped and on its own
scale. (e) True phase, the sum of (c) and (d). Each point scatterer adds the phase of its height above the ground,
\vSceneHeightPhase\,rad per meter of height in this interferogram, i.e., more than one cycle at 20\,m. This
phase per meter is \vSceneHeightPhaseRatio{} times the standard deviation of the phase per meter over the acquisitions
(Table~\ref{tab:generator} of the supplement). (f) Simulated
single-look phase. The dashed frame marks the evaluated region and the circle the search window of one center.}
\label{fig:scene}
\end{figure}

\paragraph{Scenes}
The benchmark consists of random scenes, each composed of independent
components (Table~\ref{tab:generator} of the supplement). A scene contains one to three clutter classes,
each with its own coherence $c$, backscatter, texture and point scatterers. Different kinds of land cover differ in
the same properties, but the classes are not modeled on any particular one. The true phase of a scene is the sum of a
deformation phase and an independent atmospheric phase screen for each acquisition. The deformation phase can be
discontinuous along a fault or at the edge of a region that moves while its surroundings stay still, as a
landslide body does. Point scatterers add the phase of their height. The evaluated region
contains the centers whose search window lies inside the scene. Figure~\ref{fig:scene} shows a scene and its
components.

The deformation phase of the last acquisition relative to the first comes from a single smooth source, which
forms a ramp, a bowl, a cone, a trough or a shape in between, depending on its random parameters
(Fig.~\ref{fig:scene}c, Table~\ref{tab:generator} of the supplement). The field is
scaled so that its steepest gradient in the evaluated region gives
$F_{\max}$ fringes per window diameter (Table~\ref{tab:configs} gives one fringe per window diameter in displacement
and rate), and the deformation
grows linearly in time. A window radius is one unit of the normalized distance of Eq.~\eqref{eq:normdist}, and a
window diameter two. The local fringe density of a center is that of this field at the center, and every scene
therefore contains a continuum of local fringe densities up to $F_{\max}$. In half of the scenes with more than one
clutter class, the classes also move separately, so that a class border is a motion border as well. In the
other scenes, all classes share the deformation, as a surface with several kinds of cover moves as a whole. The
atmospheric phase screens have the power-law spectrum of Kolmogorov turbulence
\citep[Eq.~(4.7.11)]{hanssen2001radar}, with the parameters given in Table~\ref{tab:generator} of the supplement. Each scene has one of three strengths (zero, moderate and strong). The strength is the root mean square
(RMS) phase difference across one window diameter in an interferogram of two
acquisitions. The moderate strength of 0.3\,rad is of the order of what the delay model of \citet{emardson2003neutral} gives
at the window extents. For distances much shorter than 1\,km they give an exponent of about 5/6 (their Eq.~(33)) with the same
coefficient as their model for 10 to 500\,km and small height variations (their Eq.~(32)), whose value they
estimate from distances of 10\,km and more. We use this coefficient, with which the two forms agree at 1\,km.
The strong one of 1\,rad is a stress case. The true phase
history of a center includes its own atmosphere, since phase linking
estimates the total phase.

The coherence of each class follows a
one-parameter exponential model \citep[cf.][Eq.~(3.6)]{rocca2007modeling} with
$c$ between consecutive acquisitions, decaying to $c^2$ with a time
constant of ten acquisitions,
\begin{equation}\label{eq:cohfamily}
 |\gamma_{st}|=c^2+(c-c^2)\,e^{-(|s-t|-1)/10},\qquad s\ne t .
\end{equation}
The images are generated in four steps. Every pixel
receives an independent draw from the $N$-dimensional zero-mean complex
Gaussian distribution whose coherence matrix is real and given by
Eq.~\eqref{eq:cohfamily} with the $c$ of its class. The draw is multiplied by
the amplitude and texture of the pixel and by its true phase, the
complex amplitudes of the point scatterers are added, and every
acquisition is filtered with the point spread function of the configuration,
a separable raised-cosine spectrum whose bandwidth gives the lag-one
speckle correlation of Table~\ref{tab:configs}. Because the phase is
applied before the point spread function, a phase gradient inside a resolution
cell reduces the coherence, and the sidelobes of a point scatterer
transfer its phase to neighboring pixels.

A small set of elementary scenes isolates single mechanisms: homogeneous clutter
at $c=0.2$, 0.5 and 0.95, phase steps of 1, 3 and 6.3\,rad at the end of
the stack at $c=0.65$ and 0.95, a phase step of 3\,rad combined with an
amplitude contrast of 6\,dB, and a coherence step between $c=0.65$ and $c=0.2$
with equal phase and amplitude. The elementary scenes serve as illustration, are never
averaged together with the random scenes, and are reported in full in Table~\ref{tab:units} of the supplement.

\paragraph{Configurations and centers}
Three configurations mirror the real stacks of
Section~\ref{sec:realdata} in stack size, search window and point spread function
(Table~\ref{tab:configs}). Each of these configurations, TSX-44, S1-79 and S1-20, has its own random scenes, drawn from the same distribution of components
(Table~\ref{tab:generator} of the supplement) and synthesized on its own pixel grid with its number of acquisitions and
point spread function. S1-20 mirrors the landslide of Section~\ref{sec:stacks}, which is processed as a substack of 20
acquisitions with the settings of this configuration, but the simulated scenes of S1-20 are drawn
independently of those of S1-79. In contrast to the real data, no mask is applied in the simulations.
In each scene, 40 centers are drawn uniformly over the evaluated region. Ten further centers are drawn from each of
four subsets of it if the scene has such pixels, so that a scene has 40 to 80 centers. The subsets are the pixels
within 0.25 window radii of a class border, within 0.25 window radii of a discontinuity, on a point scatterer, and in
the clutter one to six pixels from a scatterer. Centers that do not lie on a point scatterer are distributed
centers, and clutter is defined as in Section~\ref{sec:null}. We call the distributed centers of one value of $c$ a coherence class. With every center we store its local fringe
density, the coherence of its clutter class, its
distances to the nearest class border and discontinuity, its texture,
whether it lies on a scatterer, and the atmospheric strength of its scene.

\begin{table}[tbp]
\centering\footnotesize
\setlength{\tabcolsep}{3.7pt}
\begin{tabular}{lcccl}
\toprule
 & TSX-44 & S1-79 & S1-20 & Source\\
\midrule
\multicolumn{5}{l}{\emph{Configuration}}\\
Acquisitions $N$ & 44 & 79 & 20 & real stacks\\
Search window & $101\times101$ (8021 px) & $25\times225$ (4437 px) & $25\times225$ (4437 px) & real stacks\\
Lag-one speckle correlation (az., rg.) & \vpUsedTsxAz, \vpUsedTsxRg & \vpUsedSOneAz, \vpUsedSOneRg & \vpUsedSOneAz, \vpUsedSOneRg & predicted\\
Core $\mathcal C_p$, guard set $\mathcal G_p$ & \vcCoreTsx, \vcGuardTsx{} px & \vcCoreSOne, \vcGuardSOne{} px & \vcCoreSOne, \vcGuardSOne{} px & Eq.~\eqref{eq:rhopred}\\
One fringe per window diameter & \vfDispTSX\,mm, \vfRateTSX\,mm/yr & \vfDispSLong\,mm, \vfRateSLong\,mm/yr & \vfDispSShort\,mm, \vfRateSShort\,mm/yr & real stacks\\
\addlinespace[3pt]
\multicolumn{5}{l}{\emph{Settings}}\\
Mask threshold $h$ (Eq.~\eqref{eq:local}) & 0.25 & 0.30 & 0.30 & inspection\\
Null quantile $\eta_0$ (Eq.~\eqref{eq:scale}) & \vEtaNullTSX & \vEtaNullSLong & \vEtaNullSShort & calibrated\\
Speckle correlation bound $\rho_{\min}$ (Eq.~\eqref{eq:rhopred}) & \vtRhoMin & = & = & tuned\\
Smallest scaled window $n_{\min}$ (Eq.~\eqref{eq:scale}) & 100 & = & = & fixed\\
Significance level $\alpha$ of the SHP tests & 1\,\% & = & = & fixed\\
Thresholds of the SHP tests & per stack & per stack & per stack & calibrated\\
Minimum size $L_{\min}$ (Eq.~\eqref{eq:prefix}) & 25 & = & = & fixed\\
Reference acquisitions $R$ (Eq.~\eqref{eq:multilook}) & 8 & = & = & fixed\\
Margin factor $\kappa$ (Eq.~\eqref{eq:stop}) & \vcKappa & = & = & tuned\\
Noise multiplier $\lambda$ (Eq.~\eqref{eq:stop}) & \vcLambda & = & = & tuned\\
\bottomrule
\end{tabular}
\caption{Configurations and settings, identical in simulation and on real data except that the simulations
apply no mask. The search window is the ellipse
inscribed in the stated rectangle (azimuth $\times$ range). The window of TSX-44 (44 TerraSAR-X acquisitions) is larger because that stack
decorrelates more strongly and therefore needs more looks. The speckle correlation is the lag-one value that the processing uses (Table~\ref{tab:sensor} of the supplement),
from the prediction from the processing parameters of the sensor (Eq.~\eqref{eq:rhopred}, Sections~\ref{sec:null}
and \ref{sec:realspeckle}). The core, the guard set and the ring of the PCI follow from this prediction at every lag and from
$\rho_{\min}$. One
fringe per window diameter is given as the difference of line-of-sight displacement across the window and as its
rate over the time span of the corresponding real stack (\vfSpanTSX, \vfSpanSLong{} and \vfSpanSShort{} years for TSX-44, S1-79 and S1-20, respectively,
Section~\ref{sec:realdata}). The null quantile and the thresholds were calibrated in homogeneous clutter of the calibration
scenes, $\kappa$, $\lambda$ and $\rho_{\min}$ were tuned on the tuning scenes, and the fixed values were set without tuning and changed only in the checks of
Sections~\ref{app:settings} and~\ref{app:options} of the supplement. The mask
threshold was set by inspection of the NSPC histogram of each real stack (Section~\ref{sec:nspcmask}).
Section~\ref{app:settings} of the supplement discusses the fixed values. An equals sign repeats the value of
TSX-44.}
\label{tab:configs}
\label{tab:settings}
\end{table}

\paragraph{Methods}
Each SHP selection is followed by the estimator (Eq.~\eqref{eq:unitcov}) and by the
same phase linking step. With
$\hat\gamma_{st}=|\hat\gamma_{st}|e^{\ii\hat\phi_{st}}$ the entries of
$\widehat{\bm\Gamma}$, the phase linking estimate
$\hat{\bm\theta}=(\hat\theta_1,\dots,\hat\theta_N)$ is the
maximizer over $\bm\theta$ of
\begin{equation}\label{eq:plobjective}
 J(\bm\theta)=\sum_{s,t}|\hat\gamma_{st}|^{2}\cos\bigl(\hat\phi_{st}-\theta_s+\theta_t\bigr)
 =\bm\vartheta^{H}\bigl(|\widehat{\bm\Gamma}|\odot\widehat{\bm\Gamma}\bigr)\bm\vartheta,
 \qquad \vartheta_n=e^{\ii\theta_n},
\end{equation}
where $\odot$ is the entrywise product. This is the phase triangulation coherence maximization in the form of \citet[Eq.~(15)]{even2018insar} with
exponent 2, i.e., with the squared coherence magnitude as
weight. We maximize $J$ by cyclic coordinate ascent and
damped Newton iterations from the phases of the leading eigenvector of
$|\widehat{\bm\Gamma}|\odot\widehat{\bm\Gamma}$
(Section~\ref{supp:solver} of the supplement), with the same solver on
the real data.

The main comparison contains the three SHP selections of Section~\ref{sec:selection}, the full window, FaSHPS and
CMP, each alone and in three variants, with the settings of Table~\ref{tab:settings}. We write X+S for an SHP
selection X in the PCI-scaled window, X+T for X followed by the truncation step at every center, without the gate,
and X\WSsuffix{} for X with \WSname. With S, the full window is the whole PCI-scaled window.
The supplement reports checks of the design (Section~\ref{app:settings}), the thresholds of the
SHP tests (Section~\ref{app:amptests}) and further methods and options run under the same protocol
(Section~\ref{app:options}).

\paragraph{Errors and reference}
We compute three errors per center. The \emph{matrix error} is the RMS,
over all acquisition pairs, of the wrapped difference between the
phase of the estimated coherence matrix and the true phase
difference of the center. The \emph{phase linking error} is the same for
the phase linking estimate and is the main measure. Using all pairs makes it independent of a reference
acquisition. The
\emph{coherence error} is the RMS
distance between estimated and true complex coherence,
\begin{equation}\label{eq:cerr}
 e_{\Gamma}^2=\frac1M\sum_{s<t}\Bigl[\underbrace{\bigl(|\hat\gamma_{st}|-|\gamma_{st}|\bigr)^2}_{\text{magnitude}}
 +\underbrace{2|\hat\gamma_{st}||\gamma_{st}|\bigl(1-\cos\varepsilon_{st}\bigr)}_{\text{phase, weighted by magnitude}}\Bigr],
\end{equation}
where $\varepsilon_{st}$ is the phase error of pair $(s,t)$, and the true coherence magnitude is the coherence of the
class, reduced by the decorrelation that the difference of the local
phase gradients of the two acquisitions causes within the resolution
cell, and mapped by $g$ of Eq.~\eqref{eq:g} for the estimator (Eq.~\eqref{eq:unitcov}).
The reduction of the magnitude by Eq.~\eqref{eq:unitcov} is a property of the estimator and not of the SHP selection, and it is therefore not counted as an
error. At a point scatterer no true coherence magnitude is defined, so that the coherence error is given over the
distributed centers only.

Errors are compared through a fixed reference, the \emph{oracle}. For
every center it is the nested subset of the whole search window, in the
distance order or the ellipse order of the truncation step (Section~\ref{sec:adaptive}), at sizes from 25 pixels upward
in steps of a factor $\sqrt{1.5}$, with the lowest phase linking error.
Its size is the \emph{oracle size} and its error the \emph{oracle error}
\begin{equation}\label{eq:oracle}
 e^\star_p=\min_{o\in\{\mathrm d,\mathrm e\}}\ \min_{n\in\mathcal K_p}\ e_{\mathrm{PL}}\bigl(\mathcal P_p^{\,o}(n)\bigr),\qquad
 \mathcal K_p=\bigl\{\operatorname{round}\bigl(25\cdot1.5^{j/2}\bigr)<n_p:\ j=0,1,\dots\bigr\}\cup\{n_p\},
\end{equation}
where $\mathcal P_p^{\,o}(n)$ is the set of the first $n$ pixels of the search window of $p$ in the distance order ($o=\mathrm d$)
or the ellipse order ($o=\mathrm e$), $n_p$ the number of pixels of the window and $e_{\mathrm{PL}}$ the phase linking error of an
ensemble.
The oracle requires the true phase and is not a method. Every error is
divided by the oracle error at the same center, and we call this ratio the error relative to the oracle. Since the reference does not
depend on the methods, adding, removing or retuning one method changes
no other method's ratios. A ratio below one means that an SHP selection found a better ensemble than the best nested subset that the oracle tries. At centers with zero coherence
the oracle picks the nested subset whose noise happens to lie closest to the
true phase, so that a ratio above one there does not indicate room for
improvement.

Summaries are weighted geometric means of this ratio over a stated set of centers. The weight of a center is the
uniform density over the evaluated region divided by the density with which the draws of its scene sample its
position, normalized to a sum of one over the centers of the scene. The weighted centers of a scene thus represent
its area, and every scene counts equally. Centers on point scatterers, for example, make up \vsPsCentresMin{} to \vsPsCentresMax\,\% of the centers but
\vsPsWeight\,\% of the weight in each configuration. A comparison of two methods is given as the ratio of their
means. This
ratio equals the weighted geometric mean of the ratio of their errors and
therefore does not depend on the reference. Section~\ref{supp:pooling} of the supplement gives confidence intervals of
these ratios over all centers from a bootstrap resampling of the scenes, and the main results with
the arithmetic mean and the root mean square of the error in place of the geometric mean. An
average over all centers depends on the distribution
of conditions in the scenes, and in particular on the coherence classes,
which are equally frequent here. We therefore give the results against
local fringe density and coherence, and we do not rank methods on an
overall average alone.

\paragraph{Protocol}
Three disjoint sets of scenes are used, each generated in batches from its own random seeds, from which every random
element of a scene and of its centers derives. The calibration scenes, homogeneous clutter at eight coherences from 0
to 0.95 in each configuration, give the second-pass threshold of FaSHPS and the thresholds of CMP over the PCI (Section~\ref{app:calibration} of
the supplement) and, at zero coherence, the null quantile $\eta_0$ (Section~\ref{sec:scaledwindow}). On the tuning
scenes (three batches, \vcTuneScenes{} random scenes and \vcTuneCentresMin{} to \vcTuneCentresMax{} centers per
configuration), $\rho_{\min}$, $\kappa$ and $\lambda$ are tuned in turn by a criterion fixed before the tuning runs.
That criterion is the weighted geometric mean of the phase linking error relative to the oracle over all centers,
averaged over the SHP selections with \WSname{} and over the configurations. Section~\ref{supp:tuning} of the
supplement gives the rules of this tuning and its results. The evaluation scenes (five batches, \vcEvalScenes{} random
scenes and \vcEvalCentresMin{} to \vcEvalCentresMax{} centers per configuration, and \vcUnitScenes{} elementary scenes per configuration)
give the results of Section~\ref{sec:simresults}. No setting is tuned on them. The windows of lowest error in Table~\ref{tab:windows} of the supplement are
identified in hindsight, and the variants and the fixed constants of Section~\ref{app:options} of the supplement are
compared on them.

\paragraph{Reported values}
In the results of the simulations, a value or a range refers to all centers, the three SHP selections and the
configurations TSX-44 and S1-79, unless the sentence or the parentheses after it name other centers, SHP selections or
configurations. A range gives the smallest and the largest value over them. S1-20 is reported in the tables and in the
paragraph on the smaller stack (Section~\ref{sec:simresults}). Values in the text
are computed from the unrounded summaries and can therefore differ from values computed from the rounded entries of
the tables.

\subsection{Results}
\label{sec:simresults}
\begin{figure*}[tbp]
\centering
\includegraphics[width=0.95\linewidth]{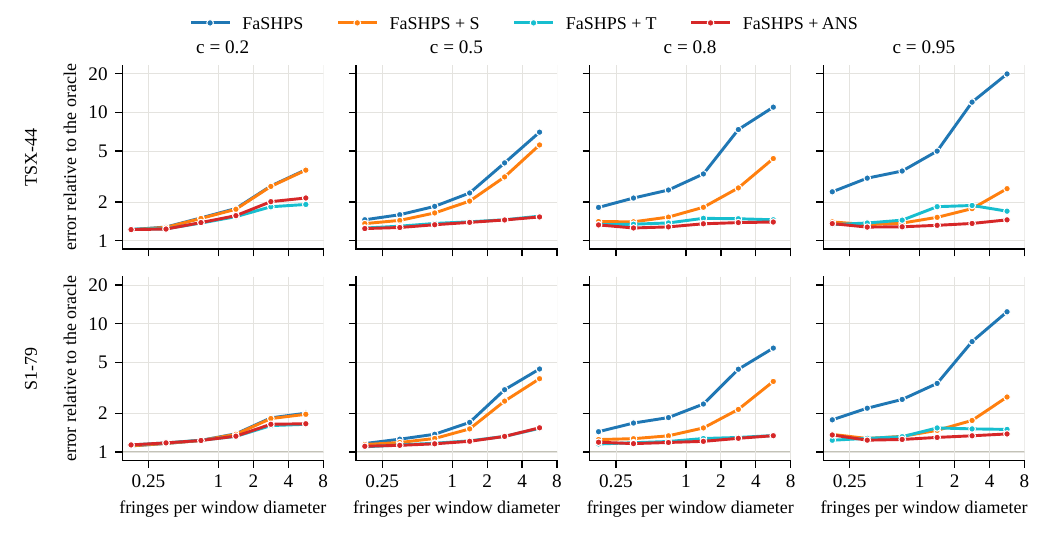}
\caption{The truncation step removes most of the growth of the phase linking error of FaSHPS with the fringe
density in coherent clutter. Phase linking error of FaSHPS relative to the oracle (Eq.~\eqref{eq:oracle}) against the
local fringe density of the deformation between the first and the last acquisition, for four coherence classes in the
configurations TSX-44 and S1-79 (44 TerraSAR-X and 79 Sentinel-1 acquisitions), weighted geometric mean over the
distributed centers of the evaluation scenes in each fringe density bin with at least 30 centers. Colors show the
variant, FaSHPS alone (blue), with the PCI-scaled window (+S, orange), with the truncation step (+T, cyan) and with both
(\WSsuffix, red). Without the truncation step the error grows steeply with the fringe density in coherent clutter, also
within the PCI-scaled window. The full window and CMP, which are not drawn, lie within \vcFigFullFaMaxDuo{} and \vcFigCmpFaMaxDuo\,\% of FaSHPS, respectively, in every bin and variant.}
\label{fig:fringes}
\end{figure*}

\paragraph{Where \WSname{} lowers the error}
\WSname{} lowers the error after every SHP selection, and most of the reduction comes from the truncation step,
which acts most strongly in coherent fringes. Over all centers it lowers the error by \vtChainGainAllDuoMin{} to \vtChainGainAllDuoMax\,\%
relative to the SHP selection alone (Table~\ref{tab:overall}). Without the truncation step the error grows steeply with
the fringe density in coherent clutter (Fig.~\ref{fig:fringes}). At two or more fringes per window diameter and
$c\ge0.5$, the truncation step lowers the error of the full window \vhTruncFrDuoMin{} to \vhTruncFrDuoMax{} times,
and the weighted median of its error from at least \vtAbsFrFullDuoLow$^\circ$ to at most \vtAbsFrSDuoUp$^\circ$. In
zero-coherence clutter it changes the error little, as intended, since there the fringes carry no information. Below
0.25 fringes per window diameter its reduction of the error of the full window grows with the coherence, because
atmospheric screens, class borders and discontinuities make a smaller ensemble preferable even without deformation
fringes.

Where nothing calls for truncation, \WSname{} can raise the error. Per center it raises the error by more than 5\,\% relative to the SHP selection alone at
\vtPerHigherDuoMin{} to \vtPerHigherDuoMax\,\% of the area, mostly in coherent clutter with few fringes, where the
error is a few degrees. In the homogeneous elementary scenes of TSX-44, where ensembles are truncated only by chance,
the truncation step alone raises the 90th percentile of the phase linking error of the full window by less than
\vuCostSUp$^\circ$. \WSname{} raises the error more than the truncation step alone in most of these scenes (Section~\ref{supp:units} of the
supplement). The coherence error and the matrix error follow the
phase linking error (Section~\ref{supp:selmore} and Table~\ref{tab:matrix} of the supplement).

\begin{table*}[tbp]
\centering\footnotesize
\setlength{\tabcolsep}{2.2pt}
\resizebox{\linewidth}{!}{%
\begin{tabular}{lc@{\hspace{8pt}}cccc@{\hspace{6pt}}cccc@{\hspace{6pt}}cccc}
\toprule
Centers & Oracle ($^\circ$) & Full & Full+S & Full+T & Full\WSsuffix & FaSHPS & FaSHPS+S & FaSHPS+T & FaSHPS\WSsuffix & CMP & CMP+S & CMP+T & CMP\WSsuffix \\
\midrule
\multicolumn{14}{l}{\emph{TSX-44}} \\
$c=0$ & 64.6 & \textbf{1.18} & \textbf{1.18} & 1.20 & 1.19 & 1.32 & 1.32 & 1.32 & 1.32 & 1.38 & 1.38 & 1.38 & 1.38 \\
$c=0.2$ & 10.0 & 1.53 & 1.52 & 1.43 & 1.44 & 1.42 & 1.41 & \textbf{1.32} & 1.33 & 1.42 & 1.42 & 1.35 & 1.35 \\
$c=0.35$ & 7.0 & 1.71 & 1.64 & 1.38 & 1.37 & 1.60 & 1.54 & 1.30 & \textbf{1.29} & 1.60 & 1.54 & 1.32 & 1.31 \\
$c=0.5$ & 6.0 & 1.94 & 1.74 & 1.37 & 1.35 & 1.85 & 1.66 & 1.31 & \textbf{1.30} & 1.85 & 1.66 & 1.33 & 1.31 \\
$c=0.65$ & 4.6 & 2.11 & 1.70 & 1.37 & 1.34 & 2.02 & 1.63 & 1.32 & \textbf{1.30} & 2.04 & 1.66 & 1.35 & 1.33 \\
$c=0.8$ & 3.5 & 2.50 & 1.63 & 1.40 & 1.35 & 2.44 & 1.60 & 1.37 & \textbf{1.32} & 2.46 & 1.62 & 1.38 & 1.33 \\
$c=0.95$ & 2.5 & 3.56 & 1.49 & 1.48 & 1.36 & 3.50 & 1.47 & 1.45 & \textbf{1.34} & 3.53 & 1.48 & 1.46 & \textbf{1.34} \\
\addlinespace[2pt]
At least one fringe & 6.7 & 3.55 & 2.26 & 1.50 & \textbf{1.43} & 3.54 & 2.26 & 1.50 & \textbf{1.43} & 3.58 & 2.29 & 1.54 & 1.46 \\
Point scatterers & 19.5 & 2.72 & 2.60 & 2.10 & 2.11 & 1.74 & 1.67 & 1.51 & 1.51 & 0.94 & 0.92 & \textbf{0.91} & \textbf{0.91} \\
\addlinespace[2pt]
All & 5.8 & 1.95 & 1.55 & 1.37 & 1.34 & 1.90 & 1.52 & 1.34 & 1.31 & 1.92 & 1.53 & 1.36 & 1.33 \\
\midrule
\multicolumn{14}{l}{\emph{S1-79}} \\
$c=0$ & 68.0 & \textbf{1.16} & \textbf{1.16} & 1.17 & 1.17 & 1.32 & 1.32 & 1.33 & 1.33 & 1.36 & 1.36 & 1.36 & 1.36 \\
$c=0.2$ & 15.6 & 1.35 & 1.35 & 1.31 & 1.32 & 1.24 & 1.24 & \textbf{1.22} & 1.23 & 1.28 & 1.28 & 1.26 & 1.26 \\
$c=0.35$ & 11.3 & 1.43 & 1.41 & 1.29 & 1.29 & 1.28 & 1.25 & \textbf{1.16} & \textbf{1.16} & 1.29 & 1.27 & 1.18 & 1.18 \\
$c=0.5$ & 9.2 & 1.57 & 1.48 & 1.26 & 1.26 & 1.44 & 1.35 & \textbf{1.16} & \textbf{1.16} & 1.46 & 1.37 & 1.18 & 1.19 \\
$c=0.65$ & 7.7 & 1.79 & 1.52 & 1.26 & 1.26 & 1.68 & 1.42 & \textbf{1.18} & \textbf{1.18} & 1.69 & 1.45 & 1.21 & 1.21 \\
$c=0.8$ & 6.1 & 1.96 & 1.49 & 1.26 & 1.26 & 1.87 & 1.42 & \textbf{1.20} & \textbf{1.20} & 1.90 & 1.44 & 1.22 & 1.23 \\
$c=0.95$ & 3.7 & 2.55 & 1.48 & 1.35 & 1.37 & 2.46 & 1.43 & \textbf{1.31} & 1.32 & 2.50 & 1.45 & 1.33 & 1.34 \\
\addlinespace[2pt]
At least one fringe & 10.8 & 2.66 & 1.91 & 1.41 & 1.38 & 2.59 & 1.85 & 1.37 & \textbf{1.34} & 2.63 & 1.89 & 1.40 & 1.37 \\
Point scatterers & 14.7 & 3.36 & 3.14 & 2.64 & 2.61 & 2.66 & 2.50 & 2.22 & 2.19 & 1.22 & 1.19 & 1.19 & \textbf{1.18} \\
\addlinespace[2pt]
All & 8.8 & 1.64 & 1.41 & 1.28 & 1.28 & 1.57 & 1.35 & 1.22 & 1.23 & 1.59 & 1.37 & 1.25 & 1.25 \\
\midrule
\multicolumn{14}{l}{\emph{S1-20}} \\
$c=0$ & 69.3 & \textbf{1.16} & \textbf{1.16} & 1.18 & 1.17 & 1.21 & 1.21 & 1.23 & 1.22 & 1.24 & 1.24 & 1.25 & 1.24 \\
$c=0.2$ & 13.0 & 1.44 & 1.43 & 1.35 & 1.39 & 1.27 & 1.27 & \textbf{1.21} & 1.23 & 1.28 & 1.27 & 1.22 & 1.24 \\
$c=0.35$ & 9.6 & 1.57 & 1.53 & 1.35 & 1.35 & 1.43 & 1.39 & \textbf{1.23} & 1.24 & 1.43 & 1.40 & 1.24 & 1.24 \\
$c=0.5$ & 7.7 & 1.74 & 1.60 & 1.33 & 1.33 & 1.60 & 1.47 & \textbf{1.23} & \textbf{1.23} & 1.61 & 1.49 & 1.25 & 1.24 \\
$c=0.65$ & 6.3 & 1.91 & 1.59 & 1.34 & 1.33 & 1.80 & 1.50 & 1.26 & \textbf{1.25} & 1.82 & 1.52 & 1.28 & 1.27 \\
$c=0.8$ & 5.0 & 2.26 & 1.61 & 1.38 & 1.35 & 2.17 & 1.54 & 1.32 & \textbf{1.30} & 2.20 & 1.57 & 1.34 & 1.32 \\
$c=0.95$ & 3.2 & 2.91 & 1.56 & 1.50 & 1.43 & 2.82 & 1.51 & 1.45 & \textbf{1.38} & 2.87 & 1.53 & 1.47 & 1.41 \\
\addlinespace[2pt]
At least one fringe & 9.0 & 3.06 & 2.11 & 1.50 & 1.46 & 2.98 & 2.05 & 1.46 & \textbf{1.41} & 3.01 & 2.07 & 1.48 & 1.43 \\
Point scatterers & 14.5 & 3.44 & 3.24 & 2.71 & 2.68 & 3.17 & 2.99 & 2.56 & 2.53 & 1.28 & 1.25 & 1.23 & \textbf{1.22} \\
\addlinespace[2pt]
All & 7.7 & 1.78 & 1.49 & 1.35 & 1.34 & 1.68 & 1.41 & 1.28 & 1.27 & 1.70 & 1.42 & 1.29 & 1.28 \\
\bottomrule
\end{tabular}}
\caption{Phase linking error relative to the oracle (weighted geometric mean, Eq.~\eqref{eq:oracle}) and weighted median oracle error, for the distributed centers of each coherence class, the distributed centers with at least one fringe per window diameter, the centers on point scatterers and all centers (evaluation scenes, amplitude-normalized estimator). The three blocks give the configurations TSX-44, S1-79 and S1-20 (44 TerraSAR-X, 79 and 20 Sentinel-1 acquisitions). Full stands for the full window, S marks the PCI-scaled window and T the truncation step at every center. \WSname{} combines both and applies the truncation step only where $\eta_p\ge\eta_0$ (Sections~\ref{sec:scaledwindow} and~\ref{sec:adaptive}). The lowest value of each row is set in bold, except in the rows over all centers.}
\label{tab:overall}
\end{table*}

\paragraph{Contributions of the two parts}
The truncation step does most of the work of \WSname. Alone, it lowers the error by \vtSGainAllDuoMin{} to
\vtSGainAllDuoMax\,\%, and the PCI-scaled window by \vtScaledGainAllDuoMin{} to \vtScaledGainAllDuoMax\,\%
(Table~\ref{tab:overall}). Together they lower it about as much as the truncation step alone over all centers, and by a
further \vtChainFrGainDuoMin{} to \vtChainFrGainDuoMax\,\% at one fringe or more per window diameter. The PCI-scaled
window thus helps where the window contains coherent fringes or an atmospheric screen (Section~\ref{supp:truncmore} of
the supplement). Where the clutter is uniform over regions much larger than the search window, as in most elementary scenes, the PCI-scaled window raises the error, since it removes looks that need not be excluded, although the errors there are small in degrees.
In the random scenes the truncation step alone is marginally better than \WSname{} in weakly coherent clutter
(Table~\ref{tab:overall}). On the real
stacks the processing with \WSname{} takes \vRtChainSaveMin{} to \vRtChainSaveMax\,\% less time than with the
truncation step alone (Section~\ref{sec:runtime}). The gate keeps the truncation step from acting on noise. In the coherence class $c=0$ of the random scenes it skips
the truncation step at \vtGatedZeroDuoMin{} to \vtGatedZeroDuoMax\,\% of the distributed centers. It rarely passes those
far from point scatterers and class borders, and often those near them (Section~\ref{supp:gate} of the supplement),
and \WSname{} leaves the error of the SHP selection nearly unchanged.

\paragraph{SHP selections and their tests}
\WSname{} improves every SHP selection, and with \WSname{} the three end within a few percent of each other
(Table~\ref{tab:overall}). FaSHPS and CMP keep an advantage of \vmSelAdvWSDuoMin{} to \vmSelAdvWSDuoMax\,\% over the full window
with \WSname{} over all centers, more than the sampling noise of the scenes except for CMP in TSX-44
(Section~\ref{supp:pooling} of the supplement). Without either step they accept most of the search window,
FaSHPS about four fifths and CMP about three quarters of it, at the backscatter contrasts of the scenes, whose classes
differ by at most 6\,dB. In zero-coherence clutter the full window has a lower error than FaSHPS and CMP, because the SHP tests exclude
coherent neighbors of other classes, which carry information on the true phase where the classes move together. On
point scatterers the test of CMP brings the largest reduction, with \vPsCmpFaMin{} to \vPsCmpFaMax{} times the error of
FaSHPS, alone and with \WSname{} (all three configurations). A different significance level of the SHP tests or their
original thresholds in place of those that depend on the PCI change the error by at most \vcSettingsMaxDuo\,\%
(Table~\ref{tab:checks} of the supplement), much less than adding \WSname{} does. A level of 0.1\,\% gives a slightly
lower error than 1\,\% here, because the scenes contain no narrow structures, such as roads a few pixels wide, at which
the additional candidates of a lower level would raise the error. We therefore keep 1\,\%.

\begin{figure}[tbp]
\centering
\includegraphics[width=\linewidth]{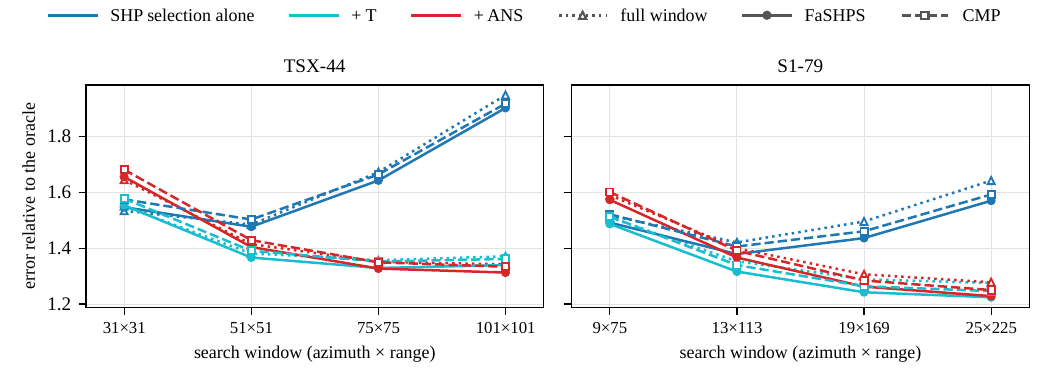}
\caption{With \WSname{} the largest window, which decorrelated areas need, has the lowest error of the tested windows for every SHP selection. Phase linking error of each window relative to the oracle in the largest window, weighted geometric mean over all centers
of the evaluation scenes, against the search window (azimuth $\times$ range, ellipse inscribed in the stated
rectangle), in the configurations TSX-44 and S1-79 (44 TerraSAR-X and 79 Sentinel-1 acquisitions). Colors show the variant, the SHP selection alone (blue), with the truncation
step (+T, cyan) and with \WSname{} (\WSsuffix, red), and line styles the SHP selection, the full window (dotted),
FaSHPS (solid) and CMP (dashed). The values are in Table~\ref{tab:windows} of the supplement. Each SHP selection is drawn separately. Without either step the error is lowest at a middle window, and with
\WSname{} at the largest. At the smaller windows the truncation step alone has a lower error than \WSname, which catches up with it at the
largest.}
\label{fig:windows}
\end{figure}

\paragraph{Size of the search window}
Without either step, every SHP selection has its lowest error at a middle window and \vtWinExcessPcNSLongMin{} to
\vtWinExcessPcNTSXMax\,\% more at the largest (Fig.~\ref{fig:windows}, Table~\ref{tab:windows} of the supplement). With
\WSname{} the largest window has the lowest error, \vwMainWSBestNDuoMin{} to \vwMainWSBestNDuoMax\,\% below that of the
same SHP selection without either step at its best window, so that a window large enough for the decorrelated parts of a scene adds little error at the
others. The search windows of Table~\ref{tab:configs} are those used on the real stacks, where decorrelated areas
require many looks. With the smaller windows the truncation step alone has a lower error than \WSname. \WSname{} catches
up with it at the two largest windows in TSX-44 and at the largest in S1-79. At the windows of Table~\ref{tab:configs}
the processing with \WSname{} takes less time than with the truncation step alone on the real stacks
(Section~\ref{sec:runtime}). Over all centers the size of the window changes the error more than the choice of SHP
selection. The error differs by at least \vwWinSpreadLow\,\% between the best and the worst tested window and by at
most \vwSelSpreadHigh\,\% between the SHP selections at one window (without either step, with the truncation step alone
and with \WSname). On point scatterers it is the other way round. There, without either step, the choice of SHP selection changes the
error by at least \vwSelSpreadPsNDuoLow\,\% and the window by at most \vwWinSpreadPsNDuoUp\,\%.

\paragraph{Smaller stack}
The smaller stack, S1-20, differs from the larger one, S1-79, only in the number of acquisitions. With 20 acquisitions
the PCI is noisier, and its null quantile is $\eta_0=\vEtaNullSShort$ against \vEtaNullSLong{} with 79
(Table~\ref{tab:configs}). The SHP selections then differ slightly more than in S1-79, and the advantage of the largest window with
\WSname{} persists (Table~\ref{tab:overall} and Table~\ref{tab:windows} of the supplement). At high coherence, the original thresholds of
FaSHPS reject fewer homogeneous candidates in excess with fewer acquisitions. In homogeneous clutter at $c=0.95$ they
reject \vcRejFaPubHighSShort\,\% of the candidates in the smaller stack and \vcRejFaPubHighSLong\,\% in the larger one,
whereas the thresholds that depend on the PCI reject at most \vcRejHomFaMax\,\% at every simulated coherence
(Fig.~\ref{fig:rejection}).

\section{Experiments on TerraSAR-X and Sentinel-1 data}
\label{sec:realdata}
The real stacks show what the mask, the PCI and \WSname{} do where no true phase is known. Section~\ref{sec:stacks} describes the datasets. We then check the speckle correlation that sets the
correlated-noise floor of NSPC (Section~\ref{sec:realspeckle}), compare the mask with layover and shadow maps
(Section~\ref{sec:realmask}) and show how \WSname{} changes the phase-linked maps (Section~\ref{sec:realmaps}). Section~\ref{sec:runtime} reports the
runtime on these stacks, and Section~\ref{sec:noquant} explains why we report no quantitative comparison of the SHP
selections on real data.

\subsection{Datasets}
\label{sec:stacks}
The TerraSAR-X stack consists of 44 coregistered acquisitions over Visp, Switzerland, in StripMap mode. The
acquisitions were taken in the summer months between 30 June 2017 and 26 September 2022. We use a scene of
$1000\times1000$ pixels. The topographic phase was simulated from the terrain model swissALTI3D at 0.5\,m spacing
\citep{swisstopo2019alti} and removed. A layover and shadow map in radar geometry, derived from the same model with
the geocoding of the GAMMA software \citep{wegmuller1999automated,frey2013dem,werner2000gamma}, is used for comparison.

The Sentinel-1 site is a section of the Hunza valley in the Karakoram (Gilgit-Baltistan, Pakistan), about 55\,km
long and 15\,km wide, along the Hunza river and the Karakoram Highway from Rahimabad to Shishkat. Its arid slopes of
bare rock and debris remain coherent over years, its extreme relief causes extended layover and shadow, and landslide
inventories along the highway, compiled or updated with InSAR, list hundreds of landslides
\citep{hussain2023landslide,su2021landslide}
(Section~\ref{supp:s1site} of the supplement). The stack consists of 79 Sentinel-1A acquisitions in interferometric
wide-swath mode between 5 January 2023 and 20 December 2025, coregistered and deramped with the GAMMA software
\citep{wegmuller2016sentinel1} (Section~\ref{supp:s1site} of the supplement). No spectral filtering was applied. The
topographic phase was simulated from the Copernicus DEM at one arc-second posting \citep{copdem30}, a surface model
derived from TanDEM-X data, and removed. The geometric layover and shadow map used below is derived from the same
model. The coregistered SLC data is oversampled by a factor of two in range. We use five test areas of $300\times2700$
pixels (Table~\ref{tab:s1crops} of the supplement). They cover the area with the strongest decorrelation of the
five, a slope with slow deformation, the slow-deformation test area, a landslide with fast deformation, the
fast-deformation test area, an area with extended layover and shadow, the layover and shadow test area, and an area that remains coherent over
the whole period, the long-term coherent test area. The test area with the strongest decorrelation is centered on Maiun,
and we refer to it as the Maiun test area. It decorrelates over long intervals and keeps coherence over short ones. The
landslide of the fast-deformation test area is coherent only over short intervals, so that we additionally use a
substack of the 20 acquisitions between 25 November 2023 and 8 September 2024, with the mask computed from the 79
acquisitions of the test area.

Every SHP selection and the truncation step use the settings of Table~\ref{tab:configs}, with the thresholds of FaSHPS
and CMP and the null quantile $\eta_0$ of the matching simulated configuration, TSX-44 for Visp, S1-79 for the
Sentinel-1 test areas and S1-20 for the landslide substack (Section~\ref{sec:simdesign}). For CMP, the decay constant of
the coherence model is fitted to each stack with the estimator used for the simulated scenes and the temporal
baselines of its acquisitions, and its thresholds are tabulated for this constant in simulated clutter of the
configuration. As in the simulations, point scatterers receive no separate treatment.

\subsection{Speckle correlation and correlated-noise floor}
\label{sec:realspeckle}
On TerraSAR-X the speckle correlation estimated from the stack agrees with the prediction from the processing
parameters. The processing uses this prediction (Section~\ref{sec:null}). Only pixels at which the data shows no
detectable coherence enter the estimate, i.e., the pixels whose PCI (Section~\ref{sec:evidence}) lies below its null
quantile (Section~\ref{sec:scaledwindow}), so that their pairs at lag one along each image axis share no signal
beyond the speckle. The mean $\bar Q_1$ of $Q_{pq}$ over these pairs estimates $g(|\rho|)^2$ by
Eq.~\eqref{eq:psfnull}. We solve
\begin{equation}\label{eq:rhoest}
 g(|\rho|)^2=\bar Q_1
\end{equation}
for $|\rho|$. Section~\ref{supp:rhoest} of the supplement describes the extension of the estimate to the other lags and the further
selection of the pairs, which avoids a bias of $\bar Q_1$ by a subtracted model phase and by the brightness of the
surface. On the Visp stack the estimate is \vdEstVispAz{} in azimuth and \vdEstVispRg{} in range, compared with the predicted \vpPredTsxAz{} and \vpPredTsxRg, respectively.

\begin{figure}[tbp]
\centering
\includegraphics[width=\linewidth]{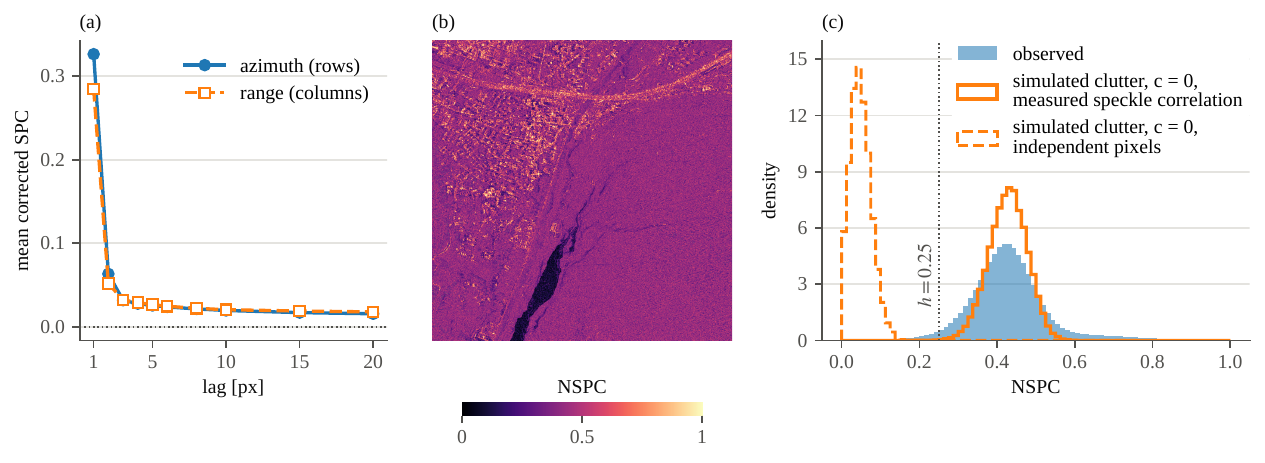}
\caption{Adjacent pixels share speckle, so that their SPC lies far
above that of independent pixels, and the NSPC histogram has its main
mode at this correlated-noise floor. Spatial correlation in the Visp stack ($1000\times1000$
pixels, $N=44$). (a) Mean corrected SPC versus lag in azimuth and
range. The dotted line marks zero, the expected value for independent pixels with zero coherence. (b) NSPC, the mean of the square root of the clipped corrected SPC over the eight direct neighbors
(Eq.~\eqref{eq:local}), after topographic phase removal. The mean
amplitude of the same scene is shown in Fig.~\ref{fig:mask}a. (c) NSPC histogram compared with simulated zero-coherence clutter
with the measured speckle correlation and with independent
pixels.}
\label{fig:psf}
\end{figure}

Adjacent pixels share speckle, so that their SPC lies far above that of independent pixels and falls quickly with
the lag (Fig.~\ref{fig:psf}a). The NSPC histogram of the Visp scene peaks at \vdVispMode, where simulated clutter with
zero coherence and the speckle correlation of this stack has its median, whereas clutter of independent pixels would
have its median at \vdSimFloorIndep{} (Fig.~\ref{fig:psf}c). Most of the scene thus agrees with its neighbors only to
the extent that shared speckle produces.

On Sentinel-1, the estimated correlation in azimuth lies below the prediction.
Equation~\eqref{eq:rhopred} with the Sentinel-1 processing parameters (Table~\ref{tab:sensor} of the supplement)
predicts \vdPredSoneRgOne{} at range lag one and \vdPredSoneAzOne{} in azimuth. The estimator gives \vdEstSoneAzMin{}
to \vdEstSoneAzMax{} in azimuth in the Maiun, fast-deformation and slow-deformation test areas and in the layover and shadow test area.
The correlation of the intensities of neighboring pixels over time, in contrast, measures $|\rho|^2$, and its
square root is close to the prediction over bright surfaces and lower in azimuth over dark surfaces, as expected for
thermal noise, which is less correlated in azimuth
than speckle (Section~\ref{supp:rhoest} of the supplement). As on
TerraSAR-X, the processing uses the prediction.

\subsection{Mask versus layover and shadow}
\label{sec:realmask}
\begin{figure}[p]
\centering
\includegraphics[width=0.92\linewidth]{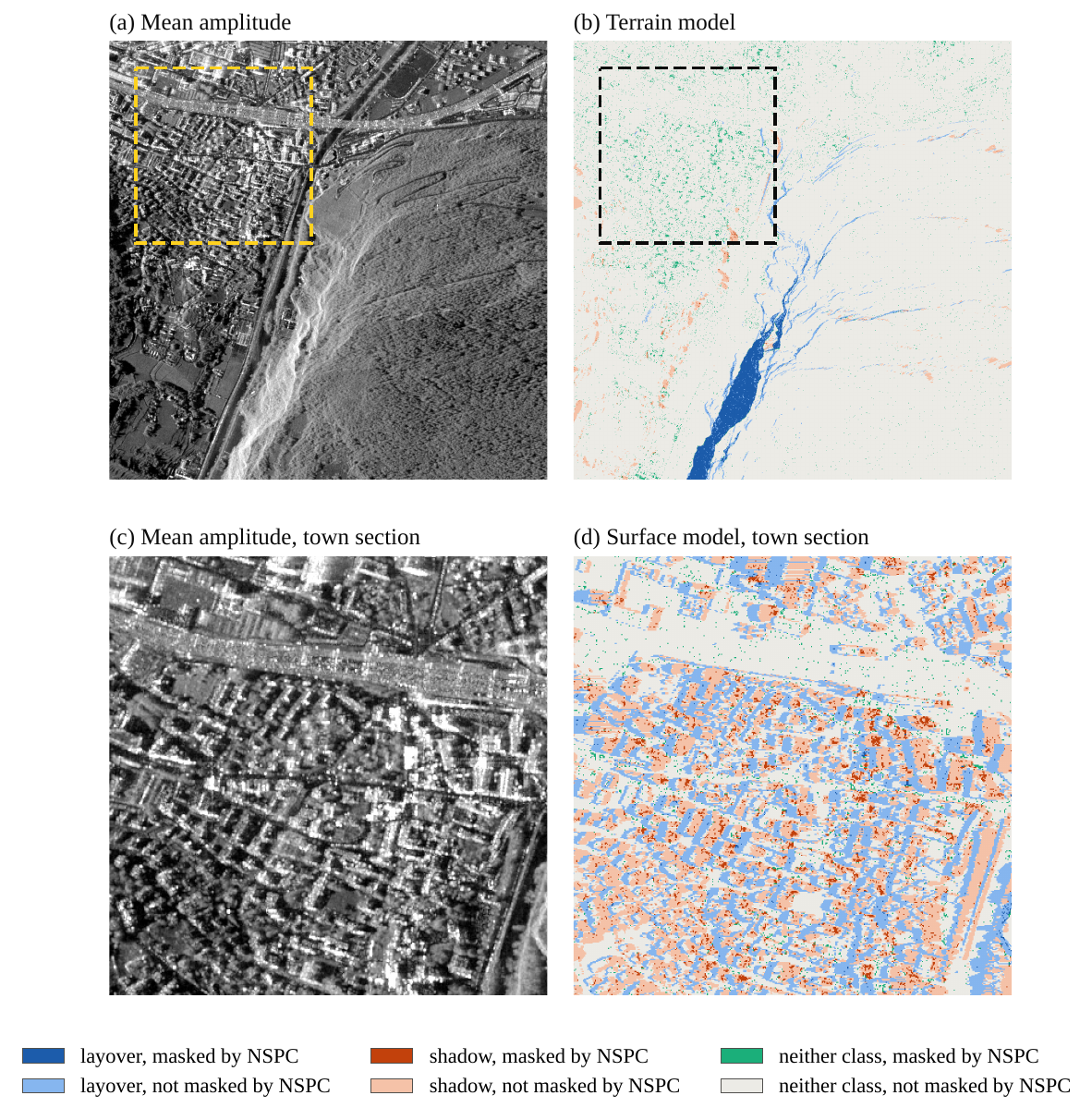}
\caption{NSPC mask (mask threshold $h=0.25$, after topographic phase removal)
compared with layover and shadow maps. Over the whole scene almost half
of the masked pixels lie in layover of the terrain model. In the town
section most of them lie in layover or shadow of the surface model, which
contains the buildings. Top: whole scene
($1000\times1000$ pixels). (a) Mean amplitude. (b) Mask by class of
the terrain model swissALTI3D. Bottom: town section (dashed box in
a and b). (c) Mean amplitude. (d) Mask by class of the surface model swissSURFACE3D Raster. Table~\ref{tab:vispmask} of the supplement gives the
shares of the masked pixels in each class and of each class that is masked. Source of the elevation models: Federal
Office of Topography swisstopo.}
\label{fig:mask}
\end{figure}

On both sensors the mask covers a large part of the layover, and on Sentinel-1 also of the shadow. On Visp it
also flags pixels in the shadow of buildings and dark pixels near bright scatterers, which the terrain model map does not
contain. We set $h$ by inspection of
the NSPC histogram of each stack, at $h=0.25$ on TerraSAR-X and $h=0.30$ on Sentinel-1. Simulated clutter with zero
coherence falls below $h=0.30$ at none of its \vkNullSLongPixels{} pixels with 79 acquisitions. On Visp, \vdVispMasked\,\% of the valid pixels lie
below $h$. They form the layover band on the valley slope and compact patches in the town section of the scene
(Figs.~\ref{fig:psf}b and~\ref{fig:mask}).

In layover, the low NSPC arises from the removal of the topographic phase. The mask covers
\vdVispLayMaskTwentyFive\,\% of the layover of the terrain model on Visp and \vdSoneLayMaskMin{} to
\vdSoneLayMaskMax\,\% of the layover in the Sentinel-1 test areas, against \vdSoneNeiMaskMin{} to \vdSoneNeiMaskMax\,\% of
the pixels in neither class there (Fig.~\ref{fig:s1mask}, Table~\ref{tab:s1mask} of the supplement). Before the
removal it covers less than 5\,\% of the layover on both sensors (Sections~\ref{supp:vispmask}
and~\ref{supp:s1mask} of the supplement). A layover pixel collects echoes from several heights, and the simulated
topographic phase, computed for one height per pixel, changes strongly between adjacent pixels where the phase of
the data, dominated by shared speckle and mixed returns, does not change with it. The mask thus flags pixels where
the topographic correction applied to the stack is inconsistent with the local phase structure of the data.

\begin{figure}[tbp]
\centering
\includegraphics[width=\linewidth]{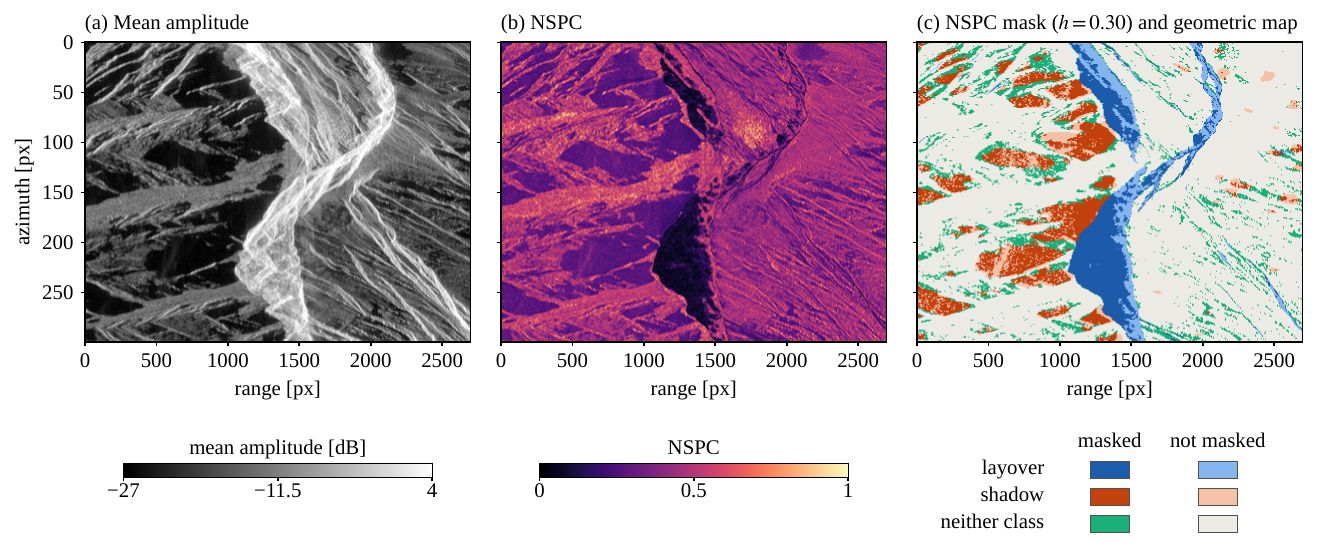}
\caption{On Sentinel-1 the NSPC mask covers a large part of the layover and of the shadow. Test area with layover and shadow
($N=79$). (a) Mean
amplitude. (b) NSPC. (c) NSPC mask at the mask threshold $h=0.30$ compared with the
geometric layover and shadow map, in the colors of Fig.~\ref{fig:mask}. Each pixel is colored by its class in the
map, layover in blue and shadow in orange, dark where the NSPC mask flags it and light where it does not. Masked
pixels outside both classes are green, the other pixels light gray. For display, (a) and (b) are reduced
to a seventh of their width with a Lanczos filter on the color image, and
(c) shows the class with the largest Lanczos weight. The range axis is
drawn compressed by a factor of seven, which approximates the aspect ratio of the test areas on the
ground.}
\label{fig:s1mask}
\end{figure}

A large part of the shadow is masked on Sentinel-1, and hardly any of the shadow of the terrain model on Visp. There the mask covers \vdVispShMaskTwentyFive\,\% of this
shadow, where NSPC stays near the correlated-noise floor, as expected for clutter with zero coherence. In
the Sentinel-1 test areas it covers \vdSoneShMaskMin{} to \vdSoneShMaskMax\,\% of the shadow. Thermal noise, which
dominates there, is less correlated in azimuth than speckle (Section~\ref{sec:realspeckle}), so that its simulated
NSPC has a lower median than that of zero-coherence clutter, closer to the threshold, though still above it at
most pixels. The NSPC of shadow lies lower still (Table~\ref{tab:s1mask} of the supplement) and goes with its low
intensity (Section~\ref{supp:s1mask} of the supplement). The masked share of shadow is
nearly the same before topographic phase removal, so that the low agreement in shadow does not arise from the
correction.

On Sentinel-1 the mask relies on the range oversampling. On the native range grid, i.e., with every second range
sample, simulated clutter with zero coherence has a median NSPC of \vdNatFloorMed{} instead of \vkNullSLongMed{}
(S1-79). A threshold at the 0.1\,\% quantile of NSPC in this clutter then masks about twice as many of the pixels in
neither class as $h=0.30$ on the oversampled grid, in each of the four test areas used for the mask, all but the long-term coherent test area
(Section~\ref{supp:s1mask} and Table~\ref{tab:s1crops} of the supplement).

The mask also flags pixels outside the layover and shadow of the terrain model. In the town section of Visp, a layover and shadow map derived in the same way from the surface model swissSURFACE3D Raster
\citep{swisstopo2022surface}, which includes buildings, trees and other objects above the ground, places
\vdTownMaskedInSurfSh\,\% of the masked pixels in shadow, which covers \vdTownSurfShArea\,\% of the area there
(Fig.~\ref{fig:mask}d). Over the whole scene, \vdCompSceneSurfNei\,\% of the masked pixels lie outside both classes
of the surface model map. They mark disagreement with the neighbors that neither map describes, in particular
near bright scatterers, since dark pixels are masked more often the closer they lie to a bright pixel
(Section~\ref{supp:vispmask} of the supplement). The sidelobes of a bright scatterer alternate in sign from one range
sample to the next beyond the main lobe, so that where a sidelobe is comparable to the clutter, two adjacent pixels
receive the scatterer with opposite signs and the clutter speckle with the same sign, and the two contributions to
their relative phasor cancel. A simulation of point scatterers in clutter with zero coherence, with the generator and
the point spread function of TSX-44, supports this role of the sidelobes (Section~\ref{supp:sidelobe} of the
supplement). On Sentinel-1 the region of low NSPC is wider than the geometric shadow, and thin lines of low NSPC appear
that the geometric map does not contain (Fig.~\ref{fig:s1mask}b and c). The mask therefore complements geometric layover
and shadow maps, also those derived from a surface model.

\subsection{Phase-linked maps}
\label{sec:realmaps}
\WSname{} changes the phase-linked maps mainly where the PCI is high. It preserves the fringes in the
slow-deformation test area and in part on the landslide, and keeps large ensembles where the coherence is low. The maps
are computed with the NSPC mask (Section~\ref{sec:nspcmask}), and masked pixels are shown in gray. On Sentinel-1 the
search window is the ellipse inscribed in $25\times225$ pixels (azimuth $\times$ range, 4437 pixels), of comparable
extent in both ground directions. FaSHPS and CMP accept fewer candidates on the real stacks than in the simulations,
on Visp \vaccFaVisp\,\% and \vaccCmpVisp\,\%, respectively, of the unmasked pixels of the $101\times101$ search window on average.
For FaSHPS the difference arises mainly at centers with a low PCI, where the mean amplitudes spread over the search
window about as much as in the simulated clutter with strong texture or more. On Visp CMP accepts fewer candidates than in
the simulations at every PCI, since the mean intensity of the candidates differs from that of its reference set more
than in the simulated scenes (Section~\ref{supp:accept} of the supplement).

\begin{figure*}[tbp]
\centering
\includegraphics[width=\linewidth]{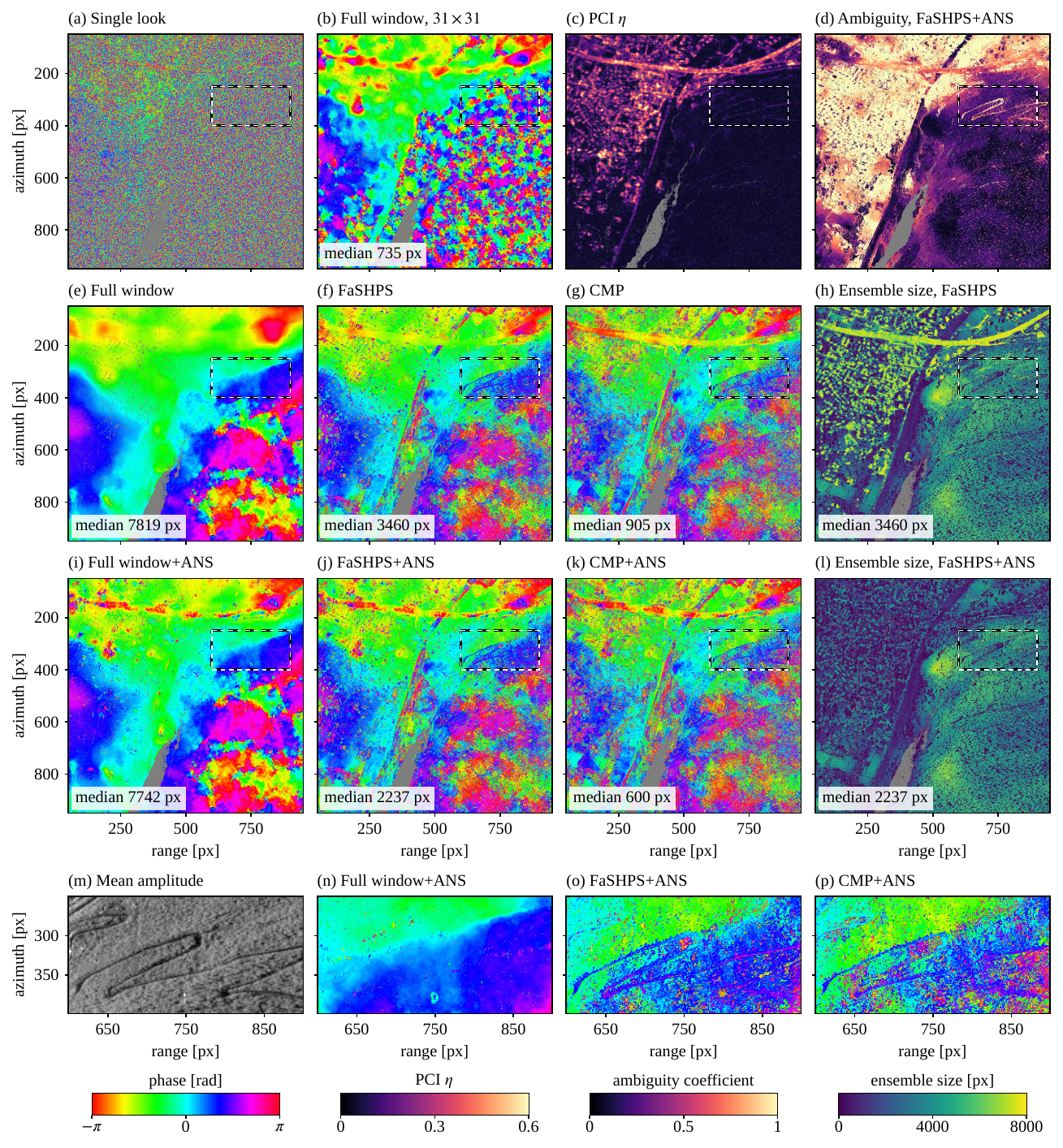}
\caption{On Visp, which has no dense fringes, \WSname{} changes the phase-linked maps mainly in the town
section and along the band of high $\eta$ in the north, where it shrinks the search window and truncates the ensembles. Phase-linked differential
interferogram of the Visp scene between the first and the last acquisition (30 June 2017 to 26 September 2022). With
the small search window the full window gives a patchwork over vegetation, and with the large one the three SHP selections give continuous fields, the smoothest with the full window. The estimate is computed for every unmasked pixel
of the area shown, and pixels masked by NSPC ($h=0.25$) are shown
in gray. The search windows are the ellipses inscribed in the stated rectangles. (a) Single-look interferogram. (b) Full window with a search window of
$31\times31$ pixels. (c) PCI $\eta$, drawn on $[0,0.6]$.
(d) Ambiguity coefficient of FaSHPS\WSsuffix{} \citep{heimpel2026heuristic}.
(e) to (g) Full window, FaSHPS and CMP with a search window of
$101\times101$ pixels. (h) Ensemble size of FaSHPS. (i) to (l) The same
with \WSname, (h) and (l) on a common scale. (m) to (p) Close-ups of the
dashed box, which contains roads crossing the forest: mean amplitude and the three SHP selections with \WSname. FaSHPS\WSsuffix{} keeps the roads as
lines of distinct phase and CMP\WSsuffix{} keeps parts of them, while the full window blends them with the
surrounding forest. The labels give the median ensemble size.}
\label{fig:vispmaps}
\end{figure*}

Visp has no dense fringes, and the decorrelation over its forested and vegetated parts is strong.
Figure~\ref{fig:vispmaps} shows the interferogram between the first and the last acquisition (single look in Fig.~\ref{fig:vispmaps}a). With a search window
of $31\times31$ pixels, the full window yields a patchwork of phase values over vegetation
(Fig.~\ref{fig:vispmaps}b). With $101\times101$ pixels the SHP selections give a continuous field
(Fig.~\ref{fig:vispmaps}e to g). A continuous field alone does not show that its phase is correct, since a large
ensemble also smooths noise. We use the larger window because in the simulations the largest tested window has the
lowest error at centers with zero coherence, for the SHP selections alone, with the truncation step and with
\WSname{} (Section~\ref{supp:windows} of the supplement), and with \WSname{} also over all simulated centers
(Fig.~\ref{fig:windows}). The advantage at zero coherence carries over to real data only where the decorrelated
pixels move with the coherent pixels within the window.

On Visp \WSname{} changes little except in the town and along a band of high $\eta$ in the north. At the
\vrVispBelow\,\% of the centers with $\eta<\eta_0$, it changes neither the search window nor the ensemble. About half
of the truncations lie in the town section, which holds a fifth of the centers and where $\eta$ is higher than
elsewhere (Fig.~\ref{fig:vispmaps}c). There the median ensemble size of FaSHPS falls to about a fifth with \WSname{}, against
about two thirds over the whole scene (Fig.~\ref{fig:vispmaps}h and l). In the town and along the band in the north, the
phase linking estimate of the SHP selections with \WSname{} varies more from pixel to pixel than without it
(Fig.~\ref{fig:vispmaps}e to g and i to k). Whether this variation is the motion of individual structures or the
noise of the smaller ensembles cannot be decided from this data. Elsewhere the maps with and without \WSname{} are
nearly the same.

The close-ups (Fig.~\ref{fig:vispmaps}m to p) show roads crossing the forest. With \WSname, FaSHPS keeps a road as a
line of distinct phase, CMP keeps parts of it, and the full window blends it with the surrounding forest.
Figure~\ref{fig:vispmaps}d shows the ambiguity coefficient of \citet{heimpel2026heuristic} for FaSHPS\WSsuffix. It
compares the maximum of Eq.~\eqref{eq:plobjective} with the largest value that a solution orthogonal to the maximizer
attains, relative to the value expected for noise (their Eq.~(3.22)). We use their dampening constant of 1/3 and fit
the noise value as in their Eq.~(3.13) for the weights of Eq.~\eqref{eq:plobjective}. It is close to one where the
best orthogonal solution attains no more than the noise value and the maximum lies clearly above that value, as on the
roads, in the town and along the railway, and low over most of the forest,
consistent with a road being a target of its own. The coefficient describes how distinct the maximum is for a
given ensemble. It does not
show whether the ensemble is homogeneous.

\begin{figure*}[tbp]
\centering
\includegraphics[width=0.93\linewidth]{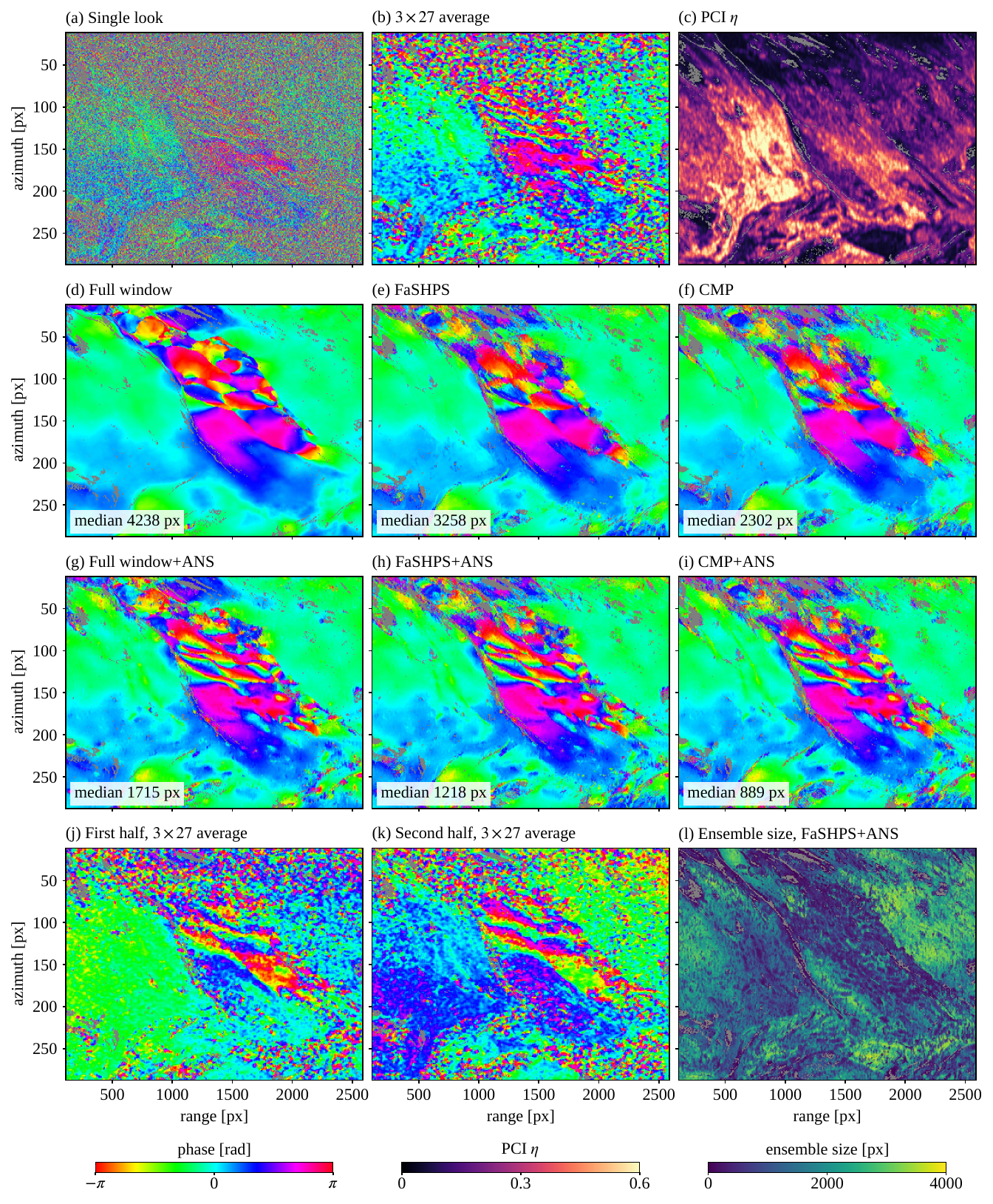}
\caption{On the slope of the Sentinel-1 slow-deformation test area, where fringes accumulate over the interval, the SHP selections without \WSname{} give broad patches separated by phase jumps, and with \WSname{} narrow bands along the
slope, with smaller ensembles on the coherent slope. Phase-linked differential interferogram between acquisitions
29 and 66 (31 December 2023 to 23 June 2025), $N=79$. For display, the color
image of each map is reduced to a seventh of its width with a Lanczos
filter, so that the color is paler where the phases within the filter
disagree, and the range axis is drawn compressed by a factor of seven. Pixels masked by NSPC ($h=0.30$) are shown in gray.
The search window is the ellipse inscribed in the stated rectangle.
(a) Single-look interferogram. (b) The same averaged over $\vfHalfBoxAz\times\vfHalfBoxRg$ pixels.
(c) PCI $\eta$, drawn on $[0,0.6]$. (d) to (f) Full window, FaSHPS and CMP with a search window of $25\times225$
pixels. (g) to (i) The same with \WSname. (j), (k) Single-look interferograms of the first half (acquisitions 29 to 48) and the second half (acquisitions 48
to 66) of the interval, averaged as in (b). (l) Ensemble size of FaSHPS\WSsuffix. The labels give the median
ensemble size.}
\label{fig:s1slow}
\end{figure*}

The slow-deformation test area represents the case for which \WSname{} was designed. The slope remains coherent over the
three years and fringes accumulate on it with time, so that the phase is flat within the window for short intervals
and varies within it for long ones. In the interferogram of Fig.~\ref{fig:s1slow}, the fringe in the lower part of
the slope can be recognized in the single look (Fig.~\ref{fig:s1slow}a). In the upper part of the slope the full
window, FaSHPS and CMP yield broad patches of nearly constant phase, separated by abrupt phase jumps
(Fig.~\ref{fig:s1slow}d to f). With \WSname{} the phase there forms narrow bands along the slope
(Fig.~\ref{fig:s1slow}g to i). Where the estimate of FaSHPS\WSsuffix{} forms these bands, the single look of the same
interval, averaged over $3\times27$ pixels, has a median of \vfSlowBandMed{} fringes per window diameter, so that the
fringes there are dense relative to the search window (Section~\ref{supp:mapsmore} of the supplement gives further values). The
single-look interferograms of the two halves of the interval, averaged over
$\vfHalfBoxAz\times\vfHalfBoxRg$ pixels, show bands of the same orientation (Fig.~\ref{fig:s1slow}j and k), and,
less clearly, so does that of the whole interval with the same average (Fig.~\ref{fig:s1slow}b). The gate passes
almost every center of this test area (Fig.~\ref{fig:s1slow}c). Where FaSHPS\WSsuffix{} forms the bands, the truncation step cuts most
ensembles, so that they are much smaller there than elsewhere (Fig.~\ref{fig:s1slow}l). With \WSname{} the median
ensemble size of FaSHPS over the test area falls from \vrSlowMedFa{} to \vrSlowMedFaWS{} pixels
(Fig.~\ref{fig:s1slow}e and~h).

\begin{figure*}[tbp]
\centering
\includegraphics[width=\linewidth]{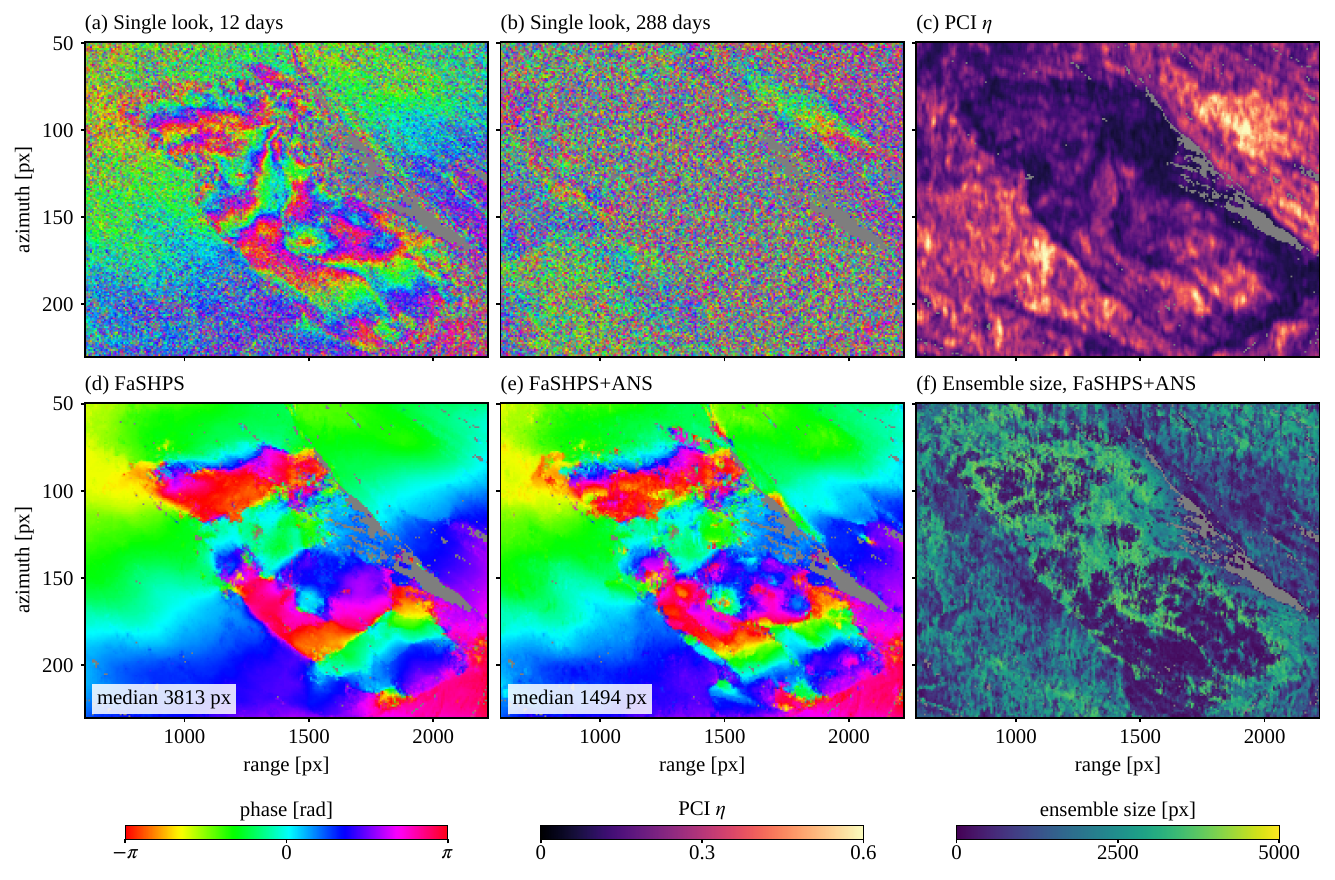}
\caption{Where dense fringes and strong decorrelation over the substack coincide, as on this landslide in the Sentinel-1
fast-deformation test area, \WSname{} recovers the fringes only in part. Phase-linked \vfLandShortDays-day interferogram of
two consecutive acquisitions in the middle of the 20-acquisition substack. FaSHPS alone smooths the dense fringes
into wider bands, and with \WSname{} it follows them where the PCI is higher, in the lower part of the landslide.
The estimate is computed for every unmasked pixel of the area shown, and the display is as in
Fig.~\ref{fig:s1slow}. Pixels masked by NSPC ($h=0.30$, from the 79 acquisitions of the test area) are shown in gray.
(a) Single-look interferogram. (b) Single-look interferogram of the first and last acquisitions of the substack. (c) PCI $\eta$, drawn on
$[0,0.6]$. (d) FaSHPS and (e) FaSHPS\WSsuffix{} for the same \vfLandShortDays-day interferogram, with a search window of $25\times225$ pixels. (f) Ensemble size of
FaSHPS\WSsuffix.} \label{fig:s1land}
\end{figure*}

Figure~\ref{fig:s1land} shows the landslide in the 20-acquisition substack, where the single-look interferograms
contain dense fringes (Fig.~\ref{fig:s1land}a). Within the lobes, FaSHPS yields smooth bands that are wider than those of the single look, and
part of the fringe pattern of the single look is missing (Fig.~\ref{fig:s1land}d). With \WSname{} it follows the
fringes of the single look in the lower part of the landslide, where the PCI is higher (Fig.~\ref{fig:s1land}c). In
the upper part, where the PCI is lowest, it recovers fewer of them, and it is noisier at the margins of the lobes
(Fig.~\ref{fig:s1land}e). Half of the FaSHPS ensembles are truncated, almost a third of these to less than a tenth of
the FaSHPS ensemble at the same center, and the median ensemble size falls to less than half (Fig.~\ref{fig:s1land}d to~f). Between the first and the last acquisition of the substack, \vfLandLongDays{}
days apart, the single look shows no fringes on the landslide (Fig.~\ref{fig:s1land}b). Where dense fringes and
strong decorrelation over the substack coincide in this way, \WSname{} recovers the fringes only in part, and a
longer wavelength would be better suited to this landslide.

On the Maiun test area, where the PCI is low, \WSname{} shrinks the search window little. The median $\eta$ of the test area,
\vrDecMedianEta, is close to that of the simulated class $c=0.35$ with 79 acquisitions (Table~\ref{tab:eta}), since
$\eta$ responds to the coherence of the short intervals. Few centers lie below the null quantile of S1-79, and at the
others the PCI-scaled window is only somewhat smaller than the search window. Over all unmasked centers of the test area
it holds a median of \vrDecMedWindow{} of the 4437 pixels of the search window. Section~\ref{supp:mapsmore} of the supplement gives
further values of the phase-linked maps.

\subsection{Runtime}
\label{sec:runtime}
\begin{table}[tbp]
\centering\small
\resizebox{\linewidth}{!}{%
\begin{tabular}{lcccc}
\toprule
 & Visp & S1 slow deformation & S1 Maiun & S1 long-term coherent\\
 & $N=44$ & $N=79$ & $N=79$ & $N=79$\\
\midrule
Full window & 1.00 (7726) & 1.00 (4146) & 1.00 (4152) & 1.00 (4105) \\
Full window + T & 1.48 (6010) & 1.07 (2541) & 1.22 (3108) & 1.10 (2591) \\
Full window + \WSname & 1.05 (6119) & 0.85 (2062) & 1.05 (2729) & 0.86 (2108) \\
\addlinespace[2pt]
FaSHPS & 0.48 (3337) & 0.72 (2829) & 0.63 (2480) & 0.68 (2635) \\
FaSHPS + T & 0.79 (2566) & 0.84 (1819) & 0.85 (1982) & 0.82 (1752) \\
FaSHPS + \WSname & 0.54 (2525) & 0.64 (1405) & 0.72 (1673) & 0.62 (1348) \\
\addlinespace[2pt]
CMP & 0.46 (1660) & 0.67 (2133) & 0.60 (1824) & 0.60 (1813) \\
CMP + T & 0.71 (1385) & 0.80 (1407) & 0.80 (1519) & 0.76 (1273) \\
CMP + \WSname & 0.52 (1366) & 0.62 (1096) & 0.69 (1298) & 0.58 (992) \\
\addlinespace[2pt]
Full window $31\times31$ & 0.10 (724) & -- & -- & -- \\
\addlinespace[2pt]
NSPC mask, per image pixel & $8\times10^{-5}$ & $9\times10^{-5}$ & $9\times10^{-5}$ & $9\times10^{-5}$ \\
\bottomrule
\end{tabular}}
\caption{Processing time per center pixel up to the coherence matrix (Section~\ref{sec:runtime}) relative to the full window of the same scene, and mean ensemble size in pixels over the timed centers (in parentheses). Search windows of 8021 pixels (Visp) and 4437 pixels (Sentinel-1). Median over five interleaved passes. T marks the truncation step, and \WSname{} the PCI-scaled window with the truncation step gated by $\eta$. The last row gives the time of the NSPC map of the mask per image pixel, relative to the full window per center pixel.}
\label{tab:runtime}
\end{table}
With \WSname{} the processing time of every SHP selection is lower than with the truncation step alone.
Table~\ref{tab:runtime} reports the processing time per center pixel on the real stacks, i.e., the time for the
PCI-scaled window where it is used, the SHP selection, the truncation step where it is used, and the coherence matrix
over the selected ensemble. Preparations that are performed once
per image pixel, such as the PCI, are included. Times are relative to the full window of the same scene, because
absolute times depend on the machine. Section~\ref{supp:runtime} of the supplement describes the measurement.

Without the truncation step, the processing with FaSHPS takes about half to three quarters of the time with the full
window, since its test takes less time than its smaller ensembles save in the coherence matrix, and that with CMP a
similar time. The truncation step adds up to half the processing time of the full window. The full window with the
small search window of $31\times31$ pixels takes a tenth of the processing time with the $101\times101$ window on
Visp (Table~\ref{tab:runtime}), but its map is a
patchwork over vegetation (Fig.~\ref{fig:vispmaps}b). Relative to the processing with the SHP selection alone, \WSname{} adds a little time
on Visp and on the Maiun test area and saves time on the slow-deformation and long-term coherent test areas, where the smaller
window and the truncated ensembles save more time than the truncation step takes. Either way the difference is within
about 15\,\%. For every SHP selection and scene,
the processing with \WSname{} thus takes \vRtChainSaveMin{} to \vRtChainSaveMax\,\% less time than with the
truncation step alone, and that with FaSHPS\WSsuffix{} less time than with the full window alone in every scene. On two other
processors, an \vRtCpuB{} and an \vRtCpuC, the processing with \WSname{} also takes less time than with the truncation step alone. The
NSPC map of the mask for the $1000\times1000$ pixels of the Visp image takes about as long as the full window at
\vRtMaskVispCentres{} center pixels, \vRtMaskVispSeconds\,s on one core of the test machine.

\subsection{On quantitative comparisons with real data}
\label{sec:noquant}
We report no quantitative comparison between the SHP selections on real data, because neither of two measures without
ground truth ranks them reliably in simulation, where the true phase is known (Section~\ref{supp:noquant} of the
supplement). The split-stack agreement is the corrected SPC, over the even acquisitions, between the single-look
phases of the center and the split-stack history, i.e., the phase history estimated from an ensemble that is
selected on the odd acquisitions without the center and the neighbors that share its speckle. Between the full window and a
pairwise SHP selection, both with \WSname, it ranks as the phase linking error of the split-stack history does in
only \vcNqHeldBetweenDuoMin{} to \vcNqHeldBetweenDuoMax\,\% of the random evaluation scenes in which these errors
differ by more than 5\,\% (TSX-44 and S1-79). It is also not independent of what it evaluates, since it is a
corrected SPC with the center, the kind of agreement by which the truncation step chooses the ensemble. The predicted
phase standard deviation, from the Cram\'er--Rao bound \citep[Eq.~(25)]{rodriguez1992theory} with the coherence and
the effective number of looks of the ensemble, describes only the variance of an average, without its bias with
respect to the center. It ranks the full window alone first in about three quarters of the random scenes, whereas
the phase linking error rarely does so.

Between an SHP selection alone and the same SHP selection with \WSname, the split-stack agreement ranks as the
phase linking error of the split-stack history does in most of the scenes in which these errors differ by more than
5\,\%. On \vcNqRealN{} regions of the real stacks (Section~\ref{supp:noquant} of the supplement), it is higher with \WSname{} in all \vhRealBig{} pairs of SHP
selection and region whose values differ by more than 5\,\%. Since the measure prefers \WSname{} in simulation
somewhat more often than the phase linking error of the split-stack history does (Section~\ref{supp:noquant} of the
supplement), this is modest support from real data for \WSname. Beyond this, the comparison of the SHP selections in this
paper rests on the simulation study of Section~\ref{sec:simulation}, and the results on real data are shown as maps.

\section{Discussion}
\label{sec:discussion}

\paragraph{Use of the mask and the PCI}
The mask and \WSname{} can be applied separately, and \WSname{} uses the PCI. The mask is computationally
inexpensive and needs no data beyond the stack after topographic phase removal, and its threshold should be set relative to the correlated-noise floor of the
stack (Section~\ref{sec:nspcmask}). Because it flags pixels in the shadow of buildings that a terrain model map misses
and disagreement that neither map describes (Section~\ref{sec:realmask}), it complements geometric layover and shadow
maps, also those derived from a surface model. The PCI rises with the local coherence and is available before any SHP selection. In this paper it sets the thresholds of the SHP tests, the scale of the search window and the
gate, and it can serve wherever another processing step depends on the local coherence.

\paragraph{Use of \WSname}
With \WSname, a search window large enough for the decorrelated areas adds little error elsewhere
(Fig.~\ref{fig:windows}), so that the user no longer has to trade the decorrelated areas against the fringes when
choosing the window. Neither part of \WSname{} compensates for a window that is too small. The truncation step of
\WSname{} can follow any SHP selection already in use. At one fringe or more per window diameter, most of the reduction by
\WSname{} comes from the truncation step (Table~\ref{tab:overall}). The PCI-scaled window gives a coherent center fewer
looks, since such a center reaches the same precision with fewer, and the gate skips the truncation step where the PCI
shows no coherence. The processing with \WSname{} is therefore faster than with the truncation step alone, at about the same error over all
centers and with less error at one fringe or more per window diameter (Sections~\ref{sec:simresults}
and~\ref{sec:runtime}). Where clutter of moderate or high coherence is uniform over large parts of the window, both parts raise the error,
\WSname{} more than the truncation step alone, since some ensembles are truncated by chance and the PCI-scaled window
gives up looks that need not be excluded (Table~\ref{tab:units} of the supplement). In absolute terms the error added
in the homogeneous elementary scenes with $c=0.5$ and $c=0.95$ is about a degree (TSX-44), against more than ten degrees that \WSname{} removes
at one fringe or more per window diameter in the random scenes. Where large uniform coherent areas dominate a scene,
the truncation step alone is preferable to \WSname, and where such areas need the highest precision, the full window alone.

\paragraph{SHP selections with \WSname}
In the simulations, FaSHPS and CMP with \WSname{} have slightly less error over all centers than the full window
with \WSname, and the choice of the search window changes the error more than the choice of SHP selection, except on point scatterers, where the choice
of SHP selection matters more (Section~\ref{sec:simresults}). The simulated classes, however, differ in backscatter by
at most 6\,dB and contain no structures a few pixels wide, so that the simulations do not penalize a permissive
significance level of the SHP tests. On Visp, the full window with \WSname{} blends roads with the surrounding forest,
whereas FaSHPS\WSsuffix{} keeps them as lines of distinct phase (Section~\ref{sec:realmaps},
Fig.~\ref{fig:vispmaps}m to p). In scenes with such structures, the choice of SHP selection can thus matter more than in
the simulations. On coherent targets, FaSHPS accepts most candidates of similar brightness, and neither SHP test
separates them by phase, since the phase term of CMP on its own rejects practically no candidate
(Section~\ref{sec:pairwise}). The discrimination among such candidates is left to the truncation step, which truncates
at a phase boundary only where the boundary lies close to the center, because the candidate ensembles are nested
around it (Section~\ref{supp:units} of the supplement).

\paragraph{Variants not adopted}
A variant of FaSHPS also accepts candidates that the amplitude test has rejected, if they lie outside the guard set
and their SPC with the center exceeds the 99th percentile of $S_{pq}$ in clutter with zero coherence. This threshold
tests only against the absence of coherence. The variant therefore accepts coherent candidates on another part of a
fringe or across a phase boundary, unless their phase difference to the center changes strongly over the acquisitions
(Eq.~\eqref{eq:ramp}). It gives FaSHPS\WSsuffix{} slightly more error (Table~\ref{tab:variants7} of the supplement). In place of the amplitude test of FaSHPS, the
Kolmogorov--Smirnov, Kuiper, Anderson--Darling and generalized likelihood ratio tests have an error within \vxDistCalPcMax\,\% of
that of FaSHPS when their thresholds depend on the PCI, and \vxDistIndDuoMin{} to \vxDistIndDuoMax{} times its error
with the thresholds for independent samples (TSX-44 and S1-79, Table~\ref{tab:disttests} of the supplement). The calibration of the threshold
thus matters more than the choice of the test. A local fringe frequency compensation before the truncation step hardly
changed the error over all centers, and the local polynomial fit with ICI applied to every interferogram \citep{katkovnik2008phase} had
a higher error than FaSHPS\WSsuffix{} (Section~\ref{app:options} of the supplement). Other variants of the truncation
step were examined and not adopted. The first is the maximum of a Fisher information $|\mathcal E|\,\Qbar/(1-\Qbar)$
over ensembles grown in order of SPC, with $\Qbar$ from Eq.~\eqref{eq:coherence}. The others are ICI
\citep{katkovnik1999new} on a reference history, proxies of the mean squared error, a window derived from the second-moment matrix
(Eq.~\eqref{eq:moment}), an order of all members by SPC, and a score computed from a full phase linking estimate in
place of the multilook (Eq.~\eqref{eq:multilook}). Of these, the rule of largest Fisher information and the order by
SPC have more than 10\,\% more error over all centers than the same SHP selection with \WSname{}
(Table~\ref{tab:variants7} of the supplement).

\paragraph{Limitations}
Several limitations remain. (i) The comparison of the methods rests on simulations with a CCG model. Neither of the two
tested measures without ground truth ranks the SHP selections reliably in simulation. On real data the phase-linked
results are therefore shown only as maps, and the split-stack agreement supports \WSname{} only modestly
(Section~\ref{sec:noquant}). (ii) The simulated phase consists of the deformation, which
grows linearly in time, the atmosphere and, at point scatterers, the phase of their height above the ground
(Section~\ref{sec:simdesign}). Residual topographic phase of the clutter, e.g., from errors of the DEM, and closure
phases, which arise, e.g., from changes of soil moisture \citep{dezan2014sar} and bias the estimated deformation
\citep{ansari2021study}, are not simulated, so the comparison of the methods does not cover them. (iii) The score of the truncation step is not sensitive enough to
truncate every ensemble that is too large, in particular at low coherence (Fig.~\ref{fig:fringes}). With \WSname, the
phase linking error of FaSHPS exceeds twice the oracle error at \vtTailWSTSX\,\% (TSX-44) and \vtTailWSSLong\,\%
(S1-79) of the area of the evaluation scenes, about a third of the share of FaSHPS alone, and a smaller margin does not remove
the ensembles that remain too large (Section~\ref{app:settings} of the supplement). (iv) The thresholds of FaSHPS and CMP are set in simulated
clutter that follows the exponential coherence model (Eq.~\eqref{eq:cohfamily}), and where the coherence of a stack
decays differently, their false rejection rate can depart from the significance level. (v) Point scatterers are not
treated (Section~\ref{sec:bright}). (vi) The NSPC and PCI maps and the other steps of Section~\ref{sec:selection} are
formulated for SLCs in radar coordinates, where the speckle correlation is the product of the correlations along the
two image axes. (vii) Stacks with fewer than 20 acquisitions were not tested. (viii) The bound $\rho_{\min}$, below
which a neighbor counts as free of shared speckle, is tuned once for all configurations. A neighbor at this bound keeps
an expected corrected SPC of $g(\rho_{\min})^2$, which does not fall with $N$, whereas the noise of one SPC value does,
so that for stacks far longer than 79 acquisitions the bound would have to be lowered. It also trades distributed centers against point scatterers
(Sections~\ref{app:settings} and~\ref{supp:sweep} of the supplement). (ix) Of the published SHP selections, only
FaSHPS and CMP are compared in full, CMP only with the single-reference interferograms of its authors
(Section~\ref{supp:cmp} of the supplement), and the distribution tests only in Section~\ref{app:options} of the
supplement. The tests on covariance matrices and the clustering by phase moments of Section~\ref{sec:pairwise} are
not compared.

\section{Conclusions}
\label{sec:conclusions}

We examined spatial phase coherence (SPC), a known pairwise quantity, as a basis for masking and for the size of the
ensembles in phase estimation. Its expectation under the CCG model shows how decorrelation and differences in phase history each
lower it. The speckle that adjacent pixels share sets a floor for it, which the processing parameters predict. Its
mean over all pairs of an ensemble equals the mean corrected squared phase-only coherence of that ensemble. A threshold on its mean over the direct neighbors of each pixel gives a mask that covers a
large part of the geometric layover on TerraSAR-X and Sentinel-1 data and, on Sentinel-1, a similar part of the
shadow. On TerraSAR-X it also flags pixels in the shadow of buildings that a terrain model misses. On both sensors it
flags further pixels that the layover and shadow maps used here do not explain. The mask thus complements
geometrically derived layover and shadow maps.
The mean of SPC over neighbors that share no speckle with the pixel, the PCI, rises with the local coherence and is
available before any SHP selection, and it can assist the calibration of SHP tests that compare temporal mean
amplitudes or intensities, such as those of FaSHPS and CMP. SPC requires no reference acquisition, no coherence model
and no external data, and its behavior under noise, including speckle shared by neighbors, can be stated explicitly.

\WSname, the PCI-scaled window with the gated truncation step, makes the size of the search window less
critical, so that the window can be chosen large enough for the decorrelated areas. In simulations (TSX-44 and S1-79), over all centers, a window large
enough for decorrelated areas has the lowest error of the tested sizes with \WSname{} and \vwMainWSBestNDuoMin{} to
\vwMainWSBestNDuoMax\,\% less error than the same SHP selection without either step at its best size, for the full
window, FaSHPS and CMP alike.
On the real stacks, the processing with \WSname{} up to the coherence matrix takes within about
15\,\% of the time with the SHP selection alone and \vRtChainSaveMin{} to \vRtChainSaveMax\,\% less than with the
truncation step alone, whose error \WSname{} matches over all centers in the simulations.
On Sentinel-1 data, \WSname{} preserves the dense fringes that accumulate on a slowly deforming slope, where the SHP
selections alone give broad patches separated by phase jumps. On a fast landslide, where dense fringes coincide
with strong decorrelation, FaSHPS\WSsuffix{} follows them only in the lower part of the landslide, where the PCI is
higher. In large uniform coherent areas the error is
small, and \WSname{} raises it by about a degree (TSX-44). There the truncation step alone is preferable, and where the
highest precision is needed, the full window without either step.

Several points remain open. A quantitative assessment on real data needs a measure without ground truth that ranks
the SHP selections as the phase linking error does (Section~\ref{sec:noquant}). The score of the truncation step should
become more sensitive at low coherence, where its noise lets ensembles that are too large pass. \WSname{} has yet to be tested on stacks of fewer than 20 acquisitions, where the score rests on few interferograms. Point scatterers, which receive no separate
treatment here, need their own handling. A PCI solved for each pixel as in \citet{vecchioli2023novel} would give the
coherence of the pixel itself instead of an average over its neighborhood. Stacks of geocoded SLCs, finally, need a
model of the speckle correlation on the map grid, where it no longer separates along the image axes.

\needspace{14\baselineskip}
\section*{Abbreviations and notation}
The lists below give the abbreviations and the mathematical notation of the paper, each symbol with the section or equation where it is defined.

\needspace{8\baselineskip}
\subsection*{Abbreviations}
{\small\raggedcolumns\interlinepenalty=10000
\begin{multicols}{2}
\raggedright
\begin{description}[style=sameline,leftmargin=4.4em,labelwidth=4em,labelsep=0.4em,font=\normalfont,itemsep=0pt,parsep=0pt,topsep=0pt]
\item[{ANS}] Adaptive Neighborhood Sizing
\item[{CCG}] Circular Complex Gaussian
\item[{CMP}] Covariance Matrix Patch \citep{zhao2022statistically}
\item[{DEM}] Digital Elevation Model
\item[{FaSHPS}] Fast Statistically Homogeneous Pixel Selection \citep{jiang2015fast}
\item[{ICI}] Intersection of Confidence Intervals
\item[{InSAR}] Interferometric Synthetic Aperture Radar
\item[{NSPC}] Neighborhood Spatial Phase Coherence
\item[{PCI}] Phase Coherence Indicator
\item[{RMS}] Root Mean Square
\item[{S1-20}] Simulated configuration of 20 Sentinel-1 acquisitions
\item[{S1-79}] Simulated configuration of 79 Sentinel-1 acquisitions
\item[{SAR}] Synthetic Aperture Radar
\item[{SHP}] Statistically Homogeneous Pixel
\item[{SLC}] Single-Look Complex
\item[{SPC}] Spatial Phase Coherence
\item[{TSX-44}] Simulated configuration of 44 TerraSAR-X acquisitions
\end{description}
\end{multicols}
}

\needspace{8\baselineskip}
\subsection*{Mathematical notation}
{\small\raggedcolumns\interlinepenalty=10000
\begin{multicols}{2}
\raggedright
\begin{description}[style=nextline,leftmargin=5.6em,labelwidth=5.2em,labelsep=0.4em,font=\normalfont,itemsep=0pt,parsep=0pt,topsep=0pt]
\item[{$N$}] Section~\ref{sec:definition}, number of acquisitions (SLC images)
\item[{$p,q$}] Section~\ref{sec:definition}, pixels, identified with their integer image coordinates
\item[{$n,s,t$}] Section~\ref{sec:definition}, acquisition indices ($r$ for a reference acquisition)
\item[{$x_p(n)$}] Section~\ref{sec:definition}, complex value of pixel $p$ in acquisition $n$
\item[{$u_p(n),\ \phi_p(n)$}] Section~\ref{sec:definition}, unit phasor $x_p(n)/|x_p(n)|$ and its phase
\item[{$\bm u_p$}] Section~\ref{sec:definition}, vector of the $N$ unit phasors of pixel $p$
\item[{$(\cdot)^{\mathsf T},\ (\cdot)^H,\ \overline{(\cdot)}$}] Section~\ref{sec:definition}, transpose, Hermitian (conjugate) transpose, complex conjugate
\item[{$\mathcal W$}] Eq.~\eqref{eq:phaseonlycoh}, spatial neighborhood
\item[{$\widehat\gamma^{\circ}_{st}(\mathcal W)$}] Eq.~\eqref{eq:phaseonlycoh}, phase-only coherence of acquisitions $s,t$ over $\mathcal W$
\item[{$S_{pq}$}] Eq.~\eqref{eq:spc}, spatial phase coherence of pixels $p$ and $q$
\item[{$Q_{pq},\ Q_{pq}^+$}] Eq.~\eqref{eq:q}, corrected SPC and its clipped version
\item[{$\Delta\phi_p^{st},\ v_p^{st}$}] Section~\ref{sec:agreement}, interferometric phase of pixel $p$ for acquisitions $s,t$ and its phasor
\item[{$M$}] Section~\ref{sec:agreement}, number of acquisition pairs, $M=\binom N2$
\item[{$\mathcal E,\ L$}] Eq.~\eqref{eq:coherence}, ensemble of pixels and its size $L=|\mathcal E|$
\item[{$\Qbar(\mathcal E)$}] Eq.~\eqref{eq:coherence}, mean pairwise corrected SPC of $\mathcal E$
\item[{$\gamma_{st},\ |\gamma_{st}|$}] Section~\ref{sec:expectation}, complex coherence of acquisitions $s,t$ and its magnitude
\item[{$g(\cdot)$}] Eq.~\eqref{eq:g}, mean cosine of the single-look interferometric phase error
\item[{$\E[\cdot]$}] Eq.~\eqref{eq:g}, expectation operator
\item[{${}_2F_1$}] Eq.~\eqref{eq:g}, Gauss hypergeometric function
\item[{$\theta_p,\ \Delta\theta_p^{st}$}] Section~\ref{sec:expectation}, true phase history of pixel $p$ and its difference between acquisitions $s,t$
\item[{$z_n,\ \xi_n$}] Eq.~\eqref{eq:EQindep}, relative phasor of two pixels in acquisition $n$ and its mean
\item[{$\bm\ell,\ \rho_{\bm\ell}$}] Section~\ref{sec:null}, spatial lag and complex speckle correlation at that lag
\item[{$\rho_{\min}$}] Section~\ref{sec:null}, speckle correlation below which two pixels are treated as sharing no speckle
\item[{$\bar Q_1$}] Eq.~\eqref{eq:rhoest}, mean corrected SPC of the pixel pairs at lag one with $\eta<\eta_0$
\item[{$W(f),\ B,\ f_{\mathrm s}$}] Eq.~\eqref{eq:rhopred}, spectral weighting, processed bandwidth and sampling rate
\item[{$\delta$}] Eq.~\eqref{eq:ramp}, phase increment per acquisition of a linear relative phase
\item[{$\psi_p(n),\ \psi_{pq}(n)$}] Section~\ref{sec:model}, subtracted model phase and its difference between pixels
\item[{$Q[e^{\ii\psi_{pq}}]$}] Eq.~\eqref{eq:attenuation}, corrected SPC of the model phase difference alone
\item[{$\mathcal V,\ \mathcal V_h$}] Section~\ref{sec:nspcmask}, valid pixels and pixels retained by the mask
\item[{$\mathcal W_p,\ \mathcal N_p$}] Section~\ref{sec:nspcmask}, neighborhood centered on $p$ and valid neighbors of $p$ in it
\item[{$\operatorname{NSPC}(p)$}] Eq.~\eqref{eq:local}, neighborhood SPC
\item[{$h$}] Section~\ref{sec:nspcmask}, mask threshold
\item[{$\mathcal F_p$}] Section~\ref{sec:evidence}, pixels beyond the speckle correlation length of $p$
\item[{$\tilde\eta_p,\ \mathcal R_p$}] Eq.~\eqref{eq:eta}, ring mean of $\sqrt{Q^+_{pq}}$, whose $5\times5$ average is the PCI, and the ring of pixels
\item[{$\eta_p$}] Section~\ref{sec:evidence}, phase coherence indicator (PCI), the mean of $\tilde\eta$ over $5\times5$ pixels
\item[{$\mathcal B_p$}] Section~\ref{sec:scaledwindow}, search window of the center $p$
\item[{$\bm d_q,\ \|\bm d_q\|_b,\ b_1,\ b_2$}] Eq.~\eqref{eq:normdist}, offset $q-p$ of pixel $q$ from the center, its normalized distance and the semiaxes of the search window
\item[{$\eta_0$}] Section~\ref{sec:scaledwindow}, null quantile of the PCI
\item[{$\eta'_p,\ \tau_p$}] Eq.~\eqref{eq:scale}, relative PCI and scale of the window
\item[{$\tau_{\min},\ n_{\min}$}] Section~\ref{sec:scaledwindow}, normalized distance of the $n_{\min}$-th nearest pixel and that number
\item[{$\mathcal B'_p$}] Section~\ref{sec:scaledwindow}, PCI-scaled window
\item[{$\bm\Gamma$}] Section~\ref{sec:amptests}, coherence matrix
\item[{$N_{\mathrm{eff}}$}] Eq.~\eqref{eq:neff}, effective sample size of the temporal mean intensity
\item[{$\bar A_q$}] Section~\ref{sec:pairwise}, temporal mean amplitude of pixel $q$
\item[{$T_{pq},\ \mu_p,\ c_A$}] Eq.~\eqref{eq:fastat}, amplitude statistic, reference mean amplitude and coefficient of variation of a Rayleigh amplitude
\item[{$\alpha$}] Section~\ref{sec:pairwise}, significance level
\item[{$z_1,\ z_\alpha(\eta)$}] Section~\ref{sec:pairwise}, first-pass threshold and second-pass threshold table of FaSHPS
\item[{$\bar I_p$}] Section~\ref{supp:cmp}, temporal mean intensity of pixel $p$
\item[{$D_{pq},\ \gamma_{\mathrm a}(n)$}] Eq.~\eqref{eq:cmpstat}, statistic of CMP and assumed coherence of interferogram $n$
\item[{$\tilde A$}] Section~\ref{supp:accept}, median of $\bar A_q$ over the unmasked pixels of the search window outside the guard set
\item[{$\bar I_{\mathrm r}$}] Section~\ref{supp:accept}, reference intensity of the second pass of CMP
\item[{$\mathcal O_p$}] Section~\ref{sec:pairwise}, ensemble of an SHP selection for the center $p$, passed to the truncation step
\item[{$L_{\min}$}] Section~\ref{sec:pairwise}, minimum ensemble size
\item[{$\bm M_p$}] Eq.~\eqref{eq:moment}, second-moment matrix of the offsets beyond the speckle correlation length, weighted with their corrected SPC with the center
\item[{$\mathcal A,\ R$}] Section~\ref{sec:adaptive}, reference acquisitions of the score and their number
\item[{$q_1,q_2,\dots$}] Section~\ref{sec:adaptive}, ordered pixels of $\mathcal O_p\setminus\mathcal G_p$
\item[{$\mathcal E_k$}] Eq.~\eqref{eq:prefix}, nested subsets of size $L_k$ evaluated by the truncation step
\item[{$\hat e_k(n,r)$}] Eq.~\eqref{eq:multilook}, multilooked interferometric phasor of acquisitions $n,r$ over $\mathcal E_k$
\item[{$\bar v_{\mathcal C}(n,r)$}] Eq.~\eqref{eq:multilook}, multilooked interferogram of the core for acquisitions $n,r$
\item[{$\bar g_{\mathcal C}(n,r),\ \varepsilon_k(n,r)$}] Eq.~\eqref{eq:scoreexp}, mean of $g$ over the core and phase error of $\hat e_k(n,r)$
\item[{$\mathcal G_p,\ \mathcal C_p,\ p'$}] Section~\ref{sec:adaptive}, guard set and core of the center $p$, and a pixel of the core
\item[{$L_k,\ L_1,\ K$}] Section~\ref{sec:adaptive}, candidate ensemble sizes, the smallest of them, $L_1=L_{\min}$, and their number
\item[{$a_k,\ a_{\max}$}] Eqs.~\eqref{eq:score} and \eqref{eq:stop}, score of $\mathcal E_k$ and its maximum
\item[{$\mathcal I,\ |\mathcal I|$}] Section~\ref{sec:adaptive}, unordered pairs of acquisitions that contain a reference acquisition, written $(n,r)$ with $r\in\mathcal A$, and their number
\item[{$y_k(n,r),\ \sigma_k$}] Eqs.~\eqref{eq:score} and~\eqref{eq:snoise}, term of the score of the pair $(n,r)$ and the noise of the score for independent terms
\item[{$\Delta a,\ \Delta y(n,r)$}] Section~\ref{sec:adaptive}, gain $a_{k_{\max}}-a_K$ of the highest score and its terms
\item[{$\mathcal I_s,\ \Delta a_{(s)},\ \overline{\Delta a},\ \sigma_{\Delta a}$}] Eq.~\eqref{eq:jackknife}, pairs that contain acquisition $s$, gain without them, their mean, and the jackknife standard deviation of $\Delta a$
\item[{$\kappa,\ \lambda$}] Section~\ref{sec:adaptive}, margin factor and noise multiplier of the truncation step
\item[{$\zeta$}] Eq.~\eqref{eq:stop}, margin of the truncation step
\item[{$\mathcal E^\star,\ k^\star$}] Eq.~\eqref{eq:stop}, returned ensemble and selected size index
\item[{$\nu(q)$}] Section~\ref{sec:adaptive}, position of $q$ in the sorted ensemble $\mathcal O_p$
\item[{$\widehat{\bm\Gamma}$}] Eq.~\eqref{eq:unitcov}, amplitude-normalized estimate of the coherence matrix
\item[{$\bm k_p(n)$}] Section~\ref{app:options}, local phase gradient at the center $p$ of acquisition $n$ relative to the first
\item[{$c$}] Eq.~\eqref{eq:cohfamily}, coherence between consecutive acquisitions in simulation
\item[{$F_{\max}$}] Section~\ref{sec:simdesign}, largest local fringe density of a scene, in fringes per window diameter between the first and the last acquisition
\item[{$J(\bm\theta)$}] Eq.~\eqref{eq:plobjective}, phase linking objective
\item[{$\hat{\bm\theta}$}] Eq.~\eqref{eq:plobjective}, phase linking estimate, the maximizer of $J$
\item[{$\bm\vartheta,\ \vartheta_n$}] Eq.~\eqref{eq:plobjective}, unit-modulus vector of the argument of $J$ and its entries, $\vartheta_n=e^{\ii\theta_n}$
\item[{$\hat\phi_{st}$}] Eq.~\eqref{eq:plobjective}, phase of the estimated coherence
\item[{$\odot$}] Eq.~\eqref{eq:plobjective}, Hadamard (entrywise) product
\item[{$e_\Gamma,\ \varepsilon_{st}$}] Eq.~\eqref{eq:cerr}, coherence error and phase error of pair $(s,t)$
\item[{$e^\star_p,\ e_{\mathrm{PL}}$}] Eq.~\eqref{eq:oracle}, oracle error of the center $p$ and phase linking error of an ensemble
\item[{$\mathcal P^{\,o}_p(n)$}] Eq.~\eqref{eq:oracle}, first $n$ pixels of the search window of $p$ in the distance order ($o=\mathrm d$) or the ellipse order ($o=\mathrm e$)
\item[{$\mathcal K_p,\ n_p$}] Eq.~\eqref{eq:oracle}, sizes of the oracle candidates and number of pixels of the search window of $p$
\end{description}
\end{multicols}
}

\section*{CRediT authorship contribution statement}
\textbf{Magnus Heimpel:} Writing -- review \& editing, Writing --
original draft, Visualization, Validation, Software, Methodology,
Investigation, Formal analysis, Data curation, Conceptualization.
\textbf{Irena Hajnsek:} Supervision, Resources. \textbf{Othmar
Frey:} Writing -- review \& editing, Validation, Supervision,
Resources, Project administration, Methodology, Funding
acquisition, Conceptualization.

\section*{Declaration of competing interest}
The authors declare that they have no known competing financial
interests or personal relationships that could have appeared to
influence the work reported in this paper.

\section*{Declaration of generative AI and AI-assisted technologies in the manuscript preparation process}
During the preparation of this work the authors used Claude
(Anthropic) in order to write and run simulation and analysis code,
to analyze the simulation results, to produce figures and tables, to
check the reference list, and to draft and revise the text of the
manuscript. The research question, the method, the data and the
decisions on what to test and report were the authors'. After using
this tool, the authors reviewed and edited the content as needed and
take full responsibility for the content of the published article.

\section*{Funding}
This work was funded by the Swiss National Science Foundation (SNSF),
\url{https://data.snf.ch/grants/grant/216526}.

\section*{Data availability}
The TerraSAR-X data of Visp used in this study, \copyright{} DLR 2024, was acquired by the TerraSAR-X
satellite of the German Aerospace Center (DLR) and provided by DLR via an announcement of opportunities
(MTH3910). The acquisition dates are listed in
\citet{heimpel2026heuristic}. The Sentinel-1 data is freely
available from the Copernicus Data Space Ecosystem (\url{https://dataspace.copernicus.eu}), and this paper contains modified Copernicus Sentinel data
(2023 to 2025). The digital elevation
models swissALTI3D and swissSURFACE3D Raster are freely available from
swisstopo (source: Federal Office of Topography swisstopo), and the Copernicus DEM from the Copernicus Data Space Ecosystem. The
topographic phase and the layover and shadow map of the Sentinel-1 test areas were produced using Copernicus WorldDEM-30
\copyright{} DLR e.V. 2010-2014 and \copyright{} Airbus Defence and Space GmbH 2014-2018 provided under COPERNICUS by
the European Union and ESA; all rights reserved.

\section*{Code availability}
The experiment code, with the mask, the PCI, the SHP selections, the PCI-scaled window, the truncation step and the scripts that produce every table and number of this
paper, will be made available upon publication.

\section*{Acknowledgments}
We would like to thank Marco Manzoni and Ehtasham Naseer for
fruitful discussions, and the German Aerospace Center (DLR) for acquiring and providing the TerraSAR-X data.

\appendix
\section{Processing of a stack}
\label{app:algorithm}
Algorithm~\ref{alg:adaptive} gives the processing of a stack with FaSHPS\WSsuffix{} (Fig.~\ref{fig:processing}). With another SHP selection, lines~\ref{alg:selfirst} to~\ref{alg:sellast} are replaced by that SHP selection. Three details do not appear in the algorithm. The PCI
(Section~\ref{sec:evidence}) is computed from all
pixels, masked ones included, except that a pixel with a zero value in some acquisition (\vdVispZero\,\% of Visp)
enters the PCI of no neighbor. Masked pixels are not part of the core of the truncation step, and the second-moment matrix that orders the members of an ensemble (Eq.~\eqref{eq:moment})
leaves out masked pixels and pixels with a zero value in some acquisition.

\begin{algorithm}[htbp]
\caption{Processing of a stack with FaSHPS\WSsuffix}
\label{alg:adaptive}
\begin{algorithmic}[1]
\Require SLC stack, simulated topographic phase, search window $\mathcal B$, ring $\mathcal R_p$ (Section~\ref{sec:evidence}), core $\mathcal C_p$ and guard set $\mathcal G_p$ (Section~\ref{sec:adaptive}) from the predicted speckle correlation, mask threshold $h$, thresholds $z_1$ and $z_\alpha(\eta)$ (Section~\ref{sec:pairwise}), null quantile $\eta_0$, settings of Table~\ref{tab:settings}
\Statex \emph{Per stack}
\State $\mathcal V\gets$ the pixels with finite, nonzero observations in all acquisitions
\State subtract the simulated topographic phase, form unit phasors $\bm u_q$ and temporal mean amplitudes $\bar A_q$
\State NSPC (Eq.~\eqref{eq:local}) for $p\in\mathcal V$ and the mask $\mathcal V_h$
\State $\tilde\eta$ (Eq.~\eqref{eq:eta}) and its $5\times5$ average $\eta$ for every pixel
\Statex \emph{Per center $p\in\mathcal V_h$}
\State $\mathcal B'_p\gets$ the PCI-scaled window (Eq.~\eqref{eq:scale}), restricted to $\mathcal V_h$
\State form the reference set of $p$ from $p$ and the $q\in\mathcal B'_p$ with $T_{pq}\le z_1$ for $\mu_p=\bar A_p$ (Eq.~\eqref{eq:fastat}), then set $\mu_p$ to the mean of $\bar A_q$ over it\label{alg:selfirst}
\State accept $q\in\mathcal B'_p$ if the second-pass statistic $T_{pq}$ is at most $z_\alpha(\eta_p)$
\State $\mathcal O_p\gets$ the eight-connected component of the accepted pixels that contains $p$\label{alg:sellast}
\If{$|\mathcal O_p|<L_{\min}$}
 \State $p$ is insufficient and keeps its single-look data
\Else
 \State $\mathcal E\gets\mathcal O_p$
 \If{$\eta_p\ge\eta_0$ and at least $\lceil1.5\,L_{\min}\rceil=38$ members of $\mathcal O_p$ lie outside $\mathcal G_p$ (Section~\ref{supp:truncdetails})}\label{alg:gate}
  \State sort $\mathcal O_p$ in the ellipse order (Eq.~\eqref{eq:moment}), or in the distance order where it is not defined (Section~\ref{supp:truncdetails} of the supplement)
  \State form the running sums of $u_q(n)\overline{u_q(r)}$ over the sorted members outside $\mathcal G_p$, for $(n,r)\in\mathcal I$ (Sections~\ref{sec:adaptive} and~\ref{sec:cost})
  \For{the sizes $L_1=25<L_2<\dots<L_K$ of Eq.~\eqref{eq:prefix} (Section~\ref{supp:truncdetails} of the supplement)}
   \State $a_k\gets$ score (Eq.~\eqref{eq:score}) of $\mathcal E_k$, and $\sigma_k$ (Eq.~\eqref{eq:snoise})
  \EndFor
  \State $k_{\max}\gets\arg\max_ka_k$, $\sigma_{\Delta a}$ by Eq.~\eqref{eq:jackknife}, $\zeta\gets\kappa\,\sigma_{\Delta a}$
  \If{$a_{k_{\max}}-a_K>\zeta$ and $a_{k_{\max}}>\lambda\,\sigma_{k_{\max}}$}
   \State $k^\star\gets\max\{k:\ a_k\ge a_{k_{\max}}-\zeta/3\}$, $\mathcal E\gets$ the sorted members up to the $L_{k^\star}$-th scored one
  \EndIf
 \EndIf
 \State $\widehat{\bm\Gamma}\gets$ Eq.~\eqref{eq:unitcov} over $\mathcal E$, $\hat{\bm\theta}\gets$ the maximizer of Eq.~\eqref{eq:plobjective}
\EndIf
\end{algorithmic}
\end{algorithm}
\bibliographystyle{plainnat}
\providecommand{\bibfont}{}\renewcommand{\bibfont}{\small}\setlength{\bibsep}{3pt}
\bibliography{references}
\end{document}


\maketitle
\noindent This supplement holds the proofs, further tables, detailed analyses and options that
complete the paper. References to sections, equations,
tables and figures without the prefix S are to the paper. TSX-44, S1-79 and S1-20 are the simulated
configurations of Section~\ref{sec:simdesign}, named by sensor and number of acquisitions.

\section{Spatial phase coherence}
\label{supp:theory}
This section completes Section~\ref{sec:theory}, Section~\ref{sec:nspcmask} and the estimate of the speckle correlation in Section~\ref{sec:realspeckle}.

\subsection{Relation to existing pairwise phase measures}
\label{app:related-measures}

This section gives the details of the relations of Section~\ref{sec:related}.

\paragraph{Zero-lag phase cross-correlation}
At zero lag, the cross-correlation criterion of
\citet[Eq.~(6)]{narayan2018novel} is $\sum_{i=1}^{N_{\mathrm I}}w_p(i)\overline{w_q(i)}$
with $w_p(i)=e^{\ii\chi_p(i)}$ the observed phasor of interferogram
$i$, where $N_{\mathrm I}$ is the number of interferograms. Its
magnitude is $N_{\mathrm I}\sqrt{S_{pq}}$ evaluated on the same
vectors. Their rule compares correlations at different lags,
whereas a threshold on SPC, the Rayleigh test of Section~\ref{sec:related},
compares the zero-lag magnitude with a fixed value. A maximum at zero lag does not by itself establish
significance.

\paragraph{Centered complex correlation}
For vectors of unit phasors, write $a_{pq}=\bm u_p^H\bm u_q/N$ and
$m_p=\sum_nu_p(n)/N$. The centered correlation coefficient of
\citet[Eq.~(1)]{shi2020potential} applied to these vectors is
\begin{equation}
 r_{pq}^{(u)}=\frac{a_{pq}-\overline{m_p}m_q}
 {\sqrt{(1-|m_p|^2)(1-|m_q|^2)}},\qquad S_{pq}=|a_{pq}|^2,
\end{equation}
provided both centered vectors are nonzero. The two coincide when
$m_p=m_q=0$ but differ in general. For a temporally constant phase
the centered coefficient is undefined, whereas SPC equals one. A
constant rotation of either pixel leaves SPC and $|r_{pq}|$
unchanged but changes $\arg r_{pq}$, so that a constraint on $\arg r_{pq}$
distinguishes offsets that SPC treats as equivalent. An acquisition-dependent rotation
common to both pixels leaves SPC unchanged but generally changes
$r_{pq}$ through the temporal means.

\paragraph{Circular variance of high-pass phases}
For each interferogram $k$, \citet[Eq.~(8)]{pepe2021adaptive} form
the high-pass phasor $\exp\bigl(\ii(\varphi_k(q)-\varphi_k^{\mathrm{LP}})\bigr)$
of pixel $q$, where $\varphi_k^{\mathrm{LP}}$ is the phase of the
mean of the unit phasors over the window (their Eqs.~(2) to (4)). The
mean resultant length of these phasors over the interferograms (their
Eq.~(9)) is $\sqrt{S}$ between the phasor vector of $q$ and the unit
phasors of the low-pass phases, and the circular variance (their
Eq.~(10)) is one minus this value. Pixels with a circular variance below
0.7 are retained, and each of them is compared with the retained pixel
of lowest circular variance by Kuiper's test \citep{kuiper1960tests} at
a significance level of 20\,\%.

\paragraph{Pairwise phase consistency and phase-locking value}
\citet[Eqs.~(14) and (15)]{vinck2010pairwise} define the pairwise phase consistency as the mean over pairs of
trials of the cosine of the difference of relative phases, which is
Eq.~\eqref{eq:agreement} with trials in place of acquisitions. They show (their Eqs.~(13) and (19))
that its expected value equals the square of the population value of the phase-locking value of
\citet[p.~195]{lachaux1999measuring} and that, unlike the squared sample
phase-locking value, its sample estimate is free of finite-sample bias.

\subsection{Proofs}
\label{app:proofs}

This section gives the proofs of Eqs.~\eqref{eq:agreement}, \eqref{eq:coherence}, \eqref{eq:EQccg} to~\eqref{eq:psfnull} and~\eqref{eq:attenuation} of Sections~\ref{sec:agreement} to~\ref{sec:model}, of the bound on the expected corrected SPC of direct neighbors in Section~\ref{sec:nspcmask}, and of the standard deviation of corrected SPC under independent uniform phases that Section~\ref{app:settings} uses. The heading of each paragraph names what it proves.

\paragraph{Agreement identity, Eq.~\eqref{eq:agreement}}
For $z_n=\overline{u_p(n)}u_q(n)$,
\begin{equation}
 N^2 S_{pq}=N+2\sum_{s<t}\operatorname{Re}(z_t\overline{z_s})
 =N+2\sum_{s<t}\cos(\Delta\phi_p^{st}-\Delta\phi_q^{st}),
\end{equation}
and rearranging gives Eq.~\eqref{eq:agreement}.

\paragraph{Coherence of an ensemble, Eq.~\eqref{eq:coherence}}
With $v_q^{st}=u_q(s)\overline{u_q(t)}$,
\begin{equation}
 \widehat\gamma^{\circ}_{st}(\mathcal E)^2
 =\frac1{L^2}\sum_{q,q'\in\mathcal E}v_q^{st}\overline{v_{q'}^{st}}
 =\frac1L+\frac1{L^2}\sum_{q\ne q'}\cos\!\left(\Delta\phi_q^{st}-\Delta\phi_{q'}^{st}\right),
\end{equation}
since the off-diagonal terms occur in conjugate pairs. Averaging
over $s<t$ and applying Eq.~\eqref{eq:agreement} to each pair gives
\begin{equation*}
 \frac1M\sum_{s<t}\widehat\gamma^{\circ}_{st}(\mathcal E)^2=\frac1L+\frac{L-1}{L}\,\Qbar(\mathcal E),
\end{equation*}
which rearranges to Eq.~\eqref{eq:coherence}.

\paragraph{Expectations, Eqs.~\eqref{eq:EQccg} to~\eqref{eq:psfnull} and~\eqref{eq:attenuation}}
Each cosine in Eq.~\eqref{eq:agreement} is
$\cos(\Delta\phi_p^{st}-\Delta\phi_q^{st})=\operatorname{Re}\,v_p^{st}\overline{v_q^{st}}$, and for independent pixels
its expectation factorizes as $\operatorname{Re}\,\E[v_p^{st}]\overline{\E[v_q^{st}]}$. For a CCG pixel, write
$\Delta\phi_p^{st}=\Delta\theta_p^{st}+\varepsilon$ with the phase error $\varepsilon=\Delta\phi_p^{st}-\arg\gamma^p_{st}$,
so that
\begin{equation*}
 \E[v_p^{st}]=e^{\ii\Delta\theta_p^{st}}\bigl(\E[\cos\varepsilon]+\ii\,\E[\sin\varepsilon]\bigr).
\end{equation*}
The acquisitions $s$ and $t$ of a CCG vector form a bivariate CCG pair with correlation $\gamma^p_{st}$, so that
$\E[\cos\varepsilon]=g(|\gamma^p_{st}|)$ by Eq.~\eqref{eq:g}. After the second acquisition is multiplied by
$e^{\ii\arg\gamma^p_{st}}$, the pair has the real correlation $|\gamma^p_{st}|$ and a real covariance matrix, so that
its distribution is invariant under complex conjugation. Since conjugation maps $\varepsilon$ to $-\varepsilon$, the
distribution of $\varepsilon$ is symmetric about zero, $\E[\sin\varepsilon]=0$, and
$\E[v_p^{st}]=g(|\gamma^p_{st}|)e^{\ii\Delta\theta_p^{st}}$ (\citealp[Eqs.~(8) and (9)]{lopezmartinez2002modeling};
\citealp[cf.][Eq.~(15)]{just1994phase}; \citealp[Eq.~(21)]{lee1994intensity}). With this for both
pixels, the expectation of each cosine is
$g(|\gamma^p_{st}|)\,g(|\gamma^q_{st}|)\cos(\Delta\theta_p^{st}-\Delta\theta_q^{st})$, and the mean over $s<t$ gives
Eq.~\eqref{eq:EQccg}. For relative phasors independent across
acquisitions,
\begin{equation*}
 \E\Bigl|\sum_nz_n\Bigr|^2=N+\sum_{n\ne m}\xi_n\overline{\xi_m}=N+\Bigl|\sum_n\xi_n\Bigr|^2-\sum_n|\xi_n|^2,
\end{equation*}
where the $N$ comes from $|z_n|=1$ and the sum over $n\ne m$ from the independence, and Eq.~\eqref{eq:EQindep} follows
from Eq.~\eqref{eq:q}. For a bivariate CCG pair with
correlation $\rho$, $\E[z_n]=g(|\rho|)e^{\ii\arg\rho}$, which gives
Eq.~\eqref{eq:psfnull}. Multiplying each $z_n$ by the deterministic
phasor $e^{-\ii\psi_{pq}(n)}$ gives Eq.~\eqref{eq:attenuation}.

\paragraph{Floor under temporal coherence, Section~\ref{sec:nspcmask}}
Eq.~\eqref{eq:EQccg} assumes independent pixels and Eq.~\eqref{eq:EQindep} independent acquisitions. Direct
neighbors with a coherence between zero and one satisfy neither: their values are correlated between the two pixels
through the shared speckle and between acquisitions through the coherence. The floor of Eq.~\eqref{eq:psfnull} remains a lower bound under the separable model of the simulations, in which the
covariance of two pixel values is the product of the temporal coherence of their acquisitions and the speckle
correlation of the two pixels. Let the pixel values $\bm x(n)=(x_p(n),x_q(n))$ be zero-mean CCG with
$\operatorname{Cov}[x_a(s),x_b(t)]=\gamma_{st}\rho_{ab}$ for $a,b\in\{p,q\}$, where $\rho_{pp}=\rho_{qq}=1$,
$\rho_{pq}=\rho$, $\gamma_{ss}=1$ and $0\le\gamma_{st}\le1$ once a true phase history common to both pixels is
removed, which leaves every $z_n$ unchanged.

By Eq.~\eqref{eq:agreement}, since $\cos(\Delta\phi_p^{st}-\Delta\phi_q^{st})=\operatorname{Re}(z_t\overline{z_s})$,
\begin{equation*}
 \E[Q_{pq}]=\frac1M\sum_{s<t}\operatorname{Re}\E[z_t\overline{z_s}],
\end{equation*}
and each term involves only the acquisitions $s$ and $t$. For one such pair with $\gamma=\gamma_{st}$, let
$\bm a=(a_p,a_q)$, $\bm b_s=(b_{s,p},b_{s,q})$ and $\bm b_t=(b_{t,p},b_{t,q})$ be independent pairs of zero-mean CCG
values with unit variance and correlation $\rho$, and
\begin{equation*}
 x_c(s)=\sqrt\gamma\,a_c+\sqrt{1-\gamma}\,b_{s,c},\qquad
 x_c(t)=\sqrt\gamma\,a_c+\sqrt{1-\gamma}\,b_{t,c},\qquad c\in\{p,q\}.
\end{equation*}
The pair $\bm a$ is common to both acquisitions, and $\bm b_s$ and $\bm b_t$ are renewed in each. These values have
unit variance, correlation $\rho$ between the two pixels in each acquisition, $\gamma$ between the two acquisitions of
one pixel and $\gamma\rho$ between different pixels in different acquisitions. They thus have the covariances of the
model and, as zero-mean CCG values, its distribution.

The phasor $z_s$ depends only on $\bm x(s)$, hence on $\bm a$ and $\bm b_s$, and $z_t$ only on $\bm a$ and $\bm b_t$.
Given $\bm a$, the two phasors are therefore independent, and they have the same conditional mean
$m(\bm a)=\E[z_s\mid\bm a]=\E[z_t\mid\bm a]$, since $\bm b_s$ and $\bm b_t$ have the same distribution. Hence
\begin{equation*}
 \E[z_t\overline{z_s}]=\E_{\bm a}\bigl[\E[z_t\mid\bm a]\,\overline{\E[z_s\mid\bm a]}\bigr]=\E_{\bm a}|m(\bm a)|^2
 =\bigl|\E_{\bm a}m(\bm a)\bigr|^2+\E_{\bm a}\bigl|m(\bm a)-\E_{\bm a}m(\bm a)\bigr|^2\ge\bigl|\E[z_s]\bigr|^2,
\end{equation*}
where the third equality splits the mean square of $m(\bm a)$ into its squared mean and its variance. Alone, the pair
$\bm x(s)$ is CCG with correlation $\rho$, so that $\E[z_s]=g(|\rho|)e^{\ii\arg\rho}$ as in the proof of
Eq.~\eqref{eq:psfnull}, and $|\E[z_s]|^2=g(|\rho|)^2$ for every pair of acquisitions. The mean over the pairs gives
\begin{equation}
 \E[Q_{pq}]\ge g(|\rho|)^2.
\end{equation}
If $\gamma_{st}=0$ for all $s\ne t$, the common part $\bm a$ drops out, $m(\bm a)$ is constant and the bound holds
with equality, which is Eq.~\eqref{eq:psfnull}. Temporal coherence can thus only raise the expected corrected SPC of
direct neighbors above the floor. Signals that do not share the speckle correlation of the clutter, such as thermal
noise or the sidelobes of a bright scatterer, are outside this model.

\paragraph{Null moments, Eq.~\eqref{eq:nullsd}}
Under independent uniform phases, the terms of $\E_0|\sum_nz_n|^4$ that pair each $z_n$ with a $\overline{z_m}$ are the
only ones with nonzero mean, which gives
\begin{equation*}
 \E_0\Bigl|\sum_nz_n\Bigr|^4=2N^2-N,\qquad\operatorname{Var}_0(S_{pq})=\frac{N-1}{N^3},
\end{equation*}
and, by Eq.~\eqref{eq:q}, the standard deviation of one corrected SPC value
\begin{equation}\label{eq:nullsd}
 \operatorname{sd}_0(Q_{pq})=\frac1{\sqrt{N(N-1)}}.
\end{equation}

\subsection{Aggregation of neighborhood SPC}
\label{app:neighborhood-aggregation}

This section completes the aggregation of SPC over the neighbors of a pixel in Section~\ref{sec:nspcmask}.

For $m=|\mathcal N_p|>0$ valid neighbors, Eq.~\eqref{eq:local} is the
first of
\begin{equation*}
\ell_p^{\mathrm{mean}}=\frac1m\sum_{q}\sqrt{Q_{pq}^+},\qquad
\ell_p^{\mathrm{rms}}=\sqrt{\max\biggl(\frac1m\sum_{q}Q_{pq},0\biggr)},
\end{equation*}
and the second reverses averaging and square root. Let $\beta_{pq}=|\E[\overline{u_p(n)}u_q(n)]|$
under independent, identically distributed sampling. As $N\to\infty$, $\ell_p^{\mathrm{mean}}\to
m^{-1}\sum_q\beta_{pq}$ and $\ell_p^{\mathrm{rms}}\to(m^{-1}\sum_q\beta_{pq}^2)^{1/2}$.
For eight neighbors, one with $\beta=1$ and seven with $\beta=0$,
the limits of the mean and the RMS form are $0.125$ and $0.354$, respectively. For eight with $\beta=0.2$,
both are $0.2$. The RMS form favors a few strongly agreeing
neighbors, whereas the mean form favors broad agreement, which is
the intended criterion for the NSPC mask. At finite $N$, the mean form
clips each term separately, so that either form can exceed the
other. If all
$Q_{pq}\ge0$, $\ell_p^{\mathrm{rms}}\ge\ell_p^{\mathrm{mean}}$.

\subsection{Pixels that enter the estimated speckle correlation}
\label{supp:rhoest}
The estimate (Eq.~\eqref{eq:rhoest}) of Section~\ref{sec:realspeckle} uses the pairs at lag one of the pixels with
$\eta<\eta_0$, so that the pairs share no signal beyond the speckle. Two further effects bias $\bar Q_1$. First, a
subtracted model phase such as the topographic phase attenuates
$Q_{pq}$ where it differs between neighbors in a way that changes
over the acquisitions (Section~\ref{sec:model}). Only pairs whose
model phase difference alone has a corrected SPC $Q[e^{\ii\psi_{pq}}]$ of at least 0.9 (Eq.~\eqref{eq:attenuation})
enter $\bar Q_1$, so that the subtraction lowers their expected corrected SPC by at most 10\,\%. Second, the correlation depends on the brightness of the
surface. Thermal noise lowers it on dark surfaces, and point
scatterers raise it on the brightest. Only pixels whose
surroundings, a $9\times9$ window without its central $3\times3$
pixels, have a mean amplitude between the median and the 90th
percentile of this quantity over the valid pixels enter. The mean amplitude of the surroundings is used rather than that
of the pixel itself, because a selection by the pixel's own amplitude
favors pixels whose speckle happens to be bright in many
acquisitions, and bright samples agree better in phase. In simulated
clutter with $N=44$, such a selection raises the estimate by \vdOwnAmpBias,
whereas the selection by the surroundings leaves it unchanged. To extend the estimate to the other lags, the ratio
$B/f_{\mathrm s}$ of a raised-cosine weighting with the weighting coefficient of the sensor is fitted to the
lag-one value of each axis, and Eq.~\eqref{eq:rhopred} gives the correlation at these lags.

On the Visp stack, \vdEstVispShare\,\% of the valid pixels have a PCI below its null quantile, and the pairs that
enter the estimate are selected among them.
Over all pairs of valid pixels of this scene, the mean corrected SPC at lag one is \vdVispQazOne{} in azimuth and
\vdVispQrgOne{} in range (Fig.~\ref{fig:psf}a), close to the floor $g(|\rho|)^2$ of Eq.~\eqref{eq:psfnull} for the
predicted correlation (Table~\ref{tab:sensor}), as expected where most pixels decorrelate over the stack. On Sentinel-1, the estimate lies below the prediction in azimuth
(Section~\ref{sec:realspeckle}). To find the cause, we use the correlation of the intensities of neighboring pixels over time,
which measures $|\rho|^2$ from the intensities alone. In the five test areas, its square root at lag one is close to the prediction over bright surfaces. Over dark
surfaces, where the mean amplitude of the surroundings lies more than 9\,dB below the median of the test area, it
decreases in azimuth to \vdIcDarkAzMin{} to \vdIcDarkAzMax, as expected for thermal noise with a flat spectrum over
the processed bandwidth. In simulated clutter with zero coherence and the predicted
correlations, NSPC has a median of \vkNullSLongMed{} (S1-79), whereas pure thermal noise over 79 acquisitions, with
the azimuth correlation measured on dark surfaces or expected for a flat noise spectrum, has a median NSPC of
\vkThermMedMin{} to \vkThermMedMax.

\section{SHP selection and adaptive sizing}
\label{supp:method}
This section completes Section~\ref{sec:selection}.

\subsection{Test of CMP}
\label{supp:cmp}
CMP \citep{zhao2022statistically} compares a center $p$ and a candidate $q$ through their temporal mean intensities
$\bar I_p$ and $\bar I_q$ and the phases $\Delta\phi_p^{1n}$ and $\Delta\phi_q^{1n}$ of the $N-1$ single-reference
interferograms, whose reference is the first acquisition. \citet[Section~2.2.1]{zhao2022statistically} note that the
interferograms can follow a single-reference or a multi-reference scheme and use single-reference interferograms, with
the first acquisition as reference in their simulations (their Section~3.1). We follow this choice. A single reference
makes every term of the statistic share the noise of that acquisition, and the first acquisition gives the
interferograms the longest temporal baselines, which under the decaying coherence of Eq.~\eqref{eq:cohfamily} lowers
their mean coherence. A central reference acquisition, several reference acquisitions or the interferograms of
consecutive acquisitions may therefore serve CMP better, and they are not tested here. The statistic of CMP is the mean over these interferograms of a symmetrized Kullback--Leibler divergence between the
$2\times2$ covariance matrices of the two pixels. Each matrix is built from the mean intensity of the pixel, its
interferometric phase and an assumed coherence $\gamma_{\mathrm a}(n)$ of interferogram $n$, and the statistic reads
\citep[Eqs.~(1), (3) to (5) and (10)]{zhao2022statistically}
\begin{equation}\label{eq:cmpstat}
 D_{pq}=\frac4\pi\,\frac{\bar I_p^2+\bar I_q^2}{\bar I_p\bar I_q}\,\Biggl(\frac1{N-1}\sum_{n=2}^{N}
 \frac{1-\gamma_{\mathrm a}^2(n)\cos\bigl(\Delta\phi_p^{1n}-\Delta\phi_q^{1n}\bigr)}{1-\gamma_{\mathrm a}^2(n)}\Biggr)
 -\frac8\pi .
\end{equation}
It is zero for equal intensities and phases and grows as the intensity ratio departs from one and with the phase
differences. The assumed coherence $\gamma_{\mathrm a}(n)$ is the product of three factors
\citep[Eqs.~(8) and (9) and Section~3.1]{zhao2022statistically}. The first is the scale 0.3, which
\citet{zhao2022statistically} set for single-look data. The second is a thermal term $1/(1+\mathrm{SNR}^{-1})$ of 0.94
for a signal-to-noise ratio SNR of 12\,dB \citep[Eq.~(4)]{zebker1992decorrelation}, against the 0.92 that
\citet{zhao2022statistically} state. The third is an exponential decay with the temporal baseline of the
interferogram. We leave out their spatial
term, which depends on the perpendicular baseline, and fit the decay constant, which \citet{zhao2022statistically} set
to 200 days, to the decay of the mean coherence of the single-reference interferograms with the temporal baseline,
with the same estimator on simulated and real stacks (Sections~\ref{app:settings} and~\ref{supp:sensor}).

A candidate passes if $D_{pq}$ is at most the threshold, which is the test of \citet[Eq.~(11)]{zhao2022statistically}
on $\exp(-D_{pq})$. The test has two passes \citep[Section~2.2.3]{zhao2022statistically}. The first compares every candidate with the
center by Eq.~\eqref{eq:cmpstat}. The center and the candidates accepted in the first pass form the reference set. The
second pass tests every candidate again, with the mean intensity of the reference set in place of $\bar I_p$ and the
phases of the multilooked interferograms of the reference set in place of $\Delta\phi_p^{1n}$. The ensemble is the
eight-connected component of the center among the candidates that pass the second pass. \citet[Fig.~2 and Eq.~(12)]{zhao2022statistically} use one threshold for both passes,
which they set from simulated data at a significance level of 5\,\% and which depends mainly on the stack size
\citep[Section~3.1.1]{zhao2022statistically}. Here each pass has its own threshold, which depends on the PCI at a
significance level of 1\,\% (Section~\ref{app:amptests}).

\subsection{Thresholds of the SHP tests}
\label{supp:rejection}
\label{app:amptests}
\label{app:calibration}
By Section~\ref{sec:amptests}, a fixed threshold rejects a fraction of
the homogeneous candidates that increases with the temporal correlation
of the speckle. The second-pass threshold of FaSHPS and the thresholds of CMP
in this paper depend on the PCI $\eta_p$ of the center (Section~\ref{sec:evidence}), which rises with the coherence
(Table~\ref{tab:eta}).

The table $z_\alpha(\eta)$ of Section~\ref{sec:pairwise} is built from
simulated homogeneous clutter without fringes, with the stack size,
search window and point spread function of the configuration, at the
coherences $c\in\{0,0.1,0.2,0.35,0.5,0.65,0.8,0.95\}$ of
Eq.~\eqref{eq:cohfamily}. For each coherence, $\eta$ is the median of the 
PCI over the scene, and $z_\alpha$ the $1-\alpha$ quantile of the
second-pass statistic (Eq.~\eqref{eq:fastat}) over the pixels outside the
guard sets of 240 randomly placed centers. Both sequences are made monotone, interpolated linearly and held at their end values outside the tabulated range. The first-pass threshold $z_1$ is
the median of the first-pass statistic at $c=0.5$. The two thresholds of
CMP are tabulated over $\eta$ in the same way, from the same clutter and
centers, for the decay constant of its coherence model.

The original thresholds of FaSHPS \citep[Eq.~(7) and Section~III-A]{jiang2015fast} are normal
quantiles of a 50\,\% interval in the first pass and of a $1-\alpha$
interval in the second, scaled with a coefficient of variation of the
temporal mean amplitude estimated from the scene (Section~\ref{sec:pairwise}). We estimate this coefficient of
variation as \citet[Sections~III-B and IV]{jiang2015fast}, from standard deviations of the temporal mean amplitude,
corrected by their Eq.~(12) for the truncation of the selected set and regressed on the mean through the origin by
iteratively reweighted least squares with Huber weights. The estimate here differs from theirs in two respects.
The standard deviations are taken over the selected sets of the two passes in 2000 search windows at random centers,
instead of in nonoverlapping blocks of $10\times10$ pixels. The interval is centered on the mean of the first pass,
instead of on the amplitude of the block's median pixel.
In the simulated scenes, each of a single coherence, this estimate follows the coherence, from \vjCvZeroTsx{} times
$c_A/\sqrt N$ at $c=0$ to \vjCvHighTsx{} times at $c=0.95$ (TSX-44, homogeneous scenes of
Table~\ref{tab:rejection}). One value per scene then fits every center, so that Fig.~\ref{fig:rejection} shows
the original thresholds in this favorable case. The original thresholds of CMP are the thresholds of
its test in homogeneous clutter whose coherence follows the thermal and temporal terms of its coherence model
without the scale 0.3, as \citet[Section~3.1]{zhao2022statistically} simulate their data.
Table~\ref{tab:rejection} completes Fig.~\ref{fig:rejection} with the share of the candidates of the same class
that each test rejects at a significance level of 1\,\%, for FaSHPS and CMP in the three configurations, in clutter
of a single coherence and at coherence steps of equal amplitude and phase. Section~\ref{sec:pairwise} gives the
rejection by FaSHPS. At a coherence step, one threshold per scene cannot fit every center. There the original
threshold of FaSHPS rejects as many of the candidates of the same class as in homogeneous clutter of the higher coherence or
more, whereas the thresholds that depend on the PCI of the center reject about as many as in homogeneous clutter. CMP with its original thresholds rejects almost none of the
homogeneous candidates up to $c=0.8$, far fewer than its significance level, and \vcRejCmpPubHighMin{}
to \vcRejCmpPubHighMax\,\% at $c=0.95$, compared with
\vcRejHomCmpHighMin{} to \vcRejHomCmpHighMax\,\% there with the thresholds that depend on $\eta$. Below $c=0.95$ the weak dependence of this rejection on the coherence agrees
with \citet[Section~3.1.1 and Fig.~4]{zhao2022statistically}, who found their threshold only slightly
affected by the coherence for two coherence matrices whose mean coherences differ by 0.14. The thresholds of CMP
that depend on $\eta$, however, reject the fewest homogeneous candidates, \vjCmpNineMin{} to \vjCmpNineMax\,\%, at
$c=0.9$, between the calibration coherences 0.8 and 0.95, where their interpolation is too permissive (three
configurations).

\begin{table*}[tbp]
\centering\footnotesize
\setlength{\tabcolsep}{2.4pt}
\resizebox{\linewidth}{!}{%
\begin{tabular}{lcccc@{\hspace{8pt}}cccc@{\hspace{8pt}}cccc}
\toprule
 & \multicolumn{4}{c}{TSX-44} & \multicolumn{4}{c}{S1-79} & \multicolumn{4}{c}{S1-20} \\
Clutter & FaSHPS & FaSHPS, orig. & CMP & CMP, orig. & FaSHPS & FaSHPS, orig. & CMP & CMP, orig. & FaSHPS & FaSHPS, orig. & CMP & CMP, orig. \\
\midrule
$c=0$ & 0.6 & 2.1 & 1.0 & 0.0 & 0.9 & 2.2 & 0.8 & 0.0 & 1.0 & 2.4 & 1.0 & 0.0 \\
$c=0.05$ & 0.9 & 2.7 & 1.0 & 0.0 & 0.8 & 1.9 & 0.8 & 0.0 & 0.9 & 2.4 & 1.0 & 0.0 \\
$c=0.1$ & 0.8 & 2.4 & 1.0 & 0.0 & 0.9 & 2.0 & 1.0 & 0.0 & 1.3 & 3.3 & 1.1 & 0.0 \\
$c=0.2$ & 0.8 & 2.4 & 1.0 & 0.0 & 1.3 & 2.5 & 1.1 & 0.0 & 1.3 & 3.1 & 1.2 & 0.0 \\
$c=0.3$ & 1.1 & 4.0 & 1.0 & 0.0 & 1.4 & 3.4 & 1.2 & 0.0 & 1.4 & 3.5 & 1.1 & 0.0 \\
$c=0.35$ & 1.0 & 3.6 & 1.0 & 0.0 & 1.6 & 4.2 & 1.3 & 0.0 & 1.5 & 3.8 & 1.1 & 0.0 \\
$c=0.5$ & 1.5 & 5.2 & 1.4 & 0.0 & 1.3 & 6.4 & 1.1 & 0.0 & 1.4 & 5.2 & 1.2 & 0.0 \\
$c=0.65$ & 1.3 & 6.0 & 1.3 & 0.0 & 1.5 & 5.9 & 1.2 & 0.0 & 1.5 & 6.4 & 1.3 & 0.0 \\
$c=0.7$ & 1.3 & 4.8 & 1.3 & 0.0 & 1.8 & 7.0 & 1.3 & 0.0 & 1.6 & 4.8 & 1.4 & 0.0 \\
$c=0.8$ & 1.3 & 5.5 & 1.2 & 0.0 & 1.6 & 11.3 & 1.2 & 0.0 & 1.5 & 3.4 & 1.3 & 0.0 \\
$c=0.9$ & 0.8 & 9.8 & 0.4 & 0.1 & 1.4 & 20.3 & 0.3 & 0.0 & 1.0 & 6.2 & 0.5 & 0.5 \\
$c=0.95$ & 1.3 & 14.5 & 1.1 & 3.7 & 1.7 & 26.9 & 1.2 & 1.6 & 1.0 & 10.3 & 1.1 & 4.0 \\
Coherence step 0.95 and 0.2 & 1.0 & 21.3 & 1.3 & 1.9 & 1.6 & 28.4 & 1.5 & 0.8 & 1.1 & 13.1 & 1.3 & 1.9 \\
Coherence step 0.95 and 0 & 1.1 & 23.3 & 1.3 & 1.8 & 1.3 & 29.2 & 1.6 & 0.8 & 1.1 & 14.6 & 1.4 & 1.8 \\
Coherence step 0.65 and 0.2 & 1.1 & 9.4 & 1.2 & 0.0 & 1.5 & 11.3 & 1.3 & 0.0 & 1.2 & 6.4 & 1.2 & 0.0 \\
\bottomrule
\end{tabular}}
\caption{Share in percent of the candidates of the same class outside the guard set (Section~\ref{sec:adaptive}) that each test rejects at a significance level of 1\,\%, in homogeneous clutter and at coherence steps, mean over scenes generated from the seeds of the evaluation scenes, for FaSHPS and CMP with the threshold that depends on the PCI of the center and with the original thresholds (orig.). A coherence step joins two classes of equal amplitude and phase but different coherence, and the candidates of the same class are those in the class of the center. The thresholds that depend on the PCI are
tabulated at the PCI of the calibration coherences 0, 0.1, 0.2, 0.35, 0.5, 0.65, 0.8 and 0.95 and interpolated
linearly between them (Section~\ref{app:calibration}).}
\label{tab:rejection}
\end{table*}

\subsection{Details of the truncation step}
\label{supp:truncdetails}
This section gives the rules of the truncation step (Section~\ref{sec:adaptive}) that Algorithm~\ref{alg:adaptive}
uses beyond the main text. The ellipse order is replaced by the distance order where $\bm M_p$ is not
positive definite, where the denominator of Eq.~\eqref{eq:moment} is at most 1\,\% of the number of its terms, or where the
smallest eigenvalue of $\bm M_p$ is at most 0.05\,pixel$^2$. Ties, such as the two pixels at offsets $\bm d_q$ and
$-\bm d_q$, are kept in a fixed order. The candidate sizes $L_k=\lfloor L_1\,1.5^{k-1}\rfloor$ of
Eq.~\eqref{eq:prefix} are used for every $k$ with $L_1\,1.5^k\le|\mathcal O_p\setminus\mathcal G_p|$, followed by
the last size $L_K=|\mathcal O_p\setminus\mathcal G_p|$. The truncation step is skipped if fewer than 38 members lie
outside the guard set, i.e., if $1.5\,L_1>|\mathcal O_p\setminus\mathcal G_p|$, since then no size below $L_K$ is a
candidate.

\section{Simulation study}
\label{supp:sim}
This section completes Section~\ref{sec:simulation}. In this section, a value or a range refers to all centers of the evaluation scenes, the full window,
FaSHPS and CMP, and the three configurations TSX-44, S1-79 and S1-20, unless the sentence or the parentheses after it
name others. A range gives the smallest and the largest value over them.

\subsection{Tuning and checks of the design}
\label{supp:tuning}
\label{app:settings}
The criterion of the tuning of Section~\ref{sec:simdesign} is the weighted geometric mean of the phase linking error
relative to the oracle over all centers of the tuning scenes, averaged over the SHP selections with \WSname{} and over
the configurations. The tuning has three steps, each of which varies one parameter with the others held fixed, and a
change counts only if it lowers the criterion by more than \vtTie\,\%. First, at $\kappa=0.75$ and $\lambda=3$,
$\rho_{\min}$ takes the value of the lowest criterion, or the largest value within \vtTie\,\% of it. Second, at this
$\rho_{\min}$, $\kappa$ moves from 0.75 to a neighboring tested value, and further in the same direction, as long as
each move counts. Third, at this $\kappa$, $\lambda$ moves from 3 by the same rule. In addition, $\lambda$ was varied at
$\kappa=0.75$ for $\rho_{\min}=0.02$, 0.03, 0.04 and 0.05, and every tested value from 0 to 4 changed the criterion by
less than \vtTie\,\% against $\lambda=3$. The tuning selected
$\rho_{\min}=\vtRhoMin$, $\kappa=\vcKappa$ and $\lambda=\vcLambda$. At $\kappa=0.75$ and $\lambda=3$, the criterion at
$\rho_{\min}=\vtRhoMin$ is \vtRhoGain\,\% lower than at $\rho_{\min}=0.05$, and that at $\rho_{\min}=\vtRhoTied$ lies
within \vtRhoTiedDiff\,\% of it. At $\rho_{\min}=\vtRhoMin$, the criterion differs by at most \vcTuneRangeMax\,\% over
the tested values of $\kappa$.

Table~\ref{tab:tuninggrid} gives the criterion on the tuning scenes, averaged over the
methods and configurations, for each bound $\rho_{\min}$ and each tested setting of the margin factor $\kappa$ and
the noise multiplier $\lambda$, and Table~\ref{tab:tuning} the criterion of each method at the selected $\rho_{\min}$
and $\lambda$. One value of each factor
applies to every method and configuration. At the tuned $\rho_{\min}=\vtRhoMin$,
$g(\rho_{\min})^2\approx\vaRhoGsq$, which is \vaRhoGsqShareMin{} to \vaRhoGsqShareMax\,\% of the standard
deviation of one corrected SPC value under independent uniform phases at $N=20$ to
79 (Eq.~\eqref{eq:nullsd}), so that the speckle that two pixels beyond the speckle correlation length still share
adds far less to their SPC than the noise of one corrected SPC value. For $\rho_{\min}$ between \vaRhoMinLo{} and \vaRhoMinHi, the core, the guard set, the ring
$\mathcal R_p$ and the pixels of the ellipse order are unchanged on both sensors, so that the tuning fixes
$\rho_{\min}$ only to within this interval.

A smaller margin does not remove the ensembles that remain too large (limitation (iii) of Section~\ref{sec:discussion}).
On the tuning scenes at the tuned bound $\rho_{\min}=\vtRhoMin$, the smallest margin factor tested there,
$\kappa=\vcOverCKappaLowDuo$, was compared with the tuned $\kappa=\vcKappa$ (\WSname, full window, FaSHPS and CMP,
TSX-44 and S1-79). The smaller factor lowered the share of the area at distributed centers whose ensemble has more than four times the oracle size by
\vcOverCDropSelDuoMin{} to \vcOverCDropSelDuoMax{} percentage points, to \vcOverCLowDuoMin{} to
\vcOverCLowDuoMax\,\%. The share of truncated ensembles rose to \vcTruncCLowDuoMin{} to \vcTruncCLowDuoMax\,\%,
compared with \vcTruncCSelDuoMin{} to \vcTruncCSelDuoMax\,\% at $\kappa=\vcKappa$. A smaller margin thus truncates
more ensembles but hardly reduces the share of those that remain too large, so that the limitation lies in the noise
of the score. With \WSname, the phase linking error of FaSHPS exceeds twice the oracle error at \vtTailWSTSX\,\%
(TSX-44) and \vtTailWSSLong\,\% (S1-79) of the area of the evaluation scenes, with the truncation step alone at \vtTailSTSX\,\%
(TSX-44) and \vtTailSSLong\,\% (S1-79), and for FaSHPS alone at \vtTailNTSX\,\% (TSX-44) and \vtTailNSLong\,\% (S1-79).

The fixed values, those of Table~\ref{tab:settings} and the other constants of Section~\ref{sec:selection}, were set
without tuning. Table~\ref{tab:checks} and Table~\ref{tab:sweep7} give the change of the error when the significance level, the percentile of
the null quantile $\eta_0$, $n_{\min}$, $L_{\min}$, $R$, the factor of the candidate sizes, the tolerance, the ring or the $5\times5$ window of the PCI is
changed. The
factor 1.5 of the candidate sizes trades the resolution of the size
against the noise of the maxima. The ring of 24 pixels has the size of a $5\times5$
neighborhood without its center.

The SHP selections use the settings of Section~\ref{sec:selection}, except the decay constant of the coherence
model of CMP, which was fitted to the mean coherence of the single-reference
interferograms over the first ten scenes of each batch of tuning scenes. The mean
coherence of an interferogram is the square root of the mean of
$\operatorname{Re}(v_p^{1n}\overline{v_q^{1n}})$ over all pixels $p$
and the pixels $q$ of their rings $\mathcal R_p$, mapped through the
inverse of $g$. The fit gave \vcCmpDecayTSX, \vcCmpDecaySLong{}
and \vcCmpDecaySShort{} days (TSX-44, S1-79 and S1-20, respectively).

Table~\ref{tab:checks} gives checks of the design on the evaluation
scenes, each as the ratio of the error over all centers to that of the
same method with \WSname. The truncation step gated by $\eta$ on the search
window, without the PCI-scaled window, has \vcGoMin{} to
\vcGoMax{} times the error of \WSname, so that the PCI-scaled window hardly changes the error over all centers and pays
off in time (Section~\ref{sec:runtime}). Conversely, a window that shrinks much faster than the PCI-scaled window, without the truncation step, tests
whether a smaller window alone could replace the truncation step. Its semiaxes shrink linearly with the PCI, from the
search window at $\eta_0$ to the smallest ellipse of \vcWinStrongSize{} pixels, the minimum ensemble size, already at
the median PCI of homogeneous clutter with $c=0.95$. Those of the PCI-scaled window reach the \vnMin{} nearest
pixels only at $\eta_p=1$ (Section~\ref{sec:scaledwindow}). This window has \vcWinAloneMin{} to \vcWinAloneMax{} times
the error of \WSname{} (full window and FaSHPS), so that a smaller window alone cannot replace the truncation step. The PCI-scaled window alone, without the truncation step, is the
variant +S of Table~\ref{tab:overall}. The distance order and the margin $1.5/N$ have a higher error than the ellipse
order and the jackknife margin. For the distance order the excess lies mostly at the centers with four to eight
fringes per window diameter (\vcDistFrMin{} to \vcDistFrMax{} times). The percentile of the null quantile $\eta_0$ changes the error
by at most \vcEtaPctAbs\,\%. The rows on the significance levels and the original thresholds of
the SHP tests are discussed in Section~\ref{sec:simresults}.

\begin{table}[tbp]
\centering\footnotesize
\setlength{\tabcolsep}{3.5pt}
\begin{tabular}{lccccccccc|ccc}
\toprule
$\kappa$ & 0.75 & 0.5 & 1 & 0.375 & 0.25 & 0.5 & 0.5 & 0.5 & 0.5 & 0.75 & 0.75 & 0.75 \\
$\lambda$ & 3 & 3 & 3 & 3 & 3 & 1 & 2 & 4 & 5 & 0 & 2 & 4 \\
\midrule
$\rho_{\min}=0.05$ & 1.329 & 1.333 & 1.332 & -- & -- & -- & -- & -- & -- & 1.329 & 1.329 & 1.329 \\
$\rho_{\min}=0.04$ & 1.316 & 1.316 & 1.321 & -- & -- & -- & -- & -- & -- & 1.317 & 1.317 & 1.317 \\
$\rho_{\min}=0.03$ & 1.309 & 1.307 & 1.316 & -- & -- & -- & -- & -- & -- & 1.309 & 1.309 & 1.309 \\
$\rho_{\min}=0.025$ & 1.300 & 1.297$^{\ast}$ & 1.309 & 1.297 & -- & 1.297 & 1.297 & 1.297 & 1.298 & -- & -- & -- \\
$\rho_{\min}=0.02$ & 1.301 & 1.296 & 1.309 & 1.296 & 1.297 & -- & -- & -- & -- & 1.301 & 1.301 & 1.301 \\
\bottomrule
\end{tabular}
\caption{Tuning of the truncation step on the tuning scenes: criterion (Section~\ref{supp:tuning}, column All of Table~\ref{tab:tuning}) for each speckle correlation bound $\rho_{\min}$ and each setting of the margin factor $\kappa$ and the noise multiplier $\lambda$ that was run. The tuning chose $\rho_{\min}$ at $\kappa=0.75$ and $\lambda=3$ (first column), then $\kappa$ at the chosen $\rho_{\min}$ with $\lambda=3$, then $\lambda$ at the chosen $\kappa$, and accepted a change only where it lowered the criterion by more than \vtTie\,\%. The three columns right of the vertical rule show the effect of $\lambda$ at $\kappa=0.75$ for the other bounds, and $\kappa=0.25$ was run at $\rho_{\min}=0.02$ only. A dash marks a setting that was not run, and $^{\ast}$ the selected setting.}
\label{tab:tuninggrid}
\end{table}

\begin{table}[tbp]
\centering\footnotesize
\setlength{\tabcolsep}{3pt}
\begin{tabular}{lcccccccccc}
\toprule
 & \multicolumn{3}{c}{TSX-44} & \multicolumn{3}{c}{S1-79} & \multicolumn{3}{c}{S1-20} & \\
$\kappa$ & Full & FaSHPS & CMP & Full & FaSHPS & CMP & Full & FaSHPS & CMP & All \\
\midrule
none & 1.986 & 1.932 & 1.959 & 1.661 & 1.593 & 1.613 & 1.775 & 1.666 & 1.678 & 1.756 \\
0.75 & 1.353 & 1.318 & 1.341 & 1.285 & 1.240 & 1.261 & 1.354 & 1.272 & 1.282 & 1.300 \\
0.5$^{\ast}$ & 1.351 & 1.316 & 1.341 & 1.279 & 1.236 & 1.257 & 1.348 & 1.269 & 1.279 & 1.297 \\
0.375 & 1.353 & 1.319 & 1.346 & 1.277 & 1.235 & 1.255 & 1.347 & 1.268 & 1.279 & 1.297 \\
\bottomrule
\end{tabular}
\caption{Tuning of the truncation step on the tuning scenes at the tuned bound $\rho_{\min}=\vtRhoMin$. Weighted geometric mean of the phase linking error relative to the oracle over all centers of each method with the PCI-scaled window and the gated truncation step, for the margin factor $\kappa$ at the selected noise multiplier $\lambda=\vcLambda$, and the criterion, its geometric mean over the methods and configurations (All). Full is the full window, and the first row (none) has neither the truncation step nor the PCI-scaled window. $^{\ast}$ marks the selected $\kappa$.}
\label{tab:tuning}
\end{table}

\begin{table*}[tbp]
\centering\footnotesize
\setlength{\tabcolsep}{3pt}
\begin{tabular}{>{\raggedright\arraybackslash}p{4.4cm}ccc@{\hspace{8pt}}ccc@{\hspace{8pt}}ccc}
\toprule
 & \multicolumn{3}{c}{TSX-44} & \multicolumn{3}{c}{S1-79} & \multicolumn{3}{c}{S1-20} \\
Variant & Full & FaSHPS & CMP & Full & FaSHPS & CMP & Full & FaSHPS & CMP \\
\midrule
Gated truncation step on the search window & 1.022 & 1.021 & 1.021 & 0.998 & 0.996 & 0.996 & 1.011 & 1.011 & 1.011 \\
Distance order & 1.002 & 1.003 & 1.003 & 1.004 & 1.004 & 1.005 & 1.002 & 1.002 & 1.003 \\
Margin $1.5/N$ & 1.074 & 1.071 & 1.067 & 1.039 & 1.032 & 1.032 & 1.070 & 1.067 & 1.066 \\
$\eta_0$ at the 95th percentile & 0.999 & 1.000 & 1.000 & 1.000 & 1.000 & 1.000 & 0.998 & 0.999 & 0.999 \\
$\eta_0$ at the 99.9th percentile & 1.002 & 1.002 & 1.001 & 1.000 & 1.000 & 1.000 & 1.002 & 1.002 & 1.002 \\
Window shrunk to 25 pixels at the PCI of $c=0.95$ & 1.182 & 1.192 & -- & 1.132 & 1.134 & -- & 1.135 & 1.138 & -- \\
Level 0.1\,\% & -- & 0.984 & 0.979 & -- & 0.985 & 0.980 & -- & 0.996 & 0.993 \\
Level 5\,\% & -- & 1.044 & 1.035 & -- & 1.052 & 1.048 & -- & 1.031 & 1.023 \\
Original threshold, SHP selection alone & -- & 0.998 & 0.986 & -- & 1.026 & 0.993 & -- & 1.000 & 1.012 \\
Original threshold, \WSname & -- & 1.017 & 0.972 & -- & 1.067 & 0.984 & -- & 1.014 & 1.007 \\
\bottomrule
\end{tabular}
\caption{Checks of the design: phase linking error over all centers of the evaluation scenes relative to that of the same method with \WSname, ratio of the weighted geometric means on the same centers. Each row changes one element, namely the gated truncation step on the search window without the PCI-scaled window, the
distance order in place of the ellipse order, a fixed margin $\zeta=1.5/N$ in place of $\kappa\sigma_{\Delta a}$, the null quantile
$\eta_0$ at other percentiles, a window that shrinks to 25 pixels already at the PCI of $c=0.95$ without the truncation
step, other significance levels of the SHP tests, and their original thresholds (Section~\ref{app:settings}). The
original threshold with the SHP selection alone is compared with the same SHP selection alone. A dash marks a variant that was not run or that does not apply to the full window, which has no SHP test.}
\label{tab:checks}
\end{table*}

\subsection{Ordinary sample covariance}
\label{supp:raw}
This section completes Section~\ref{sec:estimate}. With the ordinary sample covariance in place of Eq.~\eqref{eq:unitcov}, and
both relative to the oracle of Eq.~\eqref{eq:unitcov}
(Table~\ref{tab:raw}), the error over the distributed centers is higher for every SHP selection, with the
truncation step and with \WSname. It is at most \vtRawSelMax\,\% higher for FaSHPS and CMP and about twice as high
for the full window (\vtRawFullWSMin{} to \vtRawFullMax{} times). A few bright pixels dominate the
sample covariance of a large ensemble, and the SHP tests exclude
them, which is why FaSHPS and CMP lose least. We use Eq.~\eqref{eq:unitcov}
throughout.

\begin{table}[tbp]
\centering\footnotesize
\setlength{\tabcolsep}{3pt}
\begin{tabular}{llcccccc}
\toprule
Configuration & Estimate & Full+T & Full\WSsuffix & FaSHPS+T & FaSHPS\WSsuffix & CMP+T & CMP\WSsuffix \\
\midrule
TSX-44 & Eq.~\eqref{eq:unitcov} & 1.37 & 1.34 & 1.34 & 1.31 & 1.36 & 1.34 \\
 & Sample covariance & 2.57 & 2.46 & 1.40 & 1.37 & 1.40 & 1.37 \\
\addlinespace[2pt]
S1-79 & Eq.~\eqref{eq:unitcov} & 1.27 & 1.27 & 1.22 & 1.23 & 1.25 & 1.25 \\
 & Sample covariance & 2.34 & 2.27 & 1.29 & 1.29 & 1.29 & 1.29 \\
\addlinespace[2pt]
S1-20 & Eq.~\eqref{eq:unitcov} & 1.34 & 1.33 & 1.27 & 1.26 & 1.29 & 1.28 \\
 & Sample covariance & 2.50 & 2.40 & 1.37 & 1.35 & 1.35 & 1.34 \\
\bottomrule
\end{tabular}
\caption{Amplitude-normalized estimator (Eq.~\eqref{eq:unitcov}) compared with the ordinary sample covariance: phase linking error relative to the oracle of the amplitude-normalized estimator, weighted geometric mean over the distributed centers of the evaluation scenes. Columns with the truncation step (+T) and with \WSname{} as in Table~\ref{tab:overall}.}
\label{tab:raw}
\end{table}

\subsection{Scene generator}
\label{supp:generator}
Table~\ref{tab:generator} lists the components of the random scenes of Section~\ref{sec:simdesign}.

\begin{table}[tbp]
\centering\small
\begin{tabularx}{\linewidth}{lX}
\toprule
Component & Values\\
\midrule
Scene & Six window radii along each axis\\
Deformation & One source, profile $\exp(-d^{\beta})$ with $d$ the distance from the center of the source in units of its semiaxes and $\beta\in[1.2,2.5]$, major semiaxis $w$ of 1 to 5 window radii (log-uniform), center 0 to $2w$ from the scene center, axis ratio 1 to 4 (log-uniform, 6 to 10 with probability 0.15, minor semiaxis at least 0.5), extents on the two sides of the center differing by a factor of 1 to 1.6, axis turning by up to 0.2\,rad over $w$. With probability 0.25 a second lobe $1.5w$ to $3w$ away, 0.5 to 1 times as wide, with 0.2 to 0.5 times the amplitude and either sign, oriented within $30^\circ$ of the first. Random sign of the field. Polynomial background of degree 1 to 3 with a steepest gradient of at most 0.15 times that of the source. A draw is repeated if its steepest gradient in the scene exceeds twice that in the evaluated region\\
Maximum fringe density & $F_{\max}\in\{0,0.5,1,2,4,8\}$ fringes per window diameter between the first and the last acquisition, equally likely\\
Atmosphere & Independent Gaussian random field per acquisition with power spectrum $(k^2+k_0^2)^{-11/6}$, which falls as $k^{-11/3}$, the spectrum of the atmospheric delay under Kolmogorov turbulence \citep[Eq.~(4.7.11)]{hanssen2001radar}, above the wavenumber $k_0$ of an outer scale $2\pi/k_0$ of 20 window radii. RMS difference over one window diameter 0, 0.3 or 1\,rad, equally likely\\
Discontinuities & Moving region, with the sources zero outside an ellipse of 1.2 to 2.4 times the semiaxes of the first source whose radius is perturbed by harmonics of order 2 to 4 by at most 15\,\%, and fault, with the field on one side of a straight or slightly curved line within 1.5 window radii of the scene center scaled by $1\pm0.2$ to $1\pm0.6$, each with probability 0.15\\
Clutter classes & One to three classes from a thresholded smooth random field, correlation length 0.5 to 2 window radii, each class at least 15\,\% of the evaluated region\\
Class motion & With probability 0.5 in scenes with two or three classes, every class after the first moves with its own rate on top of the common deformation, and this motion grows linearly to a step of 0.5 to 3\,rad of random sign at the last acquisition\\
Per class & Coherence $c\in\{0,0.2,0.35,0.5,0.65,0.8,0.95\}$, backscatter $-3$ to $+3$\,dB, texture none, mild (gamma shape 10) or strong (shape 2), correlated over 6 pixels of the TSX-44 window and scaled with the window, point scatterers none or sparse (one per $32\times32$ pixels) with probability 1/3 each, or dense (one per $8\times8$ or $12\times12$ pixels), isolated or, with probability 0.5, in groups of five within $7\times7$ pixels, signal-to-clutter ratio 0 to 25\,dB\\
Point scatterers & Height 0 to 20\,m, within 1\,m (standard deviation) in a group. Phase history of the topographic phase of the height, with 0.12\,rad per meter (standard deviation over the acquisitions), the deformation and atmosphere at the position, a constant random offset and noise of 0.1\,rad, amplitude noise of 10\,\%\\
\bottomrule
\end{tabularx}
\caption{Components of the random scenes. Lengths are in window radii except
the scatterer layout, which is in pixels. Values given
as ranges are drawn uniformly unless stated otherwise.}
\label{tab:generator}
\end{table}

\subsection{Phase linking solver}
\label{supp:solver}
The objective $J$ of Eq.~\eqref{eq:plobjective} in Section~\ref{sec:simdesign} is maximized from the phases of the leading
eigenvector of $|\widehat{\bm\Gamma}|\odot\widehat{\bm\Gamma}$. Up to ten sweeps of cyclic coordinate ascent come
first, each of which sets one phase after the other to its conditional maximum. Damped Newton iterations follow until the largest component of the gradient
falls below $10^{-6}$ times the sum of the weights. If the Newton
iterations do not converge within 100 iterations, coordinate ascent continues from their last iterate for at most
1000 sweeps, after which the last iterate is kept. On the evaluation scenes, every solution converged. The same solver is used
for the real data.

\subsection{Absolute errors}
\label{supp:absolute}
Table~\ref{tab:absolute} gives the absolute errors of FaSHPS in each coherence class.

\begin{table}[tbp]
\centering\small
\setlength{\tabcolsep}{4pt}
\begin{tabular}{lcccc}
\toprule
Centers & Oracle & FaSHPS & FaSHPS+T & FaSHPS\WSsuffix \\
\midrule
\multicolumn{5}{l}{\emph{TSX-44}} \\
$c=0$ & 64.6 & 96.5 & 97.0 & 96.6 \\
$c=0.2$ & 10.0 & 13.7 & 12.7 & 12.7 \\
$c=0.35$ & 7.0 & 11.6 & 8.9 & 8.7 \\
$c=0.5$ & 6.0 & 12.2 & 7.6 & 7.5 \\
$c=0.65$ & 4.6 & 10.8 & 6.2 & 5.9 \\
$c=0.8$ & 3.5 & 10.0 & 5.1 & 4.6 \\
$c=0.95$ & 2.5 & 10.5 & 3.8 & 3.1 \\
\addlinespace[2pt]
At least one fringe & 6.7 & 31.0 & 9.6 & 9.1 \\
Point scatterers & 19.5 & 49.6 & 36.9 & 37.0 \\
\addlinespace[2pt]
All & 5.8 & 13.6 & 7.4 & 7.1 \\
\midrule
\multicolumn{5}{l}{\emph{S1-79}} \\
$c=0$ & 68.0 & 95.6 & 96.1 & 96.0 \\
$c=0.2$ & 15.6 & 18.7 & 18.6 & 18.6 \\
$c=0.35$ & 11.3 & 14.3 & 13.0 & 12.9 \\
$c=0.5$ & 9.2 & 13.4 & 10.4 & 10.4 \\
$c=0.65$ & 7.7 & 13.5 & 8.8 & 8.7 \\
$c=0.8$ & 6.1 & 12.4 & 7.1 & 6.9 \\
$c=0.95$ & 3.7 & 9.4 & 4.9 & 4.4 \\
\addlinespace[2pt]
At least one fringe & 10.8 & 33.9 & 13.9 & 13.6 \\
Point scatterers & 14.7 & 58.5 & 42.3 & 41.4 \\
\addlinespace[2pt]
All & 8.8 & 15.1 & 10.4 & 10.2 \\
\midrule
\multicolumn{5}{l}{\emph{S1-20}} \\
$c=0$ & 69.3 & 90.2 & 90.8 & 90.4 \\
$c=0.2$ & 13.0 & 16.2 & 15.6 & 15.7 \\
$c=0.35$ & 9.6 & 13.8 & 11.8 & 11.8 \\
$c=0.5$ & 7.7 & 12.4 & 9.3 & 9.2 \\
$c=0.65$ & 6.3 & 11.9 & 7.9 & 7.7 \\
$c=0.8$ & 5.0 & 11.4 & 6.5 & 6.1 \\
$c=0.95$ & 3.2 & 9.2 & 4.7 & 4.0 \\
\addlinespace[2pt]
At least one fringe & 9.0 & 33.2 & 12.7 & 12.4 \\
Point scatterers & 14.5 & 72.1 & 53.0 & 52.2 \\
\addlinespace[2pt]
All & 7.7 & 14.4 & 9.7 & 9.4 \\
\bottomrule
\end{tabular}
\caption{Weighted median oracle error and phase linking error in degrees of FaSHPS alone, with the truncation step (T) and with the PCI-scaled window and the gated truncation step (\WSname), evaluation scenes. Rows as in Table~\ref{tab:overall}.}
\label{tab:absolute}
\end{table}

\subsection{Other summaries of the error}
\label{supp:pooling}
The weighted arithmetic mean and the weighted root mean square of the phase linking error over all centers give more
weight to the centers with large errors than the weighted geometric mean of the ratio to the oracle
(Section~\ref{sec:simdesign}). With either
summary, \WSname{} lowers the error relative to the SHP selection alone, and the largest window with \WSname{} has the
lowest error of the four windows and less error than the same SHP selection without either step at its best window, as
with the geometric mean.
The reductions are smaller with the root mean square. With the arithmetic mean and the root mean square, respectively,
\WSname{} lowers the error at the search window of Table~\ref{tab:configs} by \vplAriGainMin{} to \vplAriGainMax\,\%
and by \vplRmsGainMin{} to \vplRmsGainMax\,\%, against \vtChainGainAllMin{} to \vtChainGainAllMax\,\% with the
geometric mean. The largest window with \WSname{} has \vplAriBestNMin{} to \vplAriBestNMax\,\% and
\vplRmsBestNMin{} to \vplRmsBestNMax\,\%, respectively, less error than the same SHP selection without either step at
its best window, against
\vwMainWSBestNMin{} to \vwMainWSBestNMax\,\% (three configurations).

The 95\,\% intervals of a bootstrap resampling of the scenes \citep{efron1993introduction}, paired between the
methods, have half-widths of at most \vmBootHalfMax{} for the ratios between the SHP selections over all centers and
for those between the variants of one method. The intervals include one only for \WSname{} against the
truncation step alone in S1-79, and for CMP against the full window in TSX-44 with the truncation step alone
and with \WSname. The other ratios thus differ from one by more than the sampling noise of the scenes.

\subsection{Gate}
\label{supp:gate}
This section completes the paragraph on the contributions of the two parts in Section~\ref{sec:simresults}. At $c=0$, the gate passes \vgFarMin{} to \vgFarMax\,\% of the distributed centers farther than 20 pixels from a
point scatterer and 0.5 window radii from a class border. It passes \vgNearPsMin{} to \vgNearPsMax\,\% of those
closer than three pixels to a scatterer and \vgNearBdMin{} to \vgNearBdMax\,\% of those within 0.05 window radii of a border. The gate thus rarely
applies the truncation step in zero-coherence clutter far from scatterers and borders, and often near them. The region farther than 20 pixels from a point scatterer and 0.5 window radii from a class border holds \vgFarAreaMin{} to
\vgFarAreaMax\,\% of the area at $c=0$, so that most zero-coherence area lies near scatterers or borders. At a
coherence of 0.2, the error relative to the oracle with \WSname{} differs from that with the truncation step alone by
at most \vcLowWSDiffMax{}, so that at low coherence the gate changes the error little. At $c=0.95$ the
median of $\eta$ falls with the fringe density but stays far above $\eta_0$, at \vgEtaHiSShort{} to \vgEtaHiTSX{} at
four or more fringes per window diameter, so that the gate passes almost every coherent center in dense fringes
(at least \vgOpenHighMin\,\% of the distributed centers with $c\ge0.5$ in every fringe-density bin). At two or more fringes per window diameter, fewer FaSHPS ensembles are
truncated with \WSname{} than with the truncation step alone (\vgTrWSSLong{} to \vgTrWSTSX\,\% against
\vgTrSSLong{} to \vgTrSTSX\,\%, three configurations).

\subsection{Size of the search window}
\label{supp:windows}
This section completes the paragraph on the size of the search window in Section~\ref{sec:simresults}.
Table~\ref{tab:windows} gives the error over all centers for the four windows of Fig.~\ref{fig:windows}.

\begin{table*}[tbp]
\centering\footnotesize
\setlength{\tabcolsep}{3.2pt}
\begin{tabular}{lccc@{\hspace{8pt}}ccc@{\hspace{8pt}}ccc}
\toprule
Search window & Full & Full+T & Full\WSsuffix & FaSHPS & FaSHPS+T & FaSHPS\WSsuffix & CMP & CMP+T & CMP\WSsuffix \\
\midrule
\multicolumn{10}{l}{\emph{TSX-44}} \\
$31\times31$ & 1.53 & 1.55 & 1.65 & 1.55 & 1.55 & 1.66 & 1.57 & 1.58 & 1.68 \\
$51\times51$ & 1.48 & 1.38 & 1.41 & 1.48 & 1.37 & 1.40 & 1.50 & 1.39 & 1.43 \\
$75\times75$ & 1.67 & 1.36 & 1.35 & 1.64 & 1.33 & 1.33 & 1.66 & 1.35 & 1.35 \\
$101\times101$ & 1.95 & 1.37 & 1.34 & 1.90 & 1.34 & 1.31 & 1.92 & 1.36 & 1.33 \\
\midrule
\multicolumn{10}{l}{\emph{S1-79}} \\
$9\times75$ & 1.52 & 1.51 & 1.59 & 1.49 & 1.49 & 1.57 & 1.52 & 1.51 & 1.60 \\
$13\times113$ & 1.42 & 1.35 & 1.40 & 1.38 & 1.32 & 1.37 & 1.41 & 1.34 & 1.39 \\
$19\times169$ & 1.50 & 1.29 & 1.31 & 1.44 & 1.24 & 1.26 & 1.46 & 1.26 & 1.28 \\
$25\times225$ & 1.64 & 1.28 & 1.28 & 1.57 & 1.22 & 1.23 & 1.59 & 1.25 & 1.25 \\
\midrule
\multicolumn{10}{l}{\emph{S1-20}} \\
$9\times75$ & 1.54 & 1.53 & 1.61 & 1.49 & 1.48 & 1.56 & 1.51 & 1.50 & 1.58 \\
$13\times113$ & 1.47 & 1.40 & 1.43 & 1.41 & 1.34 & 1.37 & 1.42 & 1.35 & 1.39 \\
$19\times169$ & 1.59 & 1.35 & 1.35 & 1.50 & 1.28 & 1.29 & 1.52 & 1.30 & 1.30 \\
$25\times225$ & 1.78 & 1.35 & 1.34 & 1.68 & 1.28 & 1.27 & 1.70 & 1.29 & 1.28 \\
\bottomrule
\end{tabular}
\caption{Phase linking error relative to the oracle in the largest window, weighted geometric mean over all centers of the evaluation scenes, for four search windows (azimuth $\times$ range, ellipse inscribed in the stated rectangle). The largest is the search window of Table~\ref{tab:configs}. With \WSname{} the PCI-scaled window shrinks from the window of each row. Settings and thresholds are those of the main comparison. The columns are those of Table~\ref{tab:overall} without the PCI-scaled window alone (+S).}
\label{tab:windows}
\end{table*}

In zero-coherence clutter ($c=0$), the largest window has the lowest error of the four, for the SHP selections
alone, with the truncation step and with \WSname. There the smallest window has \vtWinZeroSmallMin{} to
\vtWinZeroSmallMax{} times the error of the largest, without and with the truncation step.

Over all centers and with \WSname, the advantage of FaSHPS and CMP over the full window is smallest at the
smallest window. There CMP\WSsuffix{} even has \vtWinSmallCmpFullWSTSX\,\% more error than the full window with
\WSname{} in TSX-44 (Table~\ref{tab:windows}).

The truncation step and \WSname{} also lower the error when every variant is used at its own best window, chosen
in hindsight on the evaluation scenes. Each SHP selection then has \vwBbFullMin{} to \vwBbFaMax\,\% less error with the
better of the two than without either step. 

Even at the window where the SHP selection alone has its lowest error, chosen in hindsight on the evaluation
scenes, the truncation step lowers the error over all centers by \vtWinBestGainSMin{} to \vtWinBestGainSMax\,\%. For FaSHPS it
lowers the error at the distributed centers with one fringe or more per window diameter by \vwGainFrMin{} to
\vwGainFrMax\,\%. \WSname{} lowers the error at these centers less than the truncation step alone, since the PCI-scaled window shrinks in
proportion to the search window.

In coherent clutter with few fringes, a truncation
raises the error at the smallest window and lowers it at the largest. At the distributed centers of coherence 0.5 to
0.8 with fewer than 0.25 fringes per diameter of the largest window, the error of the truncated FaSHPS ensembles rises
by \vtWinFlatRiseMin{} to \vtWinFlatRiseMax\,\% at the smallest window and falls by \vtWinFlatGainMainMin{} to
\vtWinFlatGainMainMax\,\% at the largest. At the largest window more of the ensembles are truncated, and to a
smaller share of their size. The smallest windows hold \vtWinSmallOffsetsTSX{} pixels (TSX-44) and
\vtWinSmallOffsetsSOne{} pixels (S1-79 and S1-20), and there the loss of looks by a truncation raises the error more
than the removal of bias lowers it.

With \WSname, the error of the four windows differs by only \vwWinSpreadEdgeWSMin{} to
\vwWinSpreadEdgeWSMax\,\% both at the centers with one fringe or more per window diameter and at those near
discontinuities, where
the truncation step acts. At $c=0$ and $c=0.2$ the smallest window has the largest error.

The largest window with \WSname{} keeps its advantage when the centers are weighted so that the distribution of
$\eta$ matches that of a real stack, a weighting that ignores the fringe density. It then has \vwRwLessMin{} to
\vwRwLessMax\,\% less error than the same SHP selection without either step at its best window. This weighting uses the real stacks of
Section~\ref{sec:realdata}, namely Visp, the slow-deformation, Maiun and long-term coherent test areas and the
landslide substack, each with its configuration. It is applied to the full window, FaSHPS and CMP.

\subsection{Elementary scenes}
\label{supp:units}
This section completes the paragraphs on where \WSname{} lowers the error and on the contributions of its two parts
in Section~\ref{sec:simresults}.
Table~\ref{tab:units} gives the error of every variant in the elementary scenes of TSX-44. In homogeneous clutter,
every SHP selection alone is within \vtUnitHomNMax\,\% of the oracle error at all three coherences (TSX-44), and the
truncation step and \WSname{} raise the error, since ensembles are truncated there only by chance. In coherent clutter
the multilooked phase of a small ensemble is already precise, so that the scores of the candidate sizes differ little,
the maximum falls on a smaller size by chance, and its excess over the score of the whole ensemble exceeds the margin
more often. At $c=0.5$, \vcUnitTruncSMidMin{} to \vcUnitTruncSMidMax\,\% of the ensembles are truncated with the
truncation step, although homogeneous clutter needs no truncation. In absolute terms, for the full window in TSX-44, the weighted median phase linking error
is \vuMedNLow$^\circ$, \vuMedNMid$^\circ$ and \vuMedNHigh$^\circ$ at $c=0.2$, 0.5 and 0.95, respectively, without the truncation step and \vuMedSLow$^\circ$, \vuMedSMid$^\circ$ and \vuMedSHigh$^\circ$ with it, compared with \vuMedBestLow$^\circ$,
\vuMedBestMid$^\circ$ and \vuMedBestHigh$^\circ$ at the oracle size. The added error lies in the truncated
ensembles, and the 90th percentile rises from \vuPnNLow$^\circ$, \vuPnNMid$^\circ$ and \vuPnNHigh$^\circ$ to
\vuPnSLow$^\circ$, \vuPnSMid$^\circ$ and \vuPnSHigh$^\circ$. The error that the truncation step adds in
homogeneous clutter is thus small in degrees.

At a phase step of 3\,rad and $c=0.95$, every SHP selection has
\vtUnitStepMin{} to \vtUnitStepMax{} times the oracle error, alone, with the truncation step and with \WSname{}
(TSX-44). Tests on amplitudes do not see a boundary in phase, and CMP gives the phase differences little
weight, since its assumed coherence is at most 0.3 (Eq.~\eqref{eq:cmpstat}). An ensemble is truncated at such a
boundary only where the boundary is close to the center. With an additional amplitude contrast of 6\,dB at a phase
step of 3\,rad and $c=0.65$, the SHP tests detect the boundary in part, CMP better than FaSHPS, and the truncation step
raises the error of CMP there. At the coherence step, the SHP selections alone mix the two sides with no loss in phase,
the truncation step raises the error about as in homogeneous clutter at $c=0.5$, and \WSname{} raises it further, since
it also reduces the window at the coherent centers.

\begin{table*}[tbp]
\centering\footnotesize
\setlength{\tabcolsep}{3.2pt}
\resizebox{\linewidth}{!}{%
\begin{tabular}{lcccc@{\hspace{8pt}}cccc@{\hspace{8pt}}ccccc}
\toprule
Elementary scene & Full & Full+S & Full+T & Full\WSsuffix & FaSHPS & FaSHPS+S & FaSHPS+T & FaSHPS\WSsuffix & CMP & CMP+S & CMP+T & CMP\WSsuffix & Oracle \\
\midrule
Homogeneous, $c=0.2$ & \textbf{1.01} & 1.02 & 1.16 & 1.13 & \textbf{1.01} & 1.02 & 1.17 & 1.13 & \textbf{1.01} & 1.02 & 1.16 & 1.13 & 4.0$^\circ$ \\
Homogeneous, $c=0.5$ & \textbf{1.01} & 1.20 & 1.34 & 1.54 & 1.03 & 1.22 & 1.35 & 1.59 & 1.03 & 1.22 & 1.35 & 1.55 & 1.8$^\circ$ \\
Homogeneous, $c=0.95$ & \textbf{1.01} & 3.70 & 1.77 & 4.07 & 1.02 & 3.75 & 1.79 & 4.17 & \textbf{1.01} & 3.70 & 1.81 & 4.10 & 0.4$^\circ$ \\
Phase step 1\,rad, $c=0.65$ & \textbf{1.30} & 1.39 & 1.50 & 1.61 & 1.31 & 1.40 & 1.50 & 1.61 & 1.31 & 1.40 & 1.49 & 1.58 & 1.7$^\circ$ \\
Phase step 3\,rad, $c=0.65$ & 1.71 & 1.65 & 1.59 & 1.69 & 1.73 & 1.67 & 1.60 & 1.68 & 1.72 & 1.66 & \textbf{1.58} & 1.67 & 1.6$^\circ$ \\
Phase step 6.3\,rad, $c=0.65$ & \textbf{1.07} & 1.32 & 1.45 & 1.76 & 1.08 & 1.34 & 1.49 & 1.77 & \textbf{1.07} & 1.34 & 1.45 & 1.74 & 1.5$^\circ$ \\
Phase step 1\,rad, $c=0.95$ & 1.86 & 2.62 & \textbf{1.79} & 2.83 & 1.87 & 2.64 & \textbf{1.79} & 2.87 & 1.86 & 2.62 & 1.80 & 2.83 & 0.6$^\circ$ \\
Phase step 3\,rad, $c=0.95$ & 2.69 & 2.97 & 1.88 & 3.07 & 2.70 & 2.99 & \textbf{1.87} & 3.09 & 2.69 & 2.97 & 1.93 & 3.08 & 0.5$^\circ$ \\
Phase step 6.3\,rad, $c=0.95$ & \textbf{1.08} & 2.88 & 1.76 & 3.33 & \textbf{1.08} & 2.90 & 1.77 & 3.35 & \textbf{1.08} & 2.88 & 1.77 & 3.30 & 0.5$^\circ$ \\
Phase step 3\,rad and 6\,dB, $c=0.65$ & 1.77 & 1.65 & 1.51 & 1.63 & 1.45 & 1.44 & 1.42 & 1.55 & \textbf{1.06} & 1.21 & 1.24 & 1.43 & 2.6$^\circ$ \\
Coherence step 0.65 and 0.2 & \textbf{1.01} & 1.16 & 1.34 & 1.44 & 1.04 & 1.20 & 1.35 & 1.45 & 1.03 & 1.19 & 1.38 & 1.49 & 2.0$^\circ$ \\
\bottomrule
\end{tabular}}
\caption{Elementary scenes of Section~\ref{sec:simdesign}, TSX-44, one kind per row: phase linking error relative to the oracle, weighted geometric mean over the centers of the elementary scenes of the row, and their median oracle error. Columns as in Table~\ref{tab:overall}. The lowest value of each row is set in bold.}
\label{tab:units}
\end{table*}

\subsection{Further results on the truncation step}
\label{supp:truncmore}
For the three SHP selections at $c=0.95$, \WSname{} raises the error relative to the truncation step alone in the
scenes without atmospheric screen, by \vtAtmNoneRiseTwoDuoMin{} to \vtAtmNoneRiseTwoDuoMax\,\% in TSX-44 and by
\vtAtmNoneRiseSLongMin{} to \vtAtmNoneRiseSLongMax\,\% in S1-79. In the scenes with a screen it lowers the error, by \vtAtmScreenGainTSXDuo\,\% in TSX-44
and by \vtAtmScreenGainSLongMin{} to \vtAtmScreenGainSLongMax\,\% in S1-79.

With few fringes, the truncation step lowers the
error most where the window contains a class border, a discontinuity or a strong atmosphere. Below 0.25 fringes per
window diameter and at $c\ge0.8$, it lowers the error of the full window by \vtLoNearGainMin{} to
\vtLoNearGainMax\,\% where a class border or a discontinuity lies within one window radius of the center or the
atmosphere is strong. These centers hold \vtLoNearShareMin{} to \vtLoNearShareMax\,\% of the area below 0.25
fringes per window diameter and at $c\ge0.8$. At the other centers the truncation step lowers the error by
\vtLoFreeLowerMin{} to \vtLoFreeLowerMax\,\%. Over all centers within 0.25 window radii of a discontinuity, at any fringe density, the error of
the full window falls from \vtDiscN{} to \vtDiscS{} times the oracle error, and over all centers of the scenes
with the strong atmosphere from \vtAtmN{} to \vtAtmS{} (TSX-44).

Below 0.25 fringes per window diameter and at $c\ge0.5$,
the weighted median phase linking error of the full window is \vtAbsLoFullMin$^\circ$ to \vtAbsLoFullMax$^\circ$,
and the truncation step lowers it to \vtAbsLoSMin$^\circ$ to \vtAbsLoSMax$^\circ$. At few fringes, the error
and its reduction by the truncation step are thus small in degrees.

\subsection{Further results on the SHP selections}
\label{supp:selmore}
Without either step, FaSHPS accepts \vmAcceptFaMin{} to \vmAcceptFaMax\,\% and CMP \vmAcceptCmpMin{} to
\vmAcceptCmpMax\,\% of the search window on average over all centers, and within the PCI-scaled window of \WSname{}
they accept about the same shares of it. With and without \WSname, at most
\vmInsuffFaMax\,\% (FaSHPS) and \vmInsuffCmpMax\,\% (CMP) of all centers are insufficient, compared with none for the
full window.

The other two errors follow the phase linking error.
The coherence error (Eq.~\eqref{eq:cerr}) over the distributed centers is \vmCohFaNTSX{} to \vmCohFaNSShort{} for
FaSHPS alone, \vmCohFaSTSX{} to \vmCohFaSSShort{} with the truncation step and \vmCohFaWSTSX{} to \vmCohFaWSSShort{}
with \WSname{}. Table~\ref{tab:matrix} gives the matrix error of the main comparison. Relative
to the matrix error at the oracle size, it is \vmxAllNMin{} to \vmxAllNMax{} over all centers for the SHP selections
alone and \vmxAllSMin{} to \vmxAllSMax{} with the truncation step or \WSname{}.

\begin{table*}[tbp]
\centering\footnotesize
\setlength{\tabcolsep}{2.6pt}
\begin{tabular}{lc@{\hspace{8pt}}ccc@{\hspace{6pt}}ccc@{\hspace{6pt}}ccc}
\toprule
Centers & Oracle ($^\circ$) & Full & Full+T & Full\WSsuffix & FaSHPS & FaSHPS+T & FaSHPS\WSsuffix & CMP & CMP+T & CMP\WSsuffix \\
\midrule
\multicolumn{11}{l}{\emph{TSX-44}} \\
$c=0$ & 80.2 & 1.00 & 1.02 & 1.01 & 1.18 & 1.18 & 1.18 & 1.23 & 1.23 & 1.23 \\
$c=0.5$ & 9.5 & 1.46 & 1.10 & 1.10 & 1.40 & 1.08 & 1.08 & 1.39 & 1.07 & 1.07 \\
$c=0.95$ & 3.2 & 3.32 & 1.30 & 1.19 & 3.25 & 1.26 & 1.16 & 3.27 & 1.26 & 1.16 \\
\addlinespace[2pt]
All & 9.2 & 1.54 & 1.12 & 1.10 & 1.55 & 1.13 & 1.11 & 1.56 & 1.14 & 1.12 \\
\midrule
\multicolumn{11}{l}{\emph{S1-79}} \\
$c=0$ & 82.0 & 1.01 & 1.03 & 1.02 & 1.20 & 1.21 & 1.21 & 1.21 & 1.22 & 1.22 \\
$c=0.5$ & 14.1 & 1.22 & 1.03 & 1.05 & 1.16 & 1.00 & 1.02 & 1.15 & 1.00 & 1.02 \\
$c=0.95$ & 4.7 & 2.43 & 1.22 & 1.23 & 2.33 & 1.18 & 1.19 & 2.36 & 1.18 & 1.19 \\
\addlinespace[2pt]
All & 15.1 & 1.34 & 1.06 & 1.07 & 1.36 & 1.08 & 1.09 & 1.36 & 1.08 & 1.09 \\
\midrule
\multicolumn{11}{l}{\emph{S1-20}} \\
$c=0$ & 80.1 & 1.02 & 1.04 & 1.03 & 1.09 & 1.11 & 1.10 & 1.12 & 1.13 & 1.13 \\
$c=0.5$ & 11.3 & 1.39 & 1.10 & 1.10 & 1.31 & 1.05 & 1.05 & 1.30 & 1.04 & 1.04 \\
$c=0.95$ & 3.9 & 2.78 & 1.36 & 1.27 & 2.69 & 1.31 & 1.23 & 2.72 & 1.32 & 1.23 \\
\addlinespace[2pt]
All & 12.1 & 1.47 & 1.12 & 1.11 & 1.43 & 1.10 & 1.08 & 1.42 & 1.09 & 1.08 \\
\bottomrule
\end{tabular}
\caption{Matrix error relative to the matrix error at the oracle size, weighted geometric mean over the distributed centers of three coherence classes and over all centers (evaluation scenes, amplitude-normalized estimator), and the weighted median matrix error at the oracle size. The oracle is chosen by the phase linking error. Columns as in Table~\ref{tab:overall}, without the PCI-scaled window alone (+S).}
\label{tab:matrix}
\end{table*}

\section{Experiments on TerraSAR-X and Sentinel-1 data}
\label{supp:real}
This section completes Section~\ref{sec:realdata}.

\subsection{Processing parameters}
\label{supp:sensor}
Table~\ref{tab:sensor} gives the parameters of Eq.~\eqref{eq:rhopred} for the speckle correlation in Sections~\ref{sec:null} and~\ref{sec:realspeckle}.

\begin{table}[htbp]
\centering\small
\begin{tabular}{lcc}
\toprule
 & TerraSAR-X (TSX-44) & Sentinel-1 (S1-79, S1-20)\\
\midrule
Range bandwidth (MHz) & 100 & 56.5\\
Range sampling rate (MHz) & 109.9 & 128.7\\
Range weighting coefficient & 0.60 & 0.75\\
Azimuth processed bandwidth (Hz) & 2765 & 327\\
Pulse repetition frequency (Hz) & 3562 & 486.5\\
Azimuth weighting coefficient & 0.60 & 0.70\\
Predicted lag-one correlation (az., rg.) & 0.695, 0.606 & 0.664, 0.817\\
\bottomrule
\end{tabular}
\caption{Processing parameters of the SLC data, with a raised-cosine (generalized Hamming) spectral weighting on both axes, and the lag-one speckle correlation that they predict through Eq.~\eqref{eq:rhopred}, which the processing uses. The range sampling rate of Sentinel-1 is that of the data oversampled by a factor of two in range.}
\label{tab:sensor}
\end{table}

The TerraSAR-X stack over Visp has a single polarization (horizontal transmit and receive), a
descending orbit and right-looking geometry, with an incidence angle of about $45^\circ$ ($43.3^\circ$ to
$45.8^\circ$ across the swath). The X-band data has a range bandwidth of 100\,MHz and a pixel spacing of 1.36\,m in
slant range and 1.98\,m in azimuth. Phases are referenced to the acquisition of 11 September 2019, and
perpendicular baselines relative to it range from $-206$ to $128$\,m.

The decay constant of the coherence model of CMP, fitted to each real stack, is \vrVispCmpDecay{} days
for Visp, and \vrSlowCmpDecay, \vrDecCmpDecay, \vrLtcCmpDecay{} and \vrLandCmpDecay{} days for the slow-deformation, Maiun and long-term coherent test areas and the landslide substack, respectively.
These constants, like those fitted in the simulations (Section~\ref{app:settings}), are much longer than the 200 days
that \citet[Section~3.1]{zhao2022statistically} set. The mean coherence of the single-reference interferograms falls
over the first part of each stack and then levels off, as the coherence of Eq.~\eqref{eq:cohfamily} does in the
simulations, and an exponential with a free level, fitted over the whole stack, follows the slow part, which gives the long
constants.

For Sentinel-1, Eq.~\eqref{eq:rhopred} predicts a speckle correlation of \vdPredSoneRgOne,
\vdPredSoneRgTwo{} and \vdPredSoneRgThree{} at range lags one to three, and \vdPredSoneAzOne, \vdPredSoneAzTwo{} and
\vdPredSoneAzThree{} in azimuth. Compared with ideal band-limited oversampling, the B-spline interpolator that oversamples the data in range
(Section~\ref{supp:s1site}) raises the range values
by at most \vgBsplineDevMax. The reference acquisition of the coregistration matches this prediction to within
\vgRefDevMax{} at range lags one to three in the Maiun and long-term coherent test areas, measured per acquisition as the
magnitude of the summed products of neighboring samples relative to the summed intensity. In the \vgNCoreg{}
coregistered acquisitions, the correlation at these lags exceeds the prediction by up to \vgRiseMax, and the
sidelobes are damped, with a median of \vgSideFiveMin{} at range lag five in both test areas against \vgPredFive{}
predicted. Both deviations are of the size that the resampling of the coregistration produces. The coregistration
resamples with a Lanczos interpolator of order \vgLanczosOrder, and at fractional range
shifts of up to half a pixel, this interpolator followed by the B-spline raises the correlation at lags one to three
by up to \vgLanczosRiseMax{} and gives \vgLanczosFiveMin{} to \vgLanczosFiveMax{} at lag five.

\subsection{Mask of the Visp stack}
\label{supp:vispmask}
This section completes the comparison of the NSPC mask with the layover
and shadow maps of Section~\ref{sec:realmask} (Fig.~\ref{fig:mask}). The shares in this section refer to the valid
pixels, since pixels of the Visp scene with a zero value in at least one acquisition, \vdVispZero\,\% of the scene,
are excluded from the outset (Section~\ref{sec:nspcmask}). Table~\ref{tab:vispmask}
gives the shares at $h=0.25$, where the mask covers \vdVispLayMaskTwentyFive\,\% of the layover of the terrain model.

In the layover of the terrain model, the median NSPC is
\vdVispLayMedRaw{} before topographic phase removal, at the correlated-noise floor, and \vdVispLayMedCor{} after it, and
\vdVispLayMaskRaw\,\% of these pixels lie below $h=0.25$ before the removal. In its shadow the median NSPC is
\vdVispShMedCor{} before and after the removal, close to the correlated-noise floor of \vkNullTSXMed.

Over the whole scene, \vdNeiSceneShare\,\% of the masked
pixels lie outside both classes of the terrain model map (Table~\ref{tab:vispmask}). Their low agreement is a property
of the data and does not arise from the topographic correction, since \vdNeiRawToo\,\% of them are masked before
it as well. They are dark, with a median amplitude of \vdNeiDb\,dB relative to the scene median, similar to
the shadow of the terrain model (\vdShDb\,dB) and unlike its layover (\vdLayDb\,dB), which argues against building layover as the main cause. They also lie close to bright pixels, i.e., the 2\,\% of pixels with the highest mean amplitude, at a
median distance of \vdNeiDist{} pixels, compared with \vdAllDist{} for all pixels. Dark pixels, i.e., those more than
3\,dB below the scene median, are masked more often the closer they lie to a bright pixel, in the town section
(\vdDarkTownFar\,\% beyond 16 pixels and \vdDarkTownNear\,\% within 4 pixels) as in the rest of the scene
(\vdDarkRestFar\,\% and \vdDarkRestNear\,\%, respectively).

The surface model map places most of the
masked pixels outside both classes of the terrain model map in its shadow. Its classes agree with the mean amplitude,
with a median of \vdSurfLayDb\,dB in layover and \vdSurfShDb\,dB in shadow relative to the scene median. Its layover
contains \vdTerrLayInSurf\,\% of the layover of the terrain model. Of the masked pixels outside both classes of the
terrain model map, it places \vdNeiSceneSurfSh\,\% in shadow and \vdNeiSceneSurfLay\,\% in layover.

Outside the layover of the terrain model, the masking of the surface model classes hardly depends on the removal
of the topographic phase. Over the whole scene, \vdSurfLayOnlyMaskRaw\,\% of the surface model layover outside that
of the terrain model lies below $h$ before the removal and \vdSurfLayOnlyMaskCor\,\% after, while the masked share of
the surface model shadow in the town section does not depend on it (\vdTownSurfShMaskedRaw\,\% before,
\vdTownSurfShMasked\,\% after).

Within the town section the mask covers the layover of the surface
model least (Table~\ref{tab:vispmask}), since bright returns keep NSPC high there (median \vdTownSurfLayMed), and being
in shadow is not sufficient for a pixel to be masked. Of the shadow pixels in the town section more than 16 pixels from
a bright pixel, \vdTownShFarMasked\,\% are masked, compared with \vdTownShNearMasked\,\% at 2 to 8 pixels. Of the masked pixels of the town section in
neither class of the surface model map, \vdTownNeiNearSh\,\% lie within two pixels of shadow. Most masked
town pixels thus lie in or beside building shadow near bright pixels. Over the forest, the surface model classes follow the
canopy at the scale of single trees, and at most \vdForestMaxMasked\,\% of the pixels of any class are masked.

\begin{table}[tbp]
\centering\footnotesize
\setlength{\tabcolsep}{4pt}
\begin{tabular}{lcccc}
\toprule
 & \multicolumn{2}{c}{Whole scene} & \multicolumn{2}{c}{Town section} \\
\cmidrule(lr){2-3}\cmidrule(lr){4-5}
 & Terrain model & Surface model & Terrain model & Surface model \\
\midrule
\multicolumn{5}{l}{\emph{Masked pixels by class of the map, \% of all masked pixels}} \\
Layover & \vdCompSceneTerrLay & \vdCompSceneSurfLay & \vdCompTownTerrLay & \vdCompTownSurfLay \\
Shadow & \vdCompSceneTerrSh & \vdCompSceneSurfSh & \vdCompTownTerrSh & \vdCompTownSurfSh \\
Neither class & \vdCompSceneTerrNei & \vdCompSceneSurfNei & \vdCompTownTerrNei & \vdCompTownSurfNei \\
\addlinespace[2pt]
\multicolumn{5}{l}{\emph{Pixels in layover or shadow of the map, \% of all pixels}} \\
Layover or shadow & \vdAreaSceneTerr & -- & \vdAreaTownTerr & \vdAreaTownSurf \\
Shadow & -- & -- & -- & \vdTownSurfShArea \\
\addlinespace[2pt]
\multicolumn{5}{l}{\emph{Masked pixels of each class, \% of the pixels of the class}} \\
Layover & \vdVispLayMaskTwentyFive & -- & -- & \vdTownSurfLayMasked \\
Shadow & \vdVispShMaskTwentyFive & -- & -- & \vdTownSurfShMasked \\
Neither class & -- & -- & -- & \vdTownSurfNeiMasked \\
\bottomrule
\end{tabular}
\caption{NSPC mask of Visp ($h=0.25$, after topographic phase removal) against the layover and shadow maps of
Fig.~\ref{fig:mask}, computed from the terrain model swissALTI3D and from the surface model swissSURFACE3D Raster. A
pixel is masked if its NSPC lies below $h$, and its class is the one that the map assigns to it. The first block
gives how the masked pixels divide among the classes of a map, the second the share of all pixels that a map puts in
layover or shadow, and the third the share of the pixels of each class that are masked. Shares refer to the valid pixels, and the town section is the dashed box of
Fig.~\ref{fig:mask}. A dash marks a value that was not computed.}
\label{tab:vispmask}
\end{table}

\subsection{Point scatterers in clutter with zero coherence}
\label{supp:sidelobe}
Table~\ref{tab:darktest} gives the simulation of Section~\ref{sec:realmask}, in which NSPC is measured at the
clutter pixels. Within two to six pixels of dense scatterers with different heights, 25 to 40\,dB above the clutter
(third row of Table~\ref{tab:darktest}), the median NSPC decreases to \vdDarkSimMedMin{} to \vdDarkSimMedMax, and \vdDarkSimShareMin{} to
\vdDarkSimShareMax\,\% of the clutter pixels fall below $h=0.25$. Among the dark pixels of the
town section of Visp within four pixels of a bright pixel, \vdDarkTownNear\,\% are masked, against \vdDarkSimShareMin{} to
\vdDarkSimShareMax\,\% of the simulated clutter pixels near dense bright scatterers.
The decrease follows from the sign of the sidelobes. Beyond the main lobe, the point spread function alternates in
sign from one sample to the next along range, at up to
\vdSideMax\,dB relative to the peak. For a sidelobe amplitude equal to that of the clutter, the expected value of
$\sqrt{Q^+}$ for two range neighbors falls from \vdSideAgrNone{} for clutter alone to \vdSideAgrOpp{} with a
sidelobe of opposite sign, and with equal signs it would rise to \vdSideAgrEq.
With scatterers of 10 to 25\,dB the decrease of NSPC is confined
to a distance of two pixels, so a wider decrease requires scatterers 25\,dB or more above the clutter. By the
sign of the sidelobes (Section~\ref{sec:realmask}), a single scatterer is sufficient to lower NSPC below the floor. Several scatterers with different phase histories, as
left by buildings of different height that the terrain model does
not contain, increase the effect. With
scatterers of 25 to 40\,dB, the masked share is \vdDarkSimRatioMin{} to \vdDarkSimRatioMax{} times that of
scatterers with one common phase history.
Around a scatterer of 10 to 25\,dB or an isolated scatterer, NSPC is lowest close to the scatterer and rises toward the floor with the
distance (first, second and last rows of Table~\ref{tab:darktest}). On the dense grid of scatterers of 25 to 40\,dB, a pixel 5
to 6 pixels from the nearest scatterer lies near the middle between four of them, where their sidelobes overlap, and
NSPC falls with the distance (third and fourth rows) and is lower with different heights than with one.

\begin{table}[tbp]
\centering\small
\begin{tabular}{llccc}
\toprule
Scatterers & Signal-to-clutter & 2 px & 3 px & 5 to 6 px\\
\midrule
Dense, different heights & 10 to 25\,dB & 0.35 (5\,\%) & 0.41 (0\,\%) & 0.42 (0\,\%) \\
Dense, one height & 10 to 25\,dB & 0.35 (4\,\%) & 0.41 (1\,\%) & 0.42 (0\,\%) \\
Dense, different heights & 25 to 40\,dB & 0.37 (16\,\%) & 0.34 (14\,\%) & 0.33 (15\,\%) \\
Dense, one height & 25 to 40\,dB & 0.43 (8\,\%) & 0.38 (5\,\%) & 0.35 (8\,\%) \\
Isolated, different heights & 25 to 40\,dB & 0.35 (13\,\%) & 0.37 (4\,\%) & 0.40 (5\,\%) \\
\bottomrule
\end{tabular}
\caption{Simulated NSPC of clutter pixels with zero coherence near point scatterers (TSX-44, $N=44$). Entries are the median and the share below $h=0.25$, by distance to the nearest scatterer. The correlated-noise floor is \vdSimFloor. Scatterers lie on a grid with a spacing of 8 pixels (dense) or 40 pixels (isolated), each shifted by up to two pixels along each axis at random. Their phase histories follow from heights between 0 and 20\,m, or from one common height.}
\label{tab:darktest}
\end{table}

\subsection{Sentinel-1 site}
\label{supp:s1site}
This section completes the description of the Sentinel-1 site of Section~\ref{sec:stacks}.

The scene of Section~\ref{sec:stacks} runs from Rahimabad ($36.12^\circ$\,N, $74.30^\circ$\,E) to Shishkat
($36.36^\circ$\,N, $74.87^\circ$\,E) on Attabad Lake. The
relief is extreme. Within the scene the valley floor rises from
about 1500\,m to about 2200\,m, and the terrain reaches about
7500\,m on the north flank of Rakaposhi (7788\,m). The slopes are
arid and mostly bare rock and debris, with vegetation confined to
irrigated terraces around the villages, so that large parts of the
scene remain coherent over years, while the steep slopes cause
extended layover and shadow. Slope instability is widespread.
Landslide inventories along the highway, compiled or updated with InSAR, list hundreds
of landslides \citep{hussain2023landslide,su2021landslide}, among them
slow-moving landslides of glacio-fluvial and moraine deposits above the villages of the
Hunza--Nagar section, the fastest with a line-of-sight rate of more than 0.1\,m per year
\citep{rehman2020analysis}.

The acquisitions are from subswath IW1, with
vertical transmit and receive polarization, on ascending orbits. The stack was coregistered to 11 May 2024 with
the coregistration procedure of the GAMMA software for data of the terrain observation with progressive scans (TOPS) mode \citep{wegmuller2016sentinel1}, with the temporally adjacent acquisition
as secondary reference for the spectral diversity, and deramped. The oversampling in range uses a B-spline
interpolator of degree \vgBsplineDegree, and the pixel spacing of the oversampled data is 1.16\,m in slant range and
13.97\,m in azimuth.

\subsection{Sentinel-1 test areas}
\label{supp:s1crops}
Table~\ref{tab:s1crops} lists the test areas of Section~\ref{sec:stacks}.

\begin{table}[tbp]
\centering\small
\begin{tabular}{lccl}
\toprule
Test area & Latitude & Longitude & Used for\\
\midrule
Layover and shadow & $36.209^\circ$\,N & $74.307^\circ$\,E & mask\\
Maiun & $36.239^\circ$\,N & $74.424^\circ$\,E & mask, \WSname, runtime, Section~\ref{sec:noquant}\\
Long-term coherent & $36.257^\circ$\,N & $74.456^\circ$\,E & \WSname, runtime, Section~\ref{sec:noquant}\\
Slow deformation & $36.294^\circ$\,N & $74.801^\circ$\,E & mask, \WSname, runtime, Section~\ref{sec:noquant}\\
Fast deformation & $36.207^\circ$\,N & $74.857^\circ$\,E & mask, \WSname, runtime, Section~\ref{sec:noquant}\\
\bottomrule
\end{tabular}
\caption{Approximate center coordinates of the Sentinel-1 test areas and
their use in this paper.}
\label{tab:s1crops}
\end{table}

\subsection{Sentinel-1 mask}
\label{supp:s1mask}
Table~\ref{tab:s1mask} gives the comparison of the NSPC mask with the geometric layover and shadow map of Section~\ref{sec:stacks}.
In the four test areas used for the mask, shadow lies \vbShDbMin{} to \vbShDbMax\,dB below the median intensity of its test area, and
layover \vbLayDbMin{} to \vbLayDbMax\,dB above it, as expected from the imaging geometry.

\begin{table}[tbp]
\centering\small
\begin{tabular}{lccccccccc}
\toprule
 & \multicolumn{3}{c}{Neither} & \multicolumn{3}{c}{Layover} & \multicolumn{3}{c}{Shadow}\\
\cmidrule(lr){2-4}\cmidrule(lr){5-7}\cmidrule(lr){8-10}
Test area & Share & NSPC & Masked & Share & NSPC & Masked & Share & NSPC & Masked\\
 & [\%] & & [\%] & [\%] & & [\%] & [\%] & & [\%]\\
\midrule
Layover and shadow & 77 & 0.44 & 13.8 & 10.4 & 0.19 & 63 & 13.0 & 0.28 & 61 \\
Maiun & 93 & 0.48 & 6.5 & 4.3 & 0.31 & 48 & 2.3 & 0.35 & 33 \\
Fast deformation & 97 & 0.47 & 6.4 & 2.1 & 0.34 & 42 & 1.1 & 0.33 & 35 \\
Slow deformation & 97 & 0.48 & 6.3 & 2.2 & 0.37 & 31 & 1.1 & 0.32 & 36 \\
\bottomrule
\end{tabular}
\caption{NSPC mask of the Sentinel-1 test areas (after topographic phase removal, on the oversampled grid) against the layover and shadow map, by class of the map: share
of the pixels of the test area in the class (Share), median NSPC of the pixels of the class (NSPC), and share of the
pixels of the class whose NSPC lies below $h=0.30$, i.e., that the mask removes (Masked). The long-term coherent test area is not part of this comparison (Table~\ref{tab:s1crops}).}
\label{tab:s1mask}
\end{table}

On the native range grid, with every second range sample, the predicted lag-one range correlation is
\vdNatRgLagOne, against \vdPredSoneRgOne{} on the oversampled grid. Simulated clutter with zero coherence then has a median NSPC of \vdNatFloorMed, against \vkNullSLongMed{} on the oversampled grid (S1-79), so that \vdNatFloorBelowThirty\,\% of it falls below
$h=0.30$. At $h=\vdNatFloorQmil$, the 0.1\,\% quantile of NSPC in this clutter, the mask covers \vdNatLayMin{} to
\vdNatLayMax\,\% of layover and \vdNatShMin{} to \vdNatShMax\,\% of shadow in the four test areas used for the mask. It
also covers \vdNatNeiMin{} to \vdNatNeiMax\,\% of the pixels in neither class, against \vdSoneNeiMaskMin{} to
\vdSoneNeiMaskMax\,\% at $h=0.30$ on the oversampled grid.

Before topographic phase removal, the mask covers \vkLayBackMaskMin{} to \vkLayBackMaskMax\,\% of layover instead of
\vkLayMaskMin{} to \vkLayMaskMax\,\%. In shadow the median NSPC is then \vkShBackMedMin{} to \vkShBackMedMax{}, and
the masked share of shadow changes by at most \vkShBackChangeMax{} percentage points. Pixels in neither class with
the same temporal mean intensity as shadow have, before the removal, a similarly low NSPC, and those with the
intensity of layover a higher NSPC than layover. The low NSPC of shadow thus goes with its low intensity, and layover
agrees with its neighbors slightly less than other pixels of the same intensity (median NSPC \vkLayBackMedMin{} to
\vkLayBackMedMax{} against \vbLayMatchMin{} to \vbLayMatchMax{} before the removal). In the layover and shadow test area,
\vdSoneLsInClass\,\% of all masked pixels belong to one of the two geometric classes, and many of the remaining ones
border them.

\subsection{Further results on the phase-linked maps}
\label{supp:mapsmore}
On Visp the gate passes \vrVispOpen\,\% of the centers, and with \WSname{} \vrVispTruncFaWS\,\% of the FaSHPS
ensembles are truncated, \vrVispTownOfTruncFaWS\,\% of them in the town section, which holds \vrVispTownShare\,\% of
the centers. On the slow-deformation test area the gate passes \vrSlowOpen\,\% of the centers and
\vrSlowTruncFaWS\,\% of the FaSHPS ensembles are truncated, and over the section of the landslide shown in
Fig.~\ref{fig:s1land} the gate passes \vrLandOpen\,\% of the centers and \vrLandTruncFaWS\,\% of the FaSHPS ensembles
are truncated. Of the centers of the Maiun test area, \vrDecBelow\,\% lie below the null quantile
$\eta_0=\vrDecEtaNull$ of S1-79.

On the slow-deformation test area, the median
ensemble size of the full window falls from \vrSlowMedFull{} to \vrSlowMedFullWS{} pixels with \WSname, and over the
landslide section \vrLandTenthPlainFaWS\,\% of the truncated FaSHPS\WSsuffix{} ensembles keep less than a tenth of the
FaSHPS ensemble at the same center. On the long-term coherent test area, where coherent clutter is uniform over large
areas, \WSname{} shrinks the search window at \vrLtcShrinks\,\% of the unmasked centers, to a median of
\vrLtcMedWindow{} of the 4437 pixels of the search window.

Below the null quantile, where FaSHPS\WSsuffix{} returns the
ensembles of FaSHPS, the truncation step alone would truncate \vrVispTruncLoFaS\,\% of the FaSHPS ensembles on Visp
($\eta_0=\vrVispEtaNull$) and \vrDecTruncLoFaS\,\% on the Maiun test area, compared with \vrVispSimBelowFa\,\% and
\vrDecSimBelowFa\,\%, respectively, at the simulated distributed centers with $\eta<\eta_0$ (TSX-44 and S1-79).

The fringe density on the slow-deformation slope is measured as in Section~\ref{sec:simdesign}, as the phase gradient
times the window diameter over $2\pi$, in the normalized coordinates of the elliptic search window of
$25\times225$ pixels. The gradients are the angles of the means of the complex products of neighbors over
$9\times81$ pixels in azimuth and in range. The estimate of FaSHPS\WSsuffix{} for the interval of Fig.~\ref{fig:s1slow} has at least one fringe per window
diameter at \vfSlowBandN{} centers with $\eta\ge0.2$. At \vfSlowBandShareOne\,\% of them, the single look of the same
interval averaged over $3\times27$ pixels, which does not depend on \WSname, has at least one as well, against \vfSlowCropShareOne\,\% of all centers of the test area with $\eta\ge0.2$.

\subsection{Acceptance of the SHP tests on the real stacks}
\label{supp:accept}
FaSHPS and CMP accept fewer candidates on the real stacks than in the simulated scenes (Section~\ref{sec:realmaps}). Table~\ref{tab:accept} gives the acceptance, i.e., the share of the unmasked pixels of the search
window in the ensemble (of all its pixels in the simulations, which have no mask), by the PCI of the center. It also gives the second-pass
threshold $z_\alpha(\eta_p)$ of
FaSHPS, the spread of the temporal mean amplitudes over the search window and the contrast in mean intensity that
CMP tests. The spread is the robust standard deviation (1.4826 times the median absolute deviation) of $\bar A_q$
over the unmasked pixels of the search window outside the guard set, in units of $c_A\,\tilde A/\sqrt N$, the scale of $T_{pq}$ in
Eq.~\eqref{eq:fastat} with the median $\tilde A$ of $\bar A_q$ in place of $\mu_p$. Where the spread exceeds $z_\alpha$ and $\mu_p$ is close to $\tilde A$,
candidates at one standard deviation from $\tilde A$ fall outside the interval of the second pass. The contrast is
the median over the same pixels of $|10\log_{10}(\bar I_q/\bar I_{\mathrm r})|$, where $\bar I_{\mathrm r}$, the
reference of the second pass of CMP, is the mean intensity of its reference set.

\begin{table}[tbp]
\centering\footnotesize\renewcommand{\arraystretch}{0.9}
\begin{tabular}{llcccc}
\toprule
 & & \multicolumn{4}{c}{PCI $\eta_p$ of the center}\\
\cmidrule(lr){3-6}
 & & $\eta_p<\eta_0$ & $\eta_0$ to 0.3 & 0.3 to 0.5 & $0.5\le\eta_p$\\
\midrule
\multicolumn{6}{l}{\textit{TSX-44} ($\eta_0=0.066$, $z_\alpha\ge\sqrt N/c_A=12.8$ from $\eta_p=0.29$)}\\
Centers [\%] & Visp & 61.2 & 32.2 & 6.1 & 0.6\\
 & Simulated & 16.4 & 42.7 & 18.1 & 22.8\\
$z_\alpha$ &  & 3.8 to 4.4 & 4.4 to 13 & 13 to 23 & 23 to 51\\
Spread & Visp & 4.6 & 7.6 & 8.1 & 8.3\\
 & Simulated, no texture & 1.3 & 2.2 & 3.3 & 5.7\\
 & Simulated, strong texture & 4.6 & 5.1 & 6.0 & 6.3\\
FaSHPS [\%] & Visp & 36 & 44 & 88 & 96\\
 & Simulated & 67 & 82 & 94 & 97\\
Contrast [dB] & Visp & 2.4 & 4.5 & 5.0 & 7.5\\
 & Simulated & 1.9 & 1.7 & 1.9 & 2.8\\
CMP [\%] & Visp & 20 & 16 & 29 & 35\\
 & Simulated & 59 & 71 & 84 & 93\\
\midrule
\multicolumn{6}{l}{\textit{S1-79} ($\eta_0=0.050$, $z_\alpha\ge\sqrt N/c_A=17.1$ from $\eta_p=0.31$)}\\
Centers [\%] & Slow deformation & 4.2 & 66.7 & 22.9 & 6.3\\
 & Maiun & 5.8 & 84.2 & 8.6 & 1.3\\
 & Long-term coherent & 3.2 & 57.4 & 34.0 & 5.4\\
 & Simulated & 12.3 & 48.5 & 17.6 & 21.6\\
$z_\alpha$ &  & 3.3 to 3.9 & 3.9 to 17 & 17 to 28 & 28 to 59\\
Spread & Slow deformation & 6.2 & 7.0 & 7.7 & 8.4\\
 & Maiun & 5.8 & 6.4 & 7.8 & 7.8\\
 & Long-term coherent & 5.0 & 7.8 & 9.6 & 8.7\\
 & Simulated, no texture & 1.4 & 2.5 & 4.0 & 8.8\\
 & Simulated, strong texture & 5.7 & 6.2 & 7.4 & 7.9\\
FaSHPS [\%] & Slow deformation & 30 & 61 & 90 & 96\\
 & Maiun & 29 & 58 & 87 & 95\\
 & Long-term coherent & 31 & 54 & 82 & 96\\
 & Simulated & 58 & 77 & 92 & 96\\
Contrast [dB] & Slow deformation & 2.1 & 2.9 & 2.9 & 3.2\\
 & Maiun & 2.2 & 2.5 & 3.0 & 3.2\\
 & Long-term coherent & 1.8 & 3.0 & 3.7 & 3.5\\
 & Simulated & 0.5 & 1.4 & 1.9 & 3.1\\
CMP [\%] & Slow deformation & 24 & 44 & 68 & 83\\
 & Maiun & 24 & 41 & 61 & 77\\
 & Long-term coherent & 26 & 36 & 53 & 73\\
 & Simulated & 53 & 68 & 79 & 92\\
\midrule
\multicolumn{6}{l}{\textit{S1-20} ($\eta_0=0.100$, $z_\alpha\ge\sqrt N/c_A=8.6$ from $\eta_p=0.24$)}\\
Centers [\%] & Landslide substack & 10.3 & 65.2 & 22.9 & 1.7\\
 & Simulated & 17.7 & 37.5 & 20.7 & 24.1\\
$z_\alpha$ &  & 3.7 to 4.4 & 4.4 to 9.9 & 9.9 to 15 & 15 to 37\\
Spread & Landslide substack & 3.5 & 3.5 & 3.5 & 3.5\\
 & Simulated, no texture & 1.3 & 1.9 & 2.7 & 3.9\\
 & Simulated, strong texture & 3.4 & 3.5 & 3.9 & 4.4\\
FaSHPS [\%] & Landslide substack & 53 & 79 & 93 & 99\\
 & Simulated & 79 & 87 & 93 & 96\\
Contrast [dB] & Landslide substack & 2.7 & 2.5 & 2.6 & 2.7\\
 & Simulated & 1.0 & 1.5 & 1.9 & 3.0\\
CMP [\%] & Landslide substack & 40 & 65 & 79 & 91\\
 & Simulated & 72 & 78 & 83 & 92\\
\bottomrule
\end{tabular}
\caption{Acceptance of FaSHPS and CMP by the PCI $\eta_p$ of the center on the real stacks and in the simulated scenes of their configuration, with the share of the centers in each bin, the second-pass threshold $z_\alpha$ of FaSHPS at the edges of the bin and the spread and the contrast. The line of each configuration gives $\eta_0$ and the PCI from which $z_\alpha\ge\sqrt N/c_A$. For each center, the spread is the robust standard deviation of the temporal mean amplitudes over the unmasked pixels of its search
window outside the guard set, relative to the scale of the FaSHPS statistic, and the contrast the median absolute difference in mean intensity, in dB, between the pixels of its search window and the reference set of CMP (Section~\ref{supp:accept}). The table gives the median of each over \vpaPerBin{} centers per bin of a real scene and over the centers of \vpaScenes{} evaluation scenes of a configuration, for the spread over the distributed centers of the texture class of the row. Shares and acceptance are taken over all centers, and every value of the simulated scenes is weighted as in Section~\ref{sec:simdesign}.}
\label{tab:accept}
\end{table}

Most Visp centers have a PCI below 0.3 (\vpaBelowVisp\,\%,
against \vpaBelowTSX\,\% in the simulated scenes of TSX-44), and there the mean amplitudes of Visp spread as much as in
simulated clutter with strong texture or more. FaSHPS accepts \vaccFaVisp\,\% of the unmasked pixels of the search window on Visp. Weighted
with the shares of the PCI bins of Visp, the acceptance in the simulated scenes is \vpaReweightTSX\,\% instead of
\vpaFaTSX\,\%, so that most of the difference arises within the bins. Above a PCI of \vpaOpenSShort{} to
\vpaOpenSLong, depending on the configuration, $z_\alpha$ exceeds $\sqrt N/c_A$. The interval
$\mu_p(1\pm z_\alpha c_A/\sqrt N)$ of the second pass then includes every mean amplitude between zero and $\mu_p$, and
FaSHPS rejects only candidates with $\bar A_q>\mu_p(1+z_\alpha c_A/\sqrt N)\ge2\mu_p$.

By Eq.~\eqref{eq:cmpstat}, $D_{pq}+8/\pi$ is $4/\pi$ times the product of an intensity term
$(\bar I_p^2+\bar I_q^2)/(\bar I_p\bar I_q)$, which is at least two, and a phase term, the mean over the
interferograms, which is at least one. In the second pass, $\bar I_{\mathrm r}$ and the phases of the multilooked interferograms of the reference set take the place of $\bar I_p$ and the phases of the center. The phase term is at most
the mean of $(1+\gamma_{\mathrm a}^2(n))/(1-\gamma_{\mathrm a}^2(n))$ over the interferograms, and with this bound,
on each stack and in each configuration, the statistic for equal intensities stays below the smallest first-pass
threshold of its own table. On its own, the phase term thus cannot reject a candidate in the first pass, and in the
second pass it rejects fewer than \vpaPhaseRejBound\,\% of the candidates in the sampled windows. The intensity term
on its own, with equal phases, accounts for \vpaIntShareReal\,\% of the rejections of the second pass on the real stacks
and \vpaIntShareSim\,\% in the simulated scenes, and the rest needs both terms. On Visp, the contrast rises with the
PCI to a median of \vpaConVisp\,dB above 0.3, against \vpaConTSX\,dB in the simulated scenes of TSX-44, and CMP
accepts \vpaCmpHiVisp\,\% of the unmasked pixels of the search window at these centers.

On the Sentinel-1 test areas, \vpaBelowSOne\,\% of the centers have a PCI below 0.3, against \vpaBelowSLong\,\%
(S1-79) and \vpaBelowSShort\,\% (S1-20) in the simulated scenes. Below 0.3, the mean amplitudes spread about as much
as in simulated clutter with strong texture or more. In each of the two bins below 0.3, FaSHPS accepts
\vpaFaGapLowSOne{} percentage points less than in the matching simulated configuration, whereas above 0.3 the two
differ by at most \vpaFaGapHiSOneMax{} points. With the bins weighted by their shares of the stack, at
least \vpaFaFracLowMin\,\% of the difference within the bins arises below 0.3 on every stack.

\subsection{Runtime}
\label{supp:runtime}
This section describes the runtime measurement of Section~\ref{sec:runtime} and gives the time of each step (Table~\ref{tab:steps}).
All timed methods are written as single-threaded C routines.
The routines are compiled with GCC 11.4 at optimization level -O3 for the processor of
the test machine, an \vRtCpu{} (-march=native), which enables automatic vectorization of the inner
loops, and with -ffast-math without the finite-math assumption. Only the score of the truncation step is compiled without
fast-math, so that its values are bitwise identical to those of the simulations.
The timed center pixels lie on a regular grid in several regions of each
stack, without the masked pixels. At
every center pixel the methods are run one after the other in a random
order, and this is repeated in five passes, so that changes of the
machine state during the measurement affect all methods alike. The NSPC map of the mask is timed per image
pixel over the same regions. We report
times relative to the full window of the same scene, because absolute
times depend on the machine. The full window computes the covariance
over the ensemble of each center directly. Running sums would make it
cheaper, but they are not possible for adaptive ensembles. The same measurement, with the same code, ran on an
\vRtCpuB{} and an \vRtCpuC{} (Section~\ref{sec:runtime}). Two complete runs of the measurement agree to
within \vRtRunDiff{} in every ratio, and to within \vRtRunDiffNext{} except for the full window with the truncation step on
Visp. Every comparison of Section~\ref{sec:runtime} between a method with \WSname{} and the same SHP selection alone, the same SHP selection with the truncation step
alone, or the full window has the same sign in both runs.

Table~\ref{tab:runtime} gives the total time of each method in this measurement divided by the number of centers.
The time of each step comes from a separate measurement, and Table~\ref{tab:steps} gives its medians over 150 center
pixels per stack, the best of three runs each, so that the steps need not add up to the times of
Table~\ref{tab:runtime}.

\begin{table}[h]
\centering\small
\begin{tabular}{lccc}
\toprule
 & TerraSAR-X & Sentinel-1 & Sentinel-1\\
 & $N=44$, 8021 px & $N=79$, 4437 px & $N=20$, 4437 px\\
\midrule
Sample covariance, full window & 1.00 & 1.00 & 1.00 \\
Sample covariance, quarter of the window & 0.26 & 0.25 & 0.27 \\
SPC with every pixel of the search window & 0.041 & 0.028 & 0.075 \\
PCI $\eta$ (per image pixel) & 0.0010 & 0.0008 & 0.0055 \\
FaSHPS statistic & 0.005 & 0.002 & 0.021 \\
FaSHPS preparation (per image pixel) & 0.005 & 0.011 & 0.012 \\
CMP & 0.20 & 0.12 & 0.42 \\
Connected component & 0.016 & 0.012 & 0.109 \\
Ordering & 0.11 & 0.11 & 0.23 \\
Truncation step & 0.49 & 0.32 & 1.01 \\
NSPC (per image pixel) & $7\times10^{-5}$ & $8\times10^{-5}$ & $2\times10^{-4}$ \\
\bottomrule
\end{tabular}
\caption{Time of the individual steps for one center pixel, or for one image pixel where marked, relative to one sample covariance over the whole search window, for Visp, the slow-deformation test area and the
landslide substack (first row, whose absolute time on the test machine, one core of an \vRtCpu, was 4.92, 7.35 and
0.79\,ms, respectively). The second row, with a quarter of the pixels, shows that the time of the sample covariance is
proportional to the ensemble size. FaSHPS preparation is the temporal mean amplitude of each pixel after the removal
of outlying amplitudes (Eq.~\eqref{eq:fastat}). All steps are single-threaded C loops compiled with the same settings and checked against the simulation code. The truncation step comprises the multilooked interferograms of the score, its evaluation at the candidate sizes, the jackknife noise and the decision. Median over 150 center pixels per stack, best of three runs each. The NSPC row comes from this separate measurement and differs slightly from the value in Table~\ref{tab:runtime} of the main text.}
\label{tab:steps}
\end{table}

Timed step by step (Table~\ref{tab:steps}), computing
$S_{pq}$ for all pixels $q$ of the search window costs \vRtStepSpcMin{} to \vRtStepSpcMax\,\% of one sample
covariance over the full window, of the order of the ratio $2/(N+1)$ of the numbers of multiplications and additions per pixel. Ordering and truncation step
together cost \vRtStepOrdSizMin{} to \vRtStepOrdSizMax{} times as much as one sample covariance, depending on the
configuration, and the PCI \vRtStepEtaMin{} to \vRtStepEtaMax{} of one sample covariance per image pixel.
The part of the time of the truncation step (Section~\ref{sec:adaptive}) that grows with $R$, i.e., the score
and its noise, is estimated from the same step timed with $R=1$, 2, 4, 8 and 16 reference acquisitions on the same
150 center pixels per stack. The time of the step is linear in $R$ to within \vRtRefsLinDev\,\%. The share of
this part at $R=8$ is eight times the slope of that line divided by the time at $R=8$. It is \vRtScoreShareMin{} to \vRtScoreShareMax\,\% on the test machine (Visp, slow-deformation test area and landslide
substack), so that most of the time of the truncation step is proportional to $R$.

The truncation step costs relatively more on Visp than on the Sentinel-1 test areas
(Table~\ref{tab:runtime}). The cost of its score grows with
$RN$ per member and that of the coherence matrix with $N^2$, so that the step costs relatively more with the 44
acquisitions of Visp than with 79 (Table~\ref{tab:steps}). On Visp the ensembles are also less often truncated than
on the Sentinel-1 test areas, so that less of the coherence matrix is saved.

Relative to the processing with the SHP selection alone, \WSname{} adds
at most \vRtWSMaiunMax{} times the processing time of the full window on Visp and the Maiun test area and saves time on the
slow-deformation and long-term coherent test areas, for the full window, FaSHPS and CMP (Table~\ref{tab:runtime}). At
the timed center pixels that pass the gate, the PCI-scaled window holds on average \vRtScaledVisp\,\% (Visp) and \vRtScaledSOneMin{} to
\vRtScaledSOneMax\,\% (Sentinel-1) of the pixels of the search window, which reduces the time of the SHP selection,
the truncation step and the coherence matrix.

\subsection{Measures without ground truth}
\label{supp:noquant}
This section gives the protocol and the remaining numbers of Section~\ref{sec:noquant}. Both measures, the
split-stack agreement and the predicted phase standard deviation, were evaluated on the first \vcNqScenesPerSeed{}
random scenes of each batch of evaluation scenes and on control scenes. The control scenes are homogeneous clutter with a planar phase ramp of at most one fringe
per window diameter, \vcNqControls{} per configuration with the exponential coherence model
(Eq.~\eqref{eq:cohfamily}) and as many with a coherence equal for every pair. In the controls with
the exponential model, the split-stack agreement prefers CMP\WSsuffix{} to the full window with \WSname{} in
\vcNqCtrlScoreMin{} to \vcNqCtrlScoreMax\,\% of the scenes, whereas the phase linking error of the split-stack
history does so in \vcNqCtrlTrueMin{} to \vcNqCtrlTrueMax\,\%. In the controls with a coherence equal for every
pair, the two shares are \vcNqEqScoreMin{} to \vcNqEqScoreMax\,\% and \vcNqEqTrueMin{} to \vcNqEqTrueMax\,\%, respectively. Where neither SHP selection should be preferred, neither
measure thus gives a clear preference. Between an SHP
selection alone and the same SHP selection with \WSname, the split-stack agreement ranks as the phase linking
error of the split-stack history does in \vcNqHeldWSDuoMin{} to \vcNqHeldWSDuoMax\,\% of the random scenes in which
these errors differ by more than 5\,\% (TSX-44 and S1-79, Section~\ref{sec:noquant}). In the random
scenes where the split-stack agreement of an SHP selection alone and of the same SHP selection with \WSname{} differs by more
than 5\,\%, the measure orders the two as the phase linking error of the split-stack history does in
\vhSimBigConcMin{} to \vhSimBigConcMax\,\% of the scenes (three configurations). Here the scenes are selected by their difference in agreement, and in Section~\ref{sec:noquant} by their difference in
phase linking error. The split-stack agreement prefers \WSname{} more often than the phase linking error of the
split-stack history does
\vhSimMorePhrase{} pairs of SHP selection and configuration, and in \vhSimScoreSTMin{} to \vhSimScoreSTMax\,\% of the
scenes against \vhSimTrueSTMin{} to \vhSimTrueSTMax\,\% over the nine pairs.

The \vcNqRealN{} regions of the real stacks are four on Visp and three on each of the Maiun, long-term coherent,
slow-deformation and fast-deformation test areas of Sentinel-1 with 79 acquisitions. There the two measures repeat the
tendencies of the simulation. The split-stack agreement ranks an SHP selection with \WSname{} first in \vcNqRealHeldWS{}
regions, and the predicted standard deviation ranks the full window alone first in all of them. Between an SHP selection alone and the same SHP selection with \WSname, the split-stack agreement is higher with \WSname{} in \vhRealST{} of the
\vhRealPairs{} pairs of SHP selection and region.

\section{Further methods and options}
\label{supp:options}
\label{app:options}
This section reports further methods and options run under the protocol of Section~\ref{sec:simulation}, to which Sections~\ref{sec:selection} and~\ref{sec:discussion} refer. In this section, a value or a range refers to all centers of the evaluation scenes, the full window,
FaSHPS and CMP, and the three configurations TSX-44, S1-79 and S1-20, unless the sentence or the parentheses after it
name others. A range gives the smallest and the largest value over them.

\paragraph{Reference acquisitions of the score}
Table~\ref{tab:options} gives the error of FaSHPS\WSsuffix{} with other
sets of reference acquisitions in the score (Eq.~\eqref{eq:score}). More references
add pairs to the score and lower the error slightly, at a proportionally higher cost. In homogeneous clutter of
coherence 0.5, $\sigma_{\Delta a}$ with 16 references is \vcJackRefSixMin{} to \vcJackRefSixMax{} times that with eight,
and with all acquisitions \vcJackRefAllMin{} to \vcJackRefAllMax{} times, so that the margin $\zeta$ falls with it.
These ratios are medians over the centers of the scenes of the paragraph on the noise of the gain below
(three configurations). The error over all centers falls by at most \vOptRAllAllMax\,\%. The
cost of the score, however, is proportional to $|\mathcal I|$, which grows from $8N-36$ to $16N-136$ and $N(N-1)/2$.
With a single reference, the first or the middle acquisition, leaving out this reference in the jackknife
(Eq.~\eqref{eq:jackknife}) leaves no pair, and its value $\Delta a_{(s)}$ is taken as $\Delta a$. The error over all
centers then rises by \vOptSingleAllMin{} to \vOptSingleAllMax\,\%, and the outcome depends on
the choice of the reference. The two single references return the same ensemble at \vOptSingleSameSizedMin{} to
\vOptSingleSameSizedMax\,\% of the centers at which the gate applies the truncation step.

\begin{table}[tbp]
\centering\footnotesize
\setlength{\tabcolsep}{3.5pt}
\begin{tabular}{lcccccc}
\toprule
Option & \multicolumn{2}{c}{TSX-44} & \multicolumn{2}{c}{S1-79} & \multicolumn{2}{c}{S1-20} \\
 & all & $\ge2$ fringes & all & $\ge2$ fringes & all & $\ge2$ fringes \\
\midrule
8 references (default) & 1.31 & 1.48 & 1.23 & 1.40 & 1.27 & 1.49 \\
16 references & 1.30 & 1.46 & 1.22 & 1.40 & 1.26 & 1.48 \\
all acquisitions & 1.29 & 1.45 & 1.22 & 1.38 & 1.26 & 1.48 \\
first acquisition only & 1.42 & 1.67 & 1.27 & 1.51 & 1.34 & 1.80 \\
middle acquisition only & 1.42 & 1.66 & 1.28 & 1.49 & 1.32 & 1.66 \\
fringe frequency compensation, safeguarded & 1.31 & 1.46 & 1.23 & 1.37 & 1.26 & 1.42 \\
fringe frequency compensation, plain & 1.99 & 1.87 & 2.08 & 1.85 & 1.61 & 1.57 \\
\bottomrule
\end{tabular}
\caption{Options of the truncation step within the PCI-scaled window and the gate, applied after FaSHPS: phase linking error relative to the oracle, weighted geometric mean over all centers of the evaluation scenes and over the distributed centers with two or more fringes per window diameter.}
\label{tab:options}
\end{table}

\paragraph{Fringe frequency compensation}
The truncation step reduces bias at the expense of looks. Where fringes are
dense and coherence is low, both a low bias and many looks are required. A known remedy for single interferograms is to estimate the local fringe frequency and to
subtract, before averaging, the local fringe pattern as a phase plane that is zero at the center,
as in the slope-compensated filter of \citet{trouve1998improving}. We examined a
simple version with FaSHPS and the PCI-scaled window, applied at the centers that pass the gate. For each interferogram of two consecutive
acquisitions, a local wavenumber is estimated along each image axis. To this end, the interferogram is multiplied by
its conjugate, shifted along the axis by more than the speckle correlation length. The product is averaged over a box
and over five consecutive interferograms, and the phase of this average, divided by the shift, is the local
wavenumber. Where the magnitude of the average lies below a threshold, so that its phase is mostly noise, the
wavenumber is set to zero. Summed over the interferograms from the first acquisition, the wavenumbers give the local
phase gradient $\bm k_p(n)$ of acquisition $n$ at center $p$, relative to the first acquisition. The data in the
PCI-scaled window of $p$ is multiplied by $\exp\{-\ii\,\bm k_p(n)\cdot(q-p)\}$, which leaves the phase of the center
unchanged. Since an
inaccurate ramp adds fringes instead of removing them, the SHP selection and the truncation step are applied to the
compensated and to the uncompensated window, and the compensated ensemble is retained only if its score
(Eq.~\eqref{eq:score}) exceeds that of the uncompensated one by more than $0.5/N$.

With this safeguard, the compensated ensemble
was kept at \vOptFcKeptAllMin{} to \vOptFcKeptAllMax\,\% of all centers. Relative to FaSHPS\WSsuffix{} without
compensation, the error over all centers changed by \vOptFcAllMin{} to \vOptFcAllMax\,\% and that at two or more
fringes per window diameter by \vOptFcFrMin{} to \vOptFcFrMax\,\% (Table~\ref{tab:options}, three configurations).
Without the safeguard, the error over all centers was \vOptPlainAllMin{} to \vOptPlainAllMax{} times that without
compensation. This ratio was largest at $c=\vOptPlainPeakAt$, at \vOptPlainPeakMin{} to \vOptPlainPeakMax{}, where the
magnitude of the average lies above the threshold but its phase is still noisy, and the sum of the wavenumbers over the stack
drifts. Since the compensation with the safeguard changes the error over all centers by less than 1\,\% and depends
on the safeguard, it is not used.

\paragraph{Local polynomial fit}
The local polynomial approximation with the intersection of confidence
intervals \citep{katkovnik2008phase} fits a phase plane, a phase and two
slopes, to the phasors of one interferogram over nested windows and keeps the largest window up to which the confidence intervals of all nested windows have a
point in common. It returns a phase per interferogram and no ensemble. We
applied it to the $N(N-1)/2$ interferograms of every center of the first batch of evaluation scenes and
linked the resulting matrix with the same phase linking (Table~\ref{tab:lpaici}). The window fractions ran from 0.1
to 1 of the search window, and the threshold factor of the confidence intervals was fixed at 2 without tuning. On the same centers, its error
was higher than those of FaSHPS\WSsuffix{} and FaSHPS+T over all centers, in each coherence class of
Table~\ref{tab:lpaici} and at two or more fringes per window diameter, in the three configurations. Over all centers it
was \vaLpaAllTSX{} to \vaLpaAllSLong{} times the oracle error, against \vaLpaFaWSSLong{} to \vaLpaFaWSTSX{} for
FaSHPS\WSsuffix, and it was highest at $c=0.2$, at \vaLpaLowMin{} to \vaLpaLowMax{} times the oracle error. Its excess
over FaSHPS+T was smallest at $c=0.95$ in TSX-44, \vaLpaHighAboveTSX\,\%, so that it has more error than FaSHPS+T in
every coherence class.

\begin{table}[tbp]
\centering\footnotesize
\setlength{\tabcolsep}{3.5pt}
\begin{tabular}{lcccccc}
\toprule
Method & all & $c=0.2$ & $c=0.5$ & $c=0.8$ & $c=0.95$ & $\ge2$ fr. \\
\midrule
\multicolumn{7}{l}{\emph{TSX-44}} \\
Local polynomial fit, ICI & 1.77 & 3.05 & 1.53 & 1.58 & 1.54 & 1.86 \\
Full window & 1.93 & 1.60 & 1.98 & 2.50 & 3.34 & 5.30 \\
Full window + T & 1.38 & 1.46 & 1.37 & 1.40 & 1.47 & 1.56 \\
Full window\WSsuffix & 1.34 & 1.48 & 1.35 & 1.33 & 1.36 & 1.51 \\
FaSHPS & 1.90 & 1.53 & 1.93 & 2.45 & 3.26 & 5.28 \\
FaSHPS + T & 1.35 & 1.39 & 1.34 & 1.37 & 1.42 & 1.56 \\
FaSHPS\WSsuffix & 1.32 & 1.41 & 1.32 & 1.31 & 1.33 & 1.51 \\
\midrule
\multicolumn{7}{l}{\emph{S1-79}} \\
Local polynomial fit, ICI & 1.92 & 4.04 & 1.56 & 1.48 & 1.52 & 1.70 \\
Full window & 1.68 & 1.33 & 1.62 & 2.06 & 2.61 & 3.36 \\
Full window + T & 1.28 & 1.29 & 1.28 & 1.28 & 1.35 & 1.44 \\
Full window\WSsuffix & 1.28 & 1.30 & 1.27 & 1.29 & 1.32 & 1.42 \\
FaSHPS & 1.60 & 1.22 & 1.45 & 1.96 & 2.52 & 3.28 \\
FaSHPS + T & 1.23 & 1.20 & 1.14 & 1.21 & 1.31 & 1.40 \\
FaSHPS\WSsuffix & 1.22 & 1.20 & 1.14 & 1.22 & 1.28 & 1.38 \\
\midrule
\multicolumn{7}{l}{\emph{S1-20}} \\
Local polynomial fit, ICI & 1.88 & 3.57 & 1.63 & 1.58 & 1.66 & 1.85 \\
Full window & 1.80 & 1.45 & 1.81 & 2.22 & 2.83 & 4.32 \\
Full window + T & 1.35 & 1.35 & 1.34 & 1.38 & 1.49 & 1.55 \\
Full window\WSsuffix & 1.34 & 1.39 & 1.34 & 1.36 & 1.40 & 1.54 \\
FaSHPS & 1.70 & 1.30 & 1.68 & 2.12 & 2.72 & 4.16 \\
FaSHPS + T & 1.28 & 1.21 & 1.26 & 1.30 & 1.43 & 1.50 \\
FaSHPS\WSsuffix & 1.26 & 1.23 & 1.25 & 1.28 & 1.35 & 1.49 \\
\bottomrule
\end{tabular}
\caption{Local polynomial fit with the intersection of confidence intervals, applied to every interferogram, compared with the full window and FaSHPS alone, with the truncation step (T) and with the PCI-scaled window and the gated truncation step (\WSname): phase linking error relative to the oracle, weighted geometric mean over the centers of the first batch of evaluation scenes (all centers, distributed centers by coherence, and distributed centers with two or more fringes per window diameter).}
\label{tab:lpaici}
\end{table}

\paragraph{Variants of the truncation step and of FaSHPS}
Table~\ref{tab:variants7} gives the error of the full window, FaSHPS and CMP with variants of \WSname, relative to the same SHP selection with \WSname{} at the selected setting, on the centers of the evaluation scenes. The
noise multiplier $\lambda$ limits truncation where the core carries almost no information about the true phase
(Section~\ref{sec:adaptive}). Where such a core happens to disagree with the true phase, smaller subsets reach a
higher score in the truncation step (Eq.~\eqref{eq:score}), because their noisier multilooked phases bring the mean
cosine closer to zero. A truncation there changes the phase linking error little, since the phases are noise either
way. With $\lambda=0$, 1, 2, 4 and 5 and the selected $\kappa$, the error over all centers changes by at most
\vxLamChangeMax\,\%, and on the tuning scenes the criterion changes by less than \vtTie\,\% for every value of
$\lambda$ in Table~\ref{tab:tuninggrid}. Ordering the members by their SPC with the center instead of the ellipse
order raises the error to \vxSpcOrderMin{} to \vxSpcOrderMax{} times that at the selected setting, and growing the
ensemble in this order up to the candidate size of largest Fisher information raises it to \vxFisherMin{} to
\vxFisherMax{} times. The variant of FaSHPS\WSsuffix{} of Section~\ref{sec:discussion} that also accepts
candidates by their SPC has \vxSpcBranchMin{} to \vxSpcBranchMax{} times the error of FaSHPS\WSsuffix.

\begin{table*}[tbp]
\centering\footnotesize
\setlength{\tabcolsep}{3pt}
\begin{tabular}{lccccccccc}
\toprule
 & \multicolumn{3}{c}{TSX-44} & \multicolumn{3}{c}{S1-79} & \multicolumn{3}{c}{S1-20} \\
\cmidrule(lr){2-4}\cmidrule(lr){5-7}\cmidrule(lr){8-10}
 & Full & FaSHPS & CMP & Full & FaSHPS & CMP & Full & FaSHPS & CMP \\
\midrule
$\lambda=0$ & 1.000 & 1.001 & 1.000 & 1.000 & 1.000 & 1.000 & 1.000 & 1.000 & 1.000 \\
$\lambda=1$ & 1.000 & 1.000 & 1.000 & 1.000 & 1.000 & 1.000 & 1.000 & 1.000 & 1.000 \\
$\lambda=2$ & 1.000 & 1.000 & 1.000 & 1.000 & 1.000 & 1.000 & 1.000 & 1.000 & 1.000 \\
$\lambda=4$ & 1.000 & 1.000 & 1.000 & 1.000 & 1.000 & 1.000 & 1.001 & 1.001 & 1.001 \\
$\lambda=5$ & 1.001 & 1.001 & 1.001 & 1.001 & 1.000 & 1.000 & 1.003 & 1.003 & 1.003 \\
\addlinespace[2pt]
SPC order of the members & 1.476 & 1.517 & 1.497 & 1.229 & 1.260 & 1.242 & 1.249 & 1.281 & 1.267 \\
Largest Fisher information & 1.272 & 1.281 & 1.270 & 1.150 & 1.143 & 1.136 & 1.215 & 1.229 & 1.220 \\
FaSHPS, also accepting by SPC & -- & 1.019 & -- & -- & 1.035 & -- & -- & 1.016 & -- \\
\bottomrule
\end{tabular}
\caption{Variants of the truncation step and of FaSHPS, each with the PCI-scaled window. Each entry is the phase linking error over all centers of the evaluation scenes relative to that of the same method with \WSname{} and the selected setting, as a ratio of the weighted geometric means on the same centers. The rows give the noise multiplier $\lambda=0$, 1, 2, 4 and 5 with the selected margin factor $\kappa$, the members ordered by their SPC with the center instead of the ellipse order, the ensemble grown in this SPC order up to the candidate size of largest Fisher information $|\mathcal E|\,\bar Q/(1-\bar Q)$, and the variant of FaSHPS of Section~\ref{sec:discussion} that also accepts candidates outside the guard set whose SPC with the center exceeds the 99th percentile of $S_{pq}$ in clutter with zero coherence. A dash marks a variant that was not run.}
\label{tab:variants7}
\end{table*}

\paragraph{Distribution tests}
Table~\ref{tab:disttests} gives the error with the Kolmogorov--Smirnov test of SqueeSAR \citep{ferretti2011new},
Kuiper's test \citep{kuiper1960tests}, the Anderson--Darling test \citep{scholz1987ksample} and the generalized
likelihood ratio test \citep{parizzi2011adaptive} in place of the amplitude test of FaSHPS, at a significance level of 1\,\%. Each test is used with a
threshold that depends on $\eta$, calibrated in homogeneous clutter like that of FaSHPS, and with the nominal threshold
for independent samples. With the threshold that depends on $\eta$, the error over all centers is \vxDistCalMin{} to
\vxDistCalMax{} times that of FaSHPS with the same variant, and with the threshold for independent samples
\vxDistIndMin{} to \vxDistIndMax{} times (alone, with the truncation step and with \WSname, three
configurations).

\begin{table*}[tbp]
\centering\footnotesize
\setlength{\tabcolsep}{3pt}
\begin{tabular}{lccccccccc}
\toprule
 & \multicolumn{3}{c}{TSX-44} & \multicolumn{3}{c}{S1-79} & \multicolumn{3}{c}{S1-20} \\
\cmidrule(lr){2-4}\cmidrule(lr){5-7}\cmidrule(lr){8-10}
 & alone & +T & \WSsuffix & alone & +T & \WSsuffix & alone & +T & \WSsuffix \\
\midrule
KS, threshold depending on $\eta$ & 0.996 & 1.001 & 1.002 & 1.003 & 1.007 & 1.008 & 1.010 & 1.011 & 1.011 \\
Kuiper, threshold depending on $\eta$ & 0.991 & 0.997 & 0.999 & 0.997 & 1.002 & 1.003 & 1.009 & 1.010 & 1.010 \\
AD, threshold depending on $\eta$ & 1.004 & 1.009 & 1.010 & 1.009 & 1.013 & 1.014 & 1.013 & 1.015 & 1.014 \\
GLRT, threshold depending on $\eta$ & 1.012 & 1.015 & 1.016 & 1.017 & 1.019 & 1.019 & 1.019 & 1.020 & 1.020 \\
\addlinespace[2pt]
KS, threshold for independent samples & 1.303 & 1.589 & 1.621 & 1.465 & 1.747 & 1.764 & 1.187 & 1.343 & 1.382 \\
Kuiper, threshold for independent samples & 1.202 & 1.433 & 1.461 & 1.359 & 1.597 & 1.615 & 1.105 & 1.215 & 1.250 \\
AD, threshold for independent samples & 1.352 & 1.667 & 1.702 & 1.592 & 1.929 & 1.941 & 1.211 & 1.380 & 1.418 \\
GLRT, threshold for independent samples & 1.217 & 1.411 & 1.442 & 1.467 & 1.749 & 1.781 & 1.122 & 1.195 & 1.232 \\
\bottomrule
\end{tabular}
\caption{Distribution tests in place of the amplitude test of FaSHPS, at a significance level of 1\,\%. Each entry is the phase linking error over all centers of the evaluation scenes relative to that of FaSHPS with the threshold that depends on $\eta$ and the same variant (alone, with the truncation step, and with \WSname), as a ratio of the weighted geometric means on the same centers. KS denotes the Kolmogorov--Smirnov test, AD the Anderson--Darling test and GLRT the generalized likelihood ratio test. The threshold that depends on $\eta$ is calibrated in homogeneous clutter like that of FaSHPS (Section~\ref{app:calibration}), and the threshold for independent samples is the nominal one.}
\label{tab:disttests}
\end{table*}

\paragraph{Fixed constants}
\label{supp:sweep}
Table~\ref{tab:sweep7} gives the error of the full window and FaSHPS with \WSname{} when one of its fixed constants (Table~\ref{tab:settings} and Section~\ref{sec:selection}) is changed, relative to the same method with \WSname{} at those values, on the centers of
the evaluation scenes. Over all
centers, the error changes by at most \vxSweepChangeMax\,\%, most
with $R=4$ reference acquisitions.  The
sweep also
covers the tuned bound $\rho_{\min}$. Relative to
$\rho_{\min}=\vtRhoMin$, the error with $\rho_{\min}=0.02$ changes by at most \vxRhoLoAllAbs\,\% over all centers
and is at most \vxRhoLoScatMax\,\% higher at point scatterers, and with $\rho_{\min}=0.03$ it is
\vxRhoHiAllMin{} to \vxRhoHiAllMax\,\% higher over all centers and \vxRhoHiScatMin{} to \vxRhoHiScatMax\,\% lower
at point scatterers, the trade-off of limitation (viii) of Section~\ref{sec:discussion}.

\begin{table}[tbp]
\centering\footnotesize
\setlength{\tabcolsep}{3pt}
\begin{tabular}{lcccccc}
\toprule
 & \multicolumn{2}{c}{TSX-44} & \multicolumn{2}{c}{S1-79} & \multicolumn{2}{c}{S1-20} \\
\cmidrule(lr){2-3}\cmidrule(lr){4-5}\cmidrule(lr){6-7}
 & Full & FaSHPS & Full & FaSHPS & Full & FaSHPS \\
\midrule
$n_{\min}=50$ & 1.001 & 1.002 & 1.003 & 1.004 & 1.003 & 1.003 \\
$n_{\min}=200$ & 0.999 & 0.999 & 0.998 & 0.997 & 0.998 & 0.998 \\
\addlinespace[2pt]
$R=4$ & 1.020 & 1.021 & 1.008 & 1.009 & 1.009 & 1.009 \\
\addlinespace[2pt]
Candidate sizes $\times1.25$ & 1.003 & 1.003 & 1.000 & 1.000 & 1.000 & 1.001 \\
Candidate sizes $\times2$ & 0.999 & 0.999 & 1.002 & 1.001 & 1.000 & 1.000 \\
\addlinespace[2pt]
Tolerance $\zeta/2$ & 1.001 & 1.001 & 1.002 & 1.001 & 1.002 & 1.002 \\
Tolerance $\zeta/6$ & 1.000 & 0.999 & 0.999 & 0.999 & 0.998 & 0.998 \\
\addlinespace[2pt]
$L_{\min}=12$ & 1.001 & 1.001 & 1.000 & 1.000 & 1.000 & 1.000 \\
$L_{\min}=50$ & 0.999 & 0.999 & 1.000 & 1.000 & 1.000 & 0.999 \\
\addlinespace[2pt]
Ring of 16 pixels & 1.002 & 1.001 & 1.000 & 1.000 & 1.000 & 1.001 \\
Ring of 40 pixels & 0.999 & 1.000 & 1.000 & 1.001 & 0.999 & 0.999 \\
\addlinespace[2pt]
Average of $\tilde\eta$ over $3\times3$ & 1.004 & 1.005 & 1.002 & 1.004 & 1.007 & 1.008 \\
Average of $\tilde\eta$ over $7\times7$ & 0.998 & 0.999 & 0.999 & 0.998 & 0.997 & 0.997 \\
\addlinespace[2pt]
$\rho_{\min}=0.02$ & 1.000 & 1.000 & 1.001 & 1.001 & 0.999 & 0.999 \\
$\rho_{\min}=0.03$ & 1.008 & 1.008 & 1.006 & 1.006 & 1.009 & 1.009 \\
\bottomrule
\end{tabular}
\caption{Sweep of the fixed constants of \WSname{} (Table~\ref{tab:settings} and Section~\ref{sec:selection}) and of the tuned bound $\rho_{\min}$, one at a time. Table~\ref{tab:options} compares further sets of reference acquisitions for FaSHPS, and Table~\ref{tab:tuninggrid} gives the tuning of $\rho_{\min}$ on the tuning scenes. Phase linking error over all centers of the evaluation scenes relative to that of the same method with \WSname{} at the values of the main comparison, ratio of the weighted geometric means on the same centers. Each row replaces one of the values of the main comparison, namely $n_{\min}=100$, $R=8$, the ratio of 1.5 between consecutive candidate sizes, the tolerance $\zeta/3$, $L_{\min}=25$, the ring of 24 pixels, the average of $\tilde\eta$ over $5\times5$ pixels and $\rho_{\min}=\vtRhoMin$. The ring, the average of $\tilde\eta$ and $\rho_{\min}$ change the PCI, and with them $\eta_0$ and the thresholds of FaSHPS are calibrated again.}
\label{tab:sweep7}
\end{table}

\paragraph{Bias of the score}
\label{supp:scorebias}
If the core pixels share the true phase history $\theta_p$ of the center, the CCG model (Section~\ref{sec:expectation})
gives $\E[\bar v_{\mathcal C}(n,r)]=\bar g_{\mathcal C}(n,r)\,e^{\ii(\theta_p(n)-\theta_p(r))}$ for the multilooked
interferogram of the core (Eq.~\eqref{eq:multilook}), where $\bar g_{\mathcal C}(n,r)$ is the mean of
$g(|\gamma^{p'}_{nr}|)$ over the core pixels $p'$. If their speckle is also independent of that of the scored pixels,
which the guard set provides, the expectation of the score (Eq.~\eqref{eq:score}) over the speckle of the core, with
$\mathcal E_k$ fixed, is therefore
\begin{equation}\label{eq:scoreexp}
 \E[a_k\mid\mathcal E_k]=\frac1{|\mathcal I|}\sum_{(n,r)\in\mathcal I}
 \bar g_{\mathcal C}(n,r)\cos\varepsilon_k(n,r),
\end{equation}
where $\varepsilon_k(n,r)=\arg\hat e_k(n,r)-\theta_p(n)+\theta_p(r)$. The score is thus a mean agreement, weighted with the coherence,
of the multilooked phases with the true phase of the center, in the form of Eq.~\eqref{eq:EQccg}. Since the
core contains the center, pixels chosen by their agreement with the phases of the center could, however, score higher
than their true phase justifies. CMP chooses them in its first pass by the single-look phase difference to the
center, and the ellipse order uses the SPC of the center only through the shape of a moment matrix.
Table~\ref{tab:scorebias} compares the score of the truncation step (Eq.~\eqref{eq:score}) with its expectation over
the speckle of the core (Eq.~\eqref{eq:scoreexp}), computed from the true phase and coherence of every core pixel
with the ensemble fixed. It uses the ensembles of the full window, FaSHPS and CMP in the search window without the
PCI-scaled window, in the ellipse order and in the distance order, on the first \vxScoreScenes{} scenes of each
batch of evaluation scenes, at the centers without a point scatterer in the core. Averaged over the candidate sizes, the
difference is \vxScoreBiasMin{} to \vxScoreBiasMax, against mean expected scores of \vxScoreMeanMin{} to
\vxScoreMeanMax{} in the ellipse order, and within each coherence class the difference for FaSHPS or CMP minus that
for the full window is \vxScoreSelDiffMin{} to \vxScoreSelDiffMax{}. With the expected
scores in place of the observed ones, the
truncation changes at up to \vxScoreChangedMax\,\% of these centers, for the full window as for FaSHPS and CMP. Since the full window selects no member by its phase, this share reflects the noise of the core. In the
simulations, the pairwise SHP selections and the ellipse order thus add no positive bias.

\begin{table*}[tbp]
\centering\footnotesize
\setlength{\tabcolsep}{3pt}
\begin{tabular}{lccccccccc}
\toprule
 & \multicolumn{3}{c}{TSX-44} & \multicolumn{3}{c}{S1-79} & \multicolumn{3}{c}{S1-20} \\
\cmidrule(lr){2-4}\cmidrule(lr){5-7}\cmidrule(lr){8-10}
 & $b$ & chg.\,\% & tr.\,\% & $b$ & chg.\,\% & tr.\,\% & $b$ & chg.\,\% & tr.\,\% \\
\midrule
Full, ellipse order & 0.000 & 22.1 & 60.6 & +0.002 & 17.6 & 52.2 & 0.000 & 19.6 & 59.1 \\
Full, distance order & 0.000 & 22.1 & 60.1 & +0.002 & 17.8 & 51.7 & 0.000 & 19.7 & 58.6 \\
FaSHPS, ellipse order & 0.000 & 21.1 & 57.6 & +0.001 & 15.3 & 47.8 & $-$0.001 & 19.1 & 57.4 \\
FaSHPS, distance order & 0.000 & 21.1 & 57.2 & +0.001 & 15.1 & 47.2 & $-$0.001 & 19.1 & 56.7 \\
CMP, ellipse order & 0.000 & 20.9 & 57.1 & +0.001 & 15.4 & 47.6 & $-$0.001 & 19.3 & 57.2 \\
CMP, distance order & 0.000 & 20.9 & 56.9 & +0.001 & 15.5 & 47.1 & $-$0.001 & 19.3 & 56.6 \\
\bottomrule
\end{tabular}
\caption{Bias of the score of the truncation step on the first \vxScoreScenes{} scenes of each batch of evaluation scenes. Observed score (Eq.~\eqref{eq:score}) minus its expectation from the truth (Eq.~\eqref{eq:scoreexp}), averaged over the candidate sizes and as a weighted mean over the centers without a point scatterer in the core ($b$), the share of these centers whose truncation changes when the expected scores decide (chg.) and the share truncated with the observed scores (tr.), for the ensembles of the search window in the ellipse order and in the distance order.}
\label{tab:scorebias}
\end{table*}

\paragraph{Noise of the gain}
\label{supp:sigmadelta}
Independent terms would give the gain $\Delta a$ of Section~\ref{sec:adaptive} the standard deviation
$|\mathcal I|^{-1}\bigl(\sum_{(n,r)\in\mathcal I}\Delta y(n,r)^2\bigr)^{1/2}$. Where the core carries information about
the true phase, terms that share an acquisition are correlated, because a deviation of the phase of that acquisition
in the core or in the ensemble shifts all of them. The jackknife over acquisitions \citep[Eqs.~(1.1) to (1.3)]{efron1981jackknife} leaves
out each acquisition in turn. With $\mathcal I_s\subset\mathcal I$ the pairs that contain acquisition $s$, it gives
\begin{equation}\label{eq:jackknife}
 \Delta a_{(s)}=\frac{1}{|\mathcal I|-|\mathcal I_s|}\sum_{(n,r)\in\mathcal I\setminus\mathcal I_s}\Delta y(n,r),\qquad
 \sigma_{\Delta a}=\Bigl(\frac{N-1}{N}\sum_{s=1}^N\bigl(\Delta a_{(s)}-\overline{\Delta a}\bigr)^2\Bigr)^{1/2},
\end{equation}
with $\overline{\Delta a}$ the mean of the $N$ values $\Delta a_{(s)}$. The jackknife treats the acquisitions as exchangeable, although a reference acquisition
belongs to $N-1$ pairs and any other to $R$, and $\sigma_{\Delta a}$ rests on $N$ values, 20 in the shortest stack. The
following values are medians over the centers at which a smaller subset has the highest score, since elsewhere
$\Delta a=0$, among 120 random centers of one scene per configuration with the full window in the ellipse order as
the ensemble. In homogeneous clutter of coherence 0.2, where no truncation is
justified, the jackknife estimate $\sigma_{\Delta a}$ is
\vcJackLowMin{} to \vcJackLowMax{} times the standard deviation for
independent terms, \vcJackMidMin{} to \vcJackMidMax{} times at a
coherence of 0.5 and \vcJackHighMin{} to \vcJackHighMax{} times at 0.95. The terms are thus correlated, and the
standard deviation for independent terms would underestimate the noise of the gain. Table~\ref{tab:sigmadelta} compares $\sigma_{\Delta a}$ with the actual
noise of $\Delta a$, i.e., its standard deviation over independent centers in homogeneous clutter, for the smallest,
the sixth and the largest candidate size $L_{K-1}$ of the distance order. The root mean square of $\sigma_{\Delta a}$
is \vsdRatioZeroMin{} to \vsdRatioZeroMax{} times that standard deviation at $c=0$,
for every candidate size. At $c\ge0.5$ it is \vsdRatioSmallMin{} to \vsdRatioSmallMax{} times for the smallest
candidate, lower than for the larger candidates at the same coherence (\vsdRatioLargerMin{} to
\vsdRatioLargerMax, the three configurations and S1-79 on the native range grid, S1-79n). Since $\kappa$
is tuned, a common factor would not matter, but the factor varies with the coherence and the candidate size. The
margin $\kappa\sigma_{\Delta a}$ is thus wider than $\kappa$ times the noise at $c=0$, where the gate skips the
truncation step at most distributed centers, and narrower for the smallest candidate in coherent clutter.

\begin{table}[tbp]
\centering\footnotesize
\setlength{\tabcolsep}{4pt}
\begin{tabular}{llcccc}
\toprule
$c$ & Candidate size & TSX-44 & S1-79 & S1-20 & S1-79n \\
\midrule
0 & smallest & 1.46 & 1.49 & 1.45 & 1.47 \\
 & sixth & 1.47 & 1.51 & 1.40 & 1.50 \\
 & largest below the whole ensemble & 1.46 & 1.47 & 1.44 & 1.47 \\
\addlinespace[2pt]
0.2 & smallest & 0.97 & 1.03 & 1.01 & 1.06 \\
 & sixth & 1.03 & 1.06 & 1.13 & 1.09 \\
 & largest below the whole ensemble & 1.33 & 1.25 & 1.38 & 1.25 \\
\addlinespace[2pt]
0.5 & smallest & 0.67 & 0.60 & 0.71 & 0.65 \\
 & sixth & 0.98 & 0.79 & 0.97 & 0.92 \\
 & largest below the whole ensemble & 1.09 & 1.05 & 1.09 & 1.02 \\
\addlinespace[2pt]
0.8 & smallest & 0.71 & 0.60 & 0.77 & 0.75 \\
 & sixth & 0.91 & 0.89 & 0.92 & 0.95 \\
 & largest below the whole ensemble & 0.92 & 0.95 & 0.96 & 0.95 \\
\addlinespace[2pt]
0.95 & smallest & 0.77 & 0.70 & 0.80 & 0.78 \\
 & sixth & 0.87 & 0.88 & 0.95 & 0.90 \\
 & largest below the whole ensemble & 0.86 & 0.89 & 0.91 & 0.89 \\
\bottomrule
\end{tabular}
\caption{Jackknife noise of the gain in homogeneous clutter of coherence $c$: root mean square of $\sigma_{\Delta a}$ over the centers divided by the standard deviation of $\Delta a$ over the same centers, for three candidate sizes of the distance order. A calibrated estimate gives one. The three sizes have 25, 189 and 4865 members in TSX-44, 25, 189 and 2162 in S1-79 and S1-20, and 25, 189 and 961
in S1-79n, S1-79 on the native range grid (Table~\ref{tab:native}). Every center has its own search window, the windows do not overlap, and the scenes are independent (1,800 centers in TSX-44, 1,800 centers in S1-79, 1,800 centers in S1-20, 1,200 centers in S1-79n per coherence).}
\label{tab:sigmadelta}
\end{table}

\paragraph{Native Sentinel-1 sampling}
Table~\ref{tab:native} repeats the main comparison of S1-79 on the native range grid, with every second range sample
of the same scenes and with calibration scenes of its own (S1-79n). On the native grid, \WSname{} lowers the error about as much as on the oversampled grid of S1-79,
over all centers by \vnSTGainAllSNatSpan\,\% relative to the SHP selection alone and at one fringe or more per window
diameter by \vnSTGainFringeSNatSpan\,\% (three SHP selections).

\begin{table*}[tbp]
\centering\footnotesize
\setlength{\tabcolsep}{2.2pt}
\resizebox{\linewidth}{!}{%
\begin{tabular}{lc@{\hspace{8pt}}cccc@{\hspace{6pt}}cccc@{\hspace{6pt}}cccc}
\toprule
Centers & Oracle ($^\circ$) & Full & Full+S & Full+T & Full\WSsuffix & FaSHPS & FaSHPS+S & FaSHPS+T & FaSHPS\WSsuffix & CMP & CMP+S & CMP+T & CMP\WSsuffix \\
\midrule
\multicolumn{14}{l}{\emph{S1-79}} \\
$c=0$ & 68.0 & 1.16 & 1.16 & 1.17 & 1.17 & 1.32 & 1.32 & 1.33 & 1.33 & 1.36 & 1.36 & 1.36 & 1.36 \\
$c=0.2$ & 15.6 & 1.35 & 1.35 & 1.31 & 1.32 & 1.24 & 1.24 & 1.22 & 1.23 & 1.28 & 1.28 & 1.26 & 1.26 \\
$c=0.35$ & 11.3 & 1.43 & 1.41 & 1.29 & 1.29 & 1.28 & 1.25 & 1.16 & 1.16 & 1.29 & 1.27 & 1.18 & 1.18 \\
$c=0.5$ & 9.2 & 1.57 & 1.48 & 1.26 & 1.26 & 1.44 & 1.35 & 1.16 & 1.16 & 1.46 & 1.37 & 1.18 & 1.19 \\
$c=0.65$ & 7.7 & 1.79 & 1.52 & 1.26 & 1.26 & 1.68 & 1.42 & 1.18 & 1.18 & 1.69 & 1.45 & 1.21 & 1.21 \\
$c=0.8$ & 6.1 & 1.96 & 1.49 & 1.26 & 1.26 & 1.87 & 1.42 & 1.20 & 1.20 & 1.90 & 1.44 & 1.22 & 1.23 \\
$c=0.95$ & 3.7 & 2.55 & 1.48 & 1.35 & 1.37 & 2.46 & 1.43 & 1.31 & 1.32 & 2.50 & 1.45 & 1.33 & 1.34 \\
\addlinespace[2pt]
At least one fringe & 10.8 & 2.66 & 1.91 & 1.41 & 1.38 & 2.59 & 1.85 & 1.37 & 1.34 & 2.63 & 1.89 & 1.40 & 1.37 \\
Point scatterers & 14.7 & 3.36 & 3.14 & 2.64 & 2.61 & 2.66 & 2.50 & 2.22 & 2.19 & 1.22 & 1.19 & 1.19 & 1.18 \\
\addlinespace[2pt]
All & 8.8 & 1.64 & 1.41 & 1.28 & 1.28 & 1.57 & 1.35 & 1.22 & 1.23 & 1.59 & 1.37 & 1.25 & 1.25 \\
\midrule
\multicolumn{14}{l}{\emph{S1-79n}} \\
$c=0$ & 69.8 & 1.15 & 1.15 & 1.16 & 1.16 & 1.31 & 1.31 & 1.32 & 1.32 & 1.35 & 1.35 & 1.35 & 1.35 \\
$c=0.2$ & 15.6 & 1.33 & 1.33 & 1.30 & 1.30 & 1.26 & 1.26 & 1.25 & 1.25 & 1.30 & 1.30 & 1.28 & 1.28 \\
$c=0.35$ & 11.4 & 1.42 & 1.40 & 1.28 & 1.28 & 1.28 & 1.26 & 1.17 & 1.17 & 1.30 & 1.28 & 1.19 & 1.20 \\
$c=0.5$ & 9.2 & 1.56 & 1.48 & 1.26 & 1.26 & 1.44 & 1.36 & 1.17 & 1.17 & 1.45 & 1.37 & 1.19 & 1.20 \\
$c=0.65$ & 7.8 & 1.79 & 1.54 & 1.26 & 1.25 & 1.68 & 1.44 & 1.19 & 1.19 & 1.69 & 1.47 & 1.21 & 1.21 \\
$c=0.8$ & 6.2 & 1.94 & 1.51 & 1.26 & 1.26 & 1.84 & 1.43 & 1.20 & 1.20 & 1.87 & 1.46 & 1.21 & 1.22 \\
$c=0.95$ & 3.9 & 2.50 & 1.50 & 1.36 & 1.35 & 2.42 & 1.45 & 1.31 & 1.31 & 2.45 & 1.47 & 1.33 & 1.32 \\
\addlinespace[2pt]
At least one fringe & 10.9 & 2.62 & 1.95 & 1.41 & 1.38 & 2.57 & 1.90 & 1.39 & 1.36 & 2.61 & 1.94 & 1.42 & 1.39 \\
Point scatterers & 18.4 & 2.63 & 2.50 & 2.10 & 2.08 & 1.92 & 1.84 & 1.67 & 1.65 & 1.02 & 1.00 & 1.00 & 0.99 \\
\addlinespace[2pt]
All & 8.8 & 1.63 & 1.41 & 1.27 & 1.27 & 1.56 & 1.36 & 1.23 & 1.23 & 1.58 & 1.38 & 1.25 & 1.25 \\
\bottomrule
\end{tabular}}
\caption{S1-79 on the oversampled grid and on the native grid of the same scenes (S1-79n, every second range sample kept, with its own calibration scenes). Phase linking error relative to the oracle of each configuration, groups, variants and columns as in Table~\ref{tab:overall}. The S1-79 block repeats Table~\ref{tab:overall}.}
\label{tab:native}
\end{table*}
\clearpage
\bibliographystyle{plainnat}
\bibliography{references}